\documentclass[structabstract]{aa}  
\usepackage{natbib}
\usepackage{lscape}
\usepackage{rotating}
\usepackage[T1]{fontenc}
\usepackage[utf8]{inputenc} 
\usepackage[toc,page]{appendix}
\usepackage{graphicx}
\usepackage{enumerate}
\usepackage{amsmath}
\usepackage{subcaption}
\usepackage{txfonts}
\usepackage{float}
\usepackage{textcomp}
\usepackage{gensymb}
\usepackage[colorlinks=true,linkcolor=blue,citecolor=blue]{hyperref}
\usepackage{xcolor}
\usepackage{mathtools}
\usepackage{ mathrsfs }
\usepackage{caption} 
\usepackage{tikz}
\usepackage[para,online,flushleft]{threeparttable}
\usepackage[normalem]{ulem}
\newcommand\boldred[1]{{\textcolor{black}{#1}}}
\newcommand\boldgreen[1]{{\textcolor{black}{#1}}}

\usepackage{xspace}

\begin{document}

  \title{X-ray luminosity function of high-mass X-ray binaries: Studying the signatures of different physical processes using detailed binary evolution calculations}

   \author{Devina\,Misra
           \inst{1,\boldred{2}}\fnmsep\thanks{e-mail: devina.misra@unige.ch},
           Konstantinos\,Kovlakas\inst{1, \boldred{3}, \boldred{4}},
           Tassos\,Fragos \inst{1,\boldred{5}},
           Margaret\,Lazzarini \inst{\boldred{6}}, 
           Simone\,S.\,Bavera \inst{1,\boldred{5}},
           Bret~D.~Lehmer \inst{\boldred{7}},
           Andreas Zezas \inst{\boldred{8,9,10}},
           Emmanouil\,Zapartas \inst{1, \boldred{11}},
           Zepei\,Xing \inst{1, \boldred{5}},
           Jeff~J.~Andrews \inst{\boldred{12,13}},
           Aaron\,Dotter \inst{\boldred{12}},
           Kyle~Akira~Rocha \inst{\boldred{12}},
           Philipp\,M.\,Srivastava \inst{\boldred{12,14}},
           Meng\,Sun \inst{\boldred{12}}
           }

   \authorrunning{Misra et al.}
    \titlerunning{Modeling the XLF of HMXBs}
   \institute{Département d'Astronomie, Université de Genève, Chemin Pegasi 51, CH-1290 Versoix, Switzerland
   \and
   \boldred{Institutt for Fysikk, Norwegian University of Science and Technology, Trondheim, Norway}
   \and
   \boldred{Institute of Space Sciences (ICE, CSIC), Campus UAB, Carrer de Magrans, 08193 Barcelona, Spain}
   \and 
   \boldred{Institut d’Estudis Espacials de Catalunya (IEEC), Carrer Gran Capit\`a, 08034 Barcelona, Spain}
   \and
    Gravitational Wave Science Center (GWSC), Université de Genève, 24 quai E. Ansermet, CH-1211 Geneva, Switzerland
   \and
   Division of Physics, Mathematics, and Astronomy, California Institute of Technology, Pasadena, CA 91125, USA
   \and Department of Physics, University of Arkansas, 226 Physics Building, 825 West Dickson Street, Fayetteville, AR 72701, USA
   \and
   Physics Department \& Institute of Theoretical \& Computational Physics, University of Crete, 71003 Heraklion, Crete, Greece
   \and
   Harvard-Smithsonian Center for Astrophysics, 60 Garden Street, Cambridge, MA02138, USA
   \and
   Institute of Astrophysics, Foundation for Research and Technology-Hellas, GR-71110 Heraklion, Greece
   \and
   IAASARS, National Observatory of Athens, Vas. Pavlou and I. Metaxa, Penteli, 15236, Greece
   \and Center for Interdisciplinary Exploration and Research in Astrophysics (CIERA), 1800 Sherman, Evanston, IL 60201, USA 
   \and Department of Physics, University of Florida, 2001 Museum Rd, Gainesville, FL 32611, USA
   \and Electrical and Computer Engineering, Northwestern University, 2145 Sheridan Road, Evanston, IL 60208, USA
     }

   \date{Accepted on February 16, 2023}

 
  \abstract
    {Many physical processes taking place during the evolution of binary stellar systems remain poorly understood. The ever-expanding observational sample of  X-ray binaries (XRBs) makes them excellent laboratories for constraining binary evolution theory. Such constraints and useful insights can be obtained by studying the effects of various physical assumptions on synthetic X-ray luminosity functions (XLFs) and comparing \boldred{them} with observed XLFs.} 
   {In this work we focus on high-mass  X-ray binaries (HMXBs) and study the effects on the XLF of various, poorly constrained assumptions regarding physical processes, such as the common-envelope phase, core collapse, and wind-fed accretion.}
   {We used the new binary population synthesis code {\tt POSYDON}, which employs extensive precomputed grids of detailed stellar structure and binary evolution models, to simulate the entire evolution of binaries. We generated 96 synthetic XRB populations corresponding to different combinations of model assumptions, including different prescriptions for supernova kicks, supernova remnant masses, common-envelope evolution, circularization at the onset of Roche-lobe overflow, and observable wind-fed accretion. }
   {The generated HMXB XLFs are feature-rich, deviating from the commonly assumed single power law. We find a break in our synthetic XLF at luminosity $\sim 10^{38}$~\boldred{erg~s$^{-1}$}, similar to observed XLFs. However, we also find a general overabundance of XRBs (up to a factor of $\sim$10 for certain model parameter combinations) driven primarily by XRBs with black hole accretors. 
   Assumptions about the transient behavior of Be XRBs, asymmetric supernova kicks, and common-envelope physics can significantly affect the shape and normalization of our synthetic XLFs. We find that less well-studied assumptions regarding the circularization of the orbit at the onset of Roche-lobe overflow and criteria for the formation of an X-ray-emitting accretion disk around wind-accreting black holes can also impact our synthetic XLFs and reduce the discrepancy with observations.}
   {Our synthetic XLFs do not always agree well with observations, especially at intermediate X-ray luminosities, which is likely due to uncertainties in the adopted physical assumptions. While some model parameters leave distinct imprints on the shape of the synthetic XLFs and can reduce this deviation, others do not have a significant effect overall. Our study reveals the importance of large-scale parameter studies, highlighting the power of XRBs in constraining binary evolution theory.}

   \keywords{X-rays: binaries -- binaries:  accretion -- stars: black holes -- neutron -- methods: numerical
               }

   \maketitle

\section{Introduction}\label{sec:intro}

X-ray binaries (XRBs) are systems with bright X-ray emission arising from the accretion of matter from \boldred{a} non-degenerate star onto a compact object (CO), such as a neutron star (NS) or a black hole (BH). Observations of XRBs with X-ray telescopes, including {\it Chandra}, {\it NuSTAR}, and {\it XMM-Newton}, have significantly contributed to our understanding of accreting CO sources \citep[e.g.,][]{2000ApJ...544L.101S, 2006ARA&A..44..323F,2012A&A...544A.118B, 2015A&A...579A..44D, 2017ARA&A..55..303K,2018ApJ...864..150V, 2019ApJ...884....2L}. Furthermore, the accumulation of large samples of extragalactic XRBs has allowed for statistical studies of their populations. X-ray luminosity functions (XLFs) are the primary tool for studying the statistical properties of observed XRBs. There is observational evidence that the properties of XRB populations (through the study of their XLFs) depend heavily on the host galaxy properties. For instance, the number and integrated luminosity of high-mass X-ray binaries  (HMXBs; XRBs with typical donor star masses $\gtrsim8\,\rm M_{\odot}$) scale with the star-formation rate (SFR) of the host galaxy due to their short lifetimes \boldgreen{of a few million years} \citep{2003MNRAS.339..793G, 2003A&A...399...39R, 2010ApJ...724..559L, 2010ApJ...716L.140A,2012MNRAS.419.2095M, 2016MNRAS.459..528A}. Low-mass X-ray binaries  (LMXBs; with typical donor star masses $\sim 1\,\rm M_{\odot}$), on the other hand, tend to have longer lifetimes ($\sim 10^9$~years) and thus do not correlate with the current local SFR. Instead\boldred{,} the integrated luminosity of LMXBs depends on the entire past star-formation history (SFH) of the host galaxy and scales with its total stellar mass \citep[integrated SFR over time;][]{2001ApJ...559L..97G,2001ApJ...559L..91P, 2002A&A...391..923G, 2004MNRAS.349..146G,2010ApJ...724..559L, 2011ApJ...729...12B}.

Observations of XRB populations in star-forming galaxies show that the number of XRBs, and their integrated X-ray luminosity, anticorrelate with the metallicity of the host galaxy \citep{2011ApJ...741...10K,2013ApJ...774..152B,2016MNRAS.457.4081B, 2016ApJ...818..140B, 2019ApJS..243....3L, 2021ApJ...907...17L, 2020MNRAS.498.4790K}. Metallicity affects the evolution of a binary through several processes. Lower metal content in stellar atmospheres leads to a decreased line-driven wind-mass loss \citep{2001A&A...369..574V}. This results, on average, in relatively heavy COs (exceeding $30$~M$_{\odot}$) at the end of the binary's life \citep[][]{2009MNRAS.395L..71M,2009MNRAS.400..677Z}. Furthermore, weaker stellar winds result in decreased orbital expansion due to orbital angular-momentum losses. Both of these factors lead to a more efficient accretion onto the COs and hence, on average, to more luminous XRB populations \citep{2013ApJ...764...41F, 2013ApJ...776L..31F,2016ApJ...818..140B}. In addition, low metallicity stars tend to expand less and later in their evolution than those with higher metallicities, reducing the chances of encountering a dynamically unstable mass-transfer episode, or if they do so, increasing their chances of surviving the common-envelope (CE) phases \citep{2010ApJ...725.1984L, 2016ApJ...818..140B}. Both effects tend to increase the formation efficiency of XRBs with decreasing metallicity. \citet{2006MNRAS.370.2079D} used Monte Carlo simulations and found the number of HMXBs increased by a factor of 3 in the Small Magellanic Cloud (SMC), with $Z/Z_{\odot}=0.2$, compared to solar \boldred{\citep[however, age effects are also in play;][]{2019ApJ...887...20A}}. Similarly, \citet{2010ApJ...725.1984L} used the {\tt StarTrack} code \citep{2008ApJS..174..223B} and found an increase of a factor of $3.5$ when going to $Z/Z_{\odot}=0.2$ for young bright HMXBs. The increased number of XRBs in low-metallicity environments, coupled with the metallicity evolution of the Universe \citep[e.g.,][]{2013ApJ...764...41F, 2013ApJ...776L..31F}, points \boldred{to} the possibility of XRBs having a significant effect on the ionization of the interstellar medium in the local Universe as well as the intergalactic medium in the early Universe \citep[e.g.,][]{2017ApJ...840...39M, Schaerer19,2022MNRAS.513.5097K,2022A&A...665A..28K}.

Generally, the XLFs of XRB populations are fitted with either a single or a broken power law depending on the relative contribution of HMXBs and LMXBs in the XRB population. Studying observations of XRBs in the Milky Way and Magellanic Clouds, \citet{2003MNRAS.339..793G} proposed first that the HMXB XLF was in the form of a single, smooth power law. They found no specific features in the XLF correlating with the Eddington limits of NSs and BHs but did find evidence of a high luminosity cutoff at $10^{40}$~erg~s$^{-1}$. \citet{2004ApJ...601L.147B} constructed the first XRB population synthesis models that closely reproduced the XLF of NGC 1569 (a starburst galaxy) when using a combination of an old (metal-poor) and young (metal-rich) stellar population. The findings of \citet{2003MNRAS.339..793G} were later confirmed by \citet{2012MNRAS.419.2095M}, who studied HMXB populations in 29 nearby star-forming galaxies. They found that the constructed XLF follows a power law with a slope of 1.6 in the X-ray luminosity range \boldred{$\log( L_{\rm X} / {\rm erg~s^{-1}}) \sim 35$ to 40}. \citet{2014MNRAS.437.1187Z} used an updated version of the \boldgreen{Binary Stellar Evolution} code \citep[{\tt BSE};][]{2000MNRAS.315..543H,2002MNRAS.329..897H,2007ChJAA...7..389L,2008MNRAS.387..121Z} to study XLFs of HMXBs in star-forming galaxies. They found that for a wide range of model parameters, the XRB population is dominated by wind-fed XRBs with BH accretors, which produces an XLF that closely resembles the observed one.

On the contrary, the XLFs of LMXB-dominated populations tend to fit better with broken power laws. \citet{Kim_2004} used a broken power law to fit the XLF of extragalactic LMXB populations observed with \textit{Chandra} and found that the best fit involved a break at $\sim 5\times10^{38}$~erg~s$^{-1}$, around the Eddington luminosity of a CO with mass $3.0$~M$_{\odot}$, and power law slopes of $\sim$1.8 and $\sim$2.8 below and above the break, respectively. 
\citet{2008ApJ...683..346F} constructed the theoretical XLFs of LMXBs and compared them with the elliptical galaxies NGC 3379 and NGC 4278 using the population synthesis code {\tt StarTrack} \citep{2008ApJS..174..223B}. They found that the XLF is dominated by binaries with NS accretors and giant (main sequence) donors above (below) luminosities of $\sim 10^{37}$~erg~s$^{-1}$. Furthermore, they concluded that although the XLFs depend on the adopted combination of the population synthesis parameters, the treatment of the transient sources is a crucial factor for the determination of the XLF -- for instance, modeling their variability instead of assuming a constant duty cycle for all sources, which would affect the number of LMXBs identified \citep{2009ApJ...702L.143F}.
Studying the XRBs in the bulge and ring of the ring galaxy NGC 1291 (which is dominated by an old stellar population with little recent star formation), \citet{2012ApJ...749..130L} reached similar conclusions. In a related study, \citet{2013ApJ...774..136T} compared synthetic with observed XLFs from a sample of 12 nearby, late-type galaxies from the \textit{Spitzer} Infrared Nearby Galaxy Survey \citep{2003PASP..115..928K}. They performed a broad model parameter space study, varying parameters such as the CE efficiency, stellar wind strengths, \boldred{and} supernova (SN) kick velocity distributions, and explored the relative contribution to the XLF from different XRB subpopulations as a function of the SFH of the parent stellar population. 

The evolution with the age of the XLF shape and integrated luminosity of an LMXB population has been a topic of both theoretical and observational studies. Theoretical studies of LMXB populations predict that their XLF flattens in younger stellar populations, while the integrated X-ray luminosity can increase by up to one order of magnitude \citep[e.g.,][]{2008ApJ...683..346F,2013ApJ...764...41F,2013ApJ...776L..31F}. Initial observational results contradicted the theoretical predictions, with \citet{2012A&A...546A..36Z} finding a higher number of LMXBs per unit of stellar mass in older elliptical galaxies compared to younger ones. This discrepancy arose from the inclusion of globular cluster LMXBs in the \citet{2012A&A...546A..36Z} study and the underlying correlation of a specific globular cluster frequency with stellar mass in the observed galaxy sample, which biased the results. Other observational investigations \citep{2009ApJ...703..829K,2014ApJ...789...52L,2020ApJS..248...31L} found that when removing the globular cluster LMXB subpopulations, more luminous LMXBs indeed dominate younger populations, confirming the theoretical predictions. \citet{2017ApJ...851...11L} studied far-UV to far-IR data of XRB sources in the spiral galaxy M51 and constructed an age-dependent XLF. They found that both single and broken power laws were good fits \boldred{for} the data. Generally, XLFs (for X-ray luminosities $\gtrsim 10^{37}$~erg~s$^{-1}$) of older populations are found to be steeper than for younger populations, which complements the fact that older populations do not reach X-ray luminosities as high as those reached by younger populations that have higher numbers of HMXBs \citep{2010ApJ...721.1523K,2014ApJ...789...52L, 2022ApJ...926...28G}. \citet{2019ApJS..243....3L} used a sample of 38 local galaxies (within a radius of $30$~Mpc) from {\it Chandra} \boldgreen{Advanced CCD Imaging Spectrometer} (ACIS) imaging data, covering a broad range of SFRs and stellar masses to constrain the XLFs of HMXBs and LMXBs. They fitted both single and broken power law models to their data. For HMXB XLFs, they found the shape to be more complex than a single power law. The XLF slope rapidly declined beyond luminosities of $10^{38}$~erg~s$^{-1}$, which is consistent with results from \citet{2013ApJ...774..136T}, \citet{2014MNRAS.437.1187Z}, and \citet{2019IAUS..346..332A}.

The aforementioned studies imply that generalizing the observed XLF of an arbitrary XRB population is a difficult task and involves many implicit assumptions about the underlying binary population. Many of these studies focused on either LMXBs or HMXBs, as they have distinct observational and physical characteristics \citep{2006csxs.book..623T}. High-mass XRBs are some of the brightest X-ray sources in the local Universe. In our study, we focus on young XRB populations that are dominated by HMXBs and the effects of various physical assumptions on their XLFs. There are generally three broad types of HMXBs: {\it (i)} binaries with a super-giant donor undergoing wind-fed accretion, {\it (ii)} binaries with a super-giant donor star undergoing Roche-lobe overflow (RLO), and {\it (iii)} binaries with a fast spinning Be-star donor. In the case of wind-fed accretion, a fraction of the mass leaving the donor star in the form of strong stellar winds is captured by the gravitational pull of the CO \citep{1944MNRAS.104..273B}. For XRBs undergoing RLO, the donor overfills its Roche lobe and transfers mass to the CO through the inner Lagrangian point. The transferred material carries enough angular momentum, which prevents it from being accreted immediately, forming an accretion disk around the CO. Be XRBs are characterized by rapidly spinning O/B donors with a disk around their equator, called the decretion disk \citep{2020MNRAS.493.2528B}. If the binary is close enough that the CO interacts with the decretion disk, the CO may accrete material from the donor without the donor filling its Roche lobe or the presence of any strong stellar wind. 

Naturally, the observed properties of XRB populations will be affected by the formation environment and subsequent evolution of these sources. Some studies have explored the effects of binary physics on XRB observables. \citet{1995MNRAS.274..461B} studied the post-SN kick velocities for XRB progenitors and found a tight correlation between the post-SN orbital periods and eccentricities of LMXBs, with LMXBs that have periods longer than a few days getting large eccentricities. They also found that a significant fraction of HMXBs would have precessing accretion disks, which would have implications on XRB observations.  \citet{2011ASPC..447..121L} and \citet{2014ApJ...797...45Z} investigated the effect of the CE phase on XRB populations and find that the constrained quantities for the CE efficiency depend on various assumptions, such as SN kicks, metallicity, or bolometric correction factors, that introduce uncertainties.

Despite the uncertainties, it is critical that theoretical models are able to reproduce observed XLFs if we aim to use them to infer the properties of the underlying XRB populations. X-ray luminosity functions contain signatures of \boldred{the} prior evolution of the XRBs that can be leveraged to constrain uncertain stellar and binary physics. The different processes involved in binary evolution directly impact the properties of additional sources that are of interest to the scientific community, such as gravitational wave sources and electromagnetic transients (e.g., gamma-ray bursts and stripped SNe), as all of them will have one or more XRB phases in their evolution \citep{2019BAAS...51c.304Z}. 

In this work we present, for the first time, population synthesis models of HMXBs that employ detailed stellar structure and binary evolution calculations throughout the entire evolution of the binaries. We carry out a parameter study of physical assumptions that affect the observed HMXB population. The different physical assumptions that we investigate iterate through various prescriptions, dictating aspects of evolution, such as SN kicks, the CE phase, and wind-fed accretion. In Sect.~\ref{sec:numericaltools} we discuss the population synthesis code used throughout this work ({\tt POSYDON}), the physics employed for the simulations, and the different model parameters we investigate. In Sect.~\ref{sec:results} we construct XLFs from the simulated populations and investigate the main effects of the different model parameters. Section~\ref{sec:discussion} discusses the comparison of different models to the observed XLF \boldred{from \citet{2019ApJS..243....3L}}, the combined effects of certain parameters on the simulated XLF, and some possible reasons for any disparity between simulations and observations. Finally\boldred{,} we present our concluding remarks in Sect.~\ref{sec:conclusion}.

\section{Method and physical assumptions}\label{sec:numericaltools}

For our study, we used \boldgreen{the POpulation SYnthesis with Detailed binary-evolution simulatiONs code} \citep[\boldgreen{{\tt POSYDON};} see][for a detailed description of the code]{2023ApJS..264...45F}, a new binary population simulation framework that combines the flexibility of parametric binary population synthesis codes with detailed stellar structure and binary evolution models. Crucial to our study, {\tt POSYDON} enables a physically accurate and self-consistent determination of the stability and mass-transfer rate evolution of mass-transfer phases thanks to extensive precomputed grids of binary-star models \citep[using the \boldgreen{Modules for Experiments in Stellar Astrophysics code or }\texttt{MESA} code;][]{2011ApJS..192....3P,2013ApJS..208....4P,2015ApJS..220...15P,2018ApJS..234...34P,2019ApJS..243...10P}. Furthermore, the available information on the internal structure of binary components enables an improved treatment of evolutionary phases, such as the CE and the core collapse. Of special interest here is also the capability of {\tt POSYDON} to track the angular momentum content of both stellar components throughout the evolution of a binary, which allows us to identify potential Be XRBs, as we describe further below.

\subsection{Initial binary properties}\label{sec:ini_properties}

The target of this study is to construct synthetic XLFs of XRB populations dominated by HMXBs \boldred{ and compare them to observed HMXB XLFs from \citet{2019ApJS..243....3L}. High-mass XRBs are associated with young ($\lesssim 100\,\rm Myr$) stellar populations, it could be argued that there would be observational effects from the wide range of ages of the consisting populations that would interfere with the comparison. While the SFH is expected to play a significant role in the XRB content of a galaxy, XLFs computed by combining observations from multiple galaxies are averaging out the effects from the varying ages in the populations. The observed XRB sample is composed of 38 galaxies, the majority of which have masses in the same order of the mass as the Milky Way \citep[$\log(M/{\rm M_{\odot}}) \approx 10.80$;][]{2015ApJ...806...96L}. These galaxies are expected to have nearly uniform SFHs, and consequently, we expect the cumulative SFH of the sample \boldred{to be} nearly constant. Now, the star formation indicators that were used to measure the SFR of the galaxies in the sample, have some implicit assumptions about the SFH of the respective populations, which was a constant SFR for a duration of 100~Myr. To compare our synthetic XLFs to the observed one in the most self-consistent way, we assumed the same SFH.} Thus, using {\tt POSYDON}, we simulate synthetic binary populations corresponding to continuous star formation for 100\,Myr. All the populations are run at solar metallicity \citep[0.0142; ][]{2009ARA&A..47..481A}.

The primary star is the initially more massive binary component at the zero-age main sequence (MS
) and is sampled from the initial mass function from \citet{2001MNRAS.322..231K} within the range $7.0$ to $120.0$~M$_{\odot}$. The secondary's mass is calculated from the mass ratio, which is sampled from a flat distribution \boldred{\citep{2013A&A...550A.107S}}. We assumed independent distributions of the mass ratio and period; however, recent studies show that there might be a dependence of mass ratios on the orbital periods \citep{2017ApJS..230...15M}. The initial orbital period is drawn from a power law in log-space \boldred{\citep{2013A&A...550A.107S}} with the same limits as the {\tt POSYDON} grids \citep{2023ApJS..264...45F} and the initial orbits are taken as circular since the current {\tt POSYDON} version only follows the mass-transfer phases for circular binaries. The binary fraction assumed is $0.7$ \citep{2012Sci...337..444S}. We normalize the generated population by $f_{\rm corr} (\approx 5.89$) to compensate for the fact that our initial parameters do not cover the entire existing masses in the Universe (see Appendix A in \boldred{\citealp{2020A&A...635A..97B}} for the detailed calculation of $f_{\rm corr}$).

\subsection{X-ray luminosity calculation}

We selected binaries that consist of a CO and a non-degenerate star and identified as XRBs those that have a calculated bolometric X-ray luminosity greater than $10^{35}$~erg~s$^{-1}$. For XRBs undergoing RLO or wind-fed mass accretion, the bolometric X-ray luminosity ($L^{\rm RLO/wind}_{\rm bolometric}$) for sub-Eddington mass-transfer rates is calculated as follows,
\begin{equation}
    L^{\rm RLO/wind}_{\rm bolometric} = \eta \dot{M}_{\rm acc} c^2, \ \ \ \ \text{if } \dot{m} \leq 1.0,
\end{equation}
where $\dot{M}_{\rm acc}$ is the mass-accretion rate, $\eta$ is the radiative efficiency of accretion, $c$ is the speed of light, and $\dot{m}\equiv \dot{M}_{\rm tr}/\dot{M}_{\rm Edd}$ is the Eddington ratio. The mass-transfer rate, $\dot{M}_{\rm tr}$, from the donor to the vicinity of the accretor is assumed to be equal to \boldred{the} accretion rate for $\dot{m} \leq 1.0$. The mass-transfer rate, for XRBs undergoing accretion from stellar wind (not including Be XRBs that are treated separately)\boldred{,} is calculated by following the \citet{1944MNRAS.104..273B} accretion model, while for XRBs undergoing RLO\boldred{,} it is computed self-consistently in our grids of detailed binary evolution simulations. The radiative efficiency, $\eta$, is the fraction of the rest mass energy of accreted matter that is radiated away and is defined as\begin{equation}
    \eta = \frac{GM_{\rm acc}}{R_{\rm acc} c^2},
\end{equation}
where $M_{\rm acc}$ is the CO mass, $R_{\rm acc}$ is the CO radius (for NS accretors) or the spin-dependent innermost stable circular orbit \citep[from][for BH accretors]{2003MNRAS.341..385P}, and $G$ is the gravitational constant.

We use the accretion disk model by \citet{1973A&A....24..337S} to define the luminosity from super-Eddington mass-transfer rates. This model describes the X-ray luminosity for an accretion disk receiving material at super-Eddington rates, with strong outflows keeping the disk locally Eddington limited. For very high mass-transfer rates, $\dot{M}_{\rm tr}>8.5\times \dot{M}_{\rm Edd}$; $\dot{M}_{\rm Edd}$ being the Eddington limit, \citet{2001ApJ...552L.109K} and \citet{2009MNRAS.393L..41K} introduced a beaming effect, due to the presence of a geometrically thick accretion disk, in the form of a beaming factor to account for the collimation of outgoing emission. Therefore, X-ray luminosity from a super-Eddington accretion disk (that includes geometrical beaming) is described as follows,
\begin{equation}\label{eq:superedd}
    L^{\rm RLO/wind}_{\rm bolometric} = \frac{L_{\mathrm{Edd}}}{b}(1 + \ln{\dot{m}}), \ \ \ \ \text{if } \dot{m} > 1.0,
\end{equation}
where $b$ is the beaming factor describing the amount of collimation to the outgoing radiation due to geometrical beaming from the accretion disk. The approximate value of $b$ following \citet{2009MNRAS.393L..41K} is defined as
\begin{equation}\label{eq:beaming}
        b= 
\begin{dcases}
    \frac{73}{\dot{m}^2},& \text{if } \dot{m}> 8.5,\\
    1,              & \text{otherwise}.
\end{dcases}
\end{equation}
The bolometric luminosity calculated by Eq.~\ref{eq:superedd} would be perceived as super-Eddington if assumed to be isotropic. Since the model described above is only approximate, very high mass-transfer rates may provide unphysical strong beaming ($b \ll 10^{-3}$). To avoid this, we set a lower limit for $b$ at $3.2\times 10^{-3}$, which roughly corresponds to a $\dot{m} \approx 150.0$ \citep[see also][]{2016A&A...587A..13L, 2017ApJ...846...17W}. 

Using similar criteria as \citet{2009ApJ...707..870B} and \citet{2014MNRAS.437.1187Z}, we identified Be XRBs by selecting binaries with fast spinning (surface velocity $\gtrsim 70\%$ of the critical surface velocity) MS donors, orbital periods in the range $10$ to $300$~days, and in a detached state (not undergoing RLO). We only included binaries with donor masses $\gtrsim 6.0$~M$_{\odot}$, which is a conservative estimate for the minimum mass of B stars \citep{2010AN....331..349H}. An additional criterion that we used to identify Be XRBs is the donor decretion disk radius exceeding the Roche-lobe radius of the Be star, at which point the CO would start to accrete matter from the decretion disk. The Roche-lobe radius of the Be stars is calculated at the periastron of their eccentric orbits. \citet{2017A&A...601A..74K} estimated the outer edges of the decretion disk at around stellar radii of $30$ to $150$~R$_{*}$, following which we use $100$~R$_{*}$ as an approximate decretion disk radius (R$_{*}$ being the donor radius). The X-ray luminosities of Be XRBs are calculated following \citet[][Eq.~11]{2006ApJ...653.1410D}, which describes the correlation between peak observed luminosities for Be XRBs and the orbital periods \citep[observational data from][]{2005A&AT...24..151R},
\begin{equation}\label{eq:be_xrb}
    \log_{10}\bigg( \frac{L^{\rm Be-XRB}_{\rm bolometric}}{10^{35} {\rm erg s}^{-1} } \bigg) = 4.53 \pm 0.66 - (1.50 \pm 0.33) \log_{10}\bigg( \frac{P_{\rm orb}}{1 \rm day} \bigg).
\end{equation}
As Be XRBs are transient in nature, we included a duty cycle of $10\%$ in our calculations \citep[][however see also \citealt{2011Ap&SS.332....1R} where duty cycles up to $30\%$ are observed]{2019IAUS..346..178S}. 

In order to compare the calculated bolometric luminosities ($L_{\rm bolometric}=\{L^{\rm RLO/wind}_{\rm bolometric},L^{\rm Be-XRB}_{\rm bolometric}\}$) to X-ray luminosities that are in the {\it Chandra} band 0.5 to 8~keV ($L_{\rm X}$), they are normalized using a bolometric normalization of $L_{\rm X,[0.5-8\,\rm KeV]}/ L_{\rm bolometric} \approx 0.5$, which is a zeroth order approximation of the {\it Chandra} bolometric correction estimates from \citet{2013ApJ...764...41F} and \citet{2022MNRAS.513.1400A}. The following subsections, Sects. \ref{sec:sn_mech} to \ref{sec:wind_hmxb_crit}, describe all the considered parameters (summarized in Table~\ref{table:params}).

\begin{table*}[]
\centering
\begin{tabular}{l|lll}
\hline\hline
Parameters                     & \multicolumn{3}{c}{Parameter options} \\ \hline
Remnant mass prescription                  & \citet{2020MNRAS.499.2803P}         & \citet{2012ApJ...749...91F} (Delayed)         &      \\
Asymmetric natal kick normalization    &    BH mass normalized kicks      & Fall-back normalized kicks       & No kick normalization      \\
Orbit circularization at RLO & Periastron         & Conserved angular momentum          &          \\
CE efficiency ($\alpha_{\rm CE}$)        & 0.3         & 1.0           &         \\
CE core-envelope boundary   & At $X_{\rm H} = 0.01$        & At $X_{\rm H} = 0.30$          &        \\
Observable wind-fed disk          & \citet{2021PASA...38...56H}         & No criterion             &         
\\\hline
\end{tabular}
\caption{All the physics parameters provided by {\tt POSYDON} that were used for this study.}
\label{table:params}
\end{table*}

\subsection{Remnant mass prescription}\label{sec:sn_mech}
At the end of its evolution, a massive star creates an iron core that collapses resulting in the formation of a NS or a BH, depending on the properties of the pre-collapse star. The grids of detailed single- and binary-star models in {\tt POSYDON} follow the stars' evolution until their core-carbon exhaustion, beyond which we employ prescriptions that dictate the core-collapse outcomes. We used the following two prescriptions to determine the ``explodability'' and final remnant properties of the pre-SN stars.

First, the prescription from \citet{2020MNRAS.499.2803P} determines the final CO by treating the carbon-oxygen core of an evolved star separately from the rest of the star. \citet{2020MNRAS.499.2803P} calculated the explodability of stripped carbon-oxygen stars, for a range of masses and compositions, and derived a prescription for the outcome of the core collapse based on the final carbon-oxygen core mass and the composition of the carbon-oxygen core at carbon ignition. \boldred{The masses of the newly formed COs are taken as the pre-SN He-core masses of the stars (accounting for any fall back, further described in the next section), assuming loss of any envelope during previous evolution.}
    
Second, \citet{2012ApJ...749...91F} \boldgreen{predicted} the core-collapse outcome based on the pre-SN carbon-oxygen core mass. \boldgreen{Following their prescription,} the newly formed CO masses are taken as the pre-SN masses of the stars, after taking account of any fall back. The authors provided ``rapid'' and ``delayed'' mechanisms for core collapse. The rapid mechanism predicts the presence of a mass gap in remnant masses in the range $2.0$ to $5.0$~M$_{\odot}$.
    The delayed mechanism produces a continuous range of NS and BH masses. Due to observations of COs that fall within the aforementioned BH mass gap \citep{2019Sci...366..637T, 2021arXiv211103606T}, we focus on the delayed mechanism in our study.

\subsection{Asymmetric natal kick normalization} \label{sec:sn_kick}
The collapse of the iron core is accompanied by a SN (unless the core implodes without any explosion taking place). If the SN explosion is asymmetric, due to either asymmetric mass loss \citep{1994A&A...290..496J, 1996PhRvL..76..352B, 2017ApJ...837...84J} or neutrino loss \citep{1993A&AT....3..287B,2017ApJ...837...84J}, the newly formed CO experiences a natal kick. We assume the magnitude of the kicks imparted are drawn from a Maxwellian distribution with the velocity dispersion $\sigma = 265.0$~km~s$^{-1}$, for core-collapse SNe \citep[from observed velocities of radio pulsars;][]{2005MNRAS.360..974H}, or $\sigma = 20.0$~km~s$^{-1}$, for electron-capture SNe \citep[][]{2019MNRAS.482.2234G}. \boldred{However, the true distribution of natal kicks is still uncertain \citep{2017A&A...608A..57V, 2021MNRAS.508.3345I, 2021ApJ...920L..37W, 2023MNRAS.519.5893K}}. 

Furthermore, we normalized the velocities based on the properties of the CO formed to calculate the final kick velocity, $v_{\rm kick}$,

\begin{equation}
    v_{\rm kick} = n v_{\rm mag},
\end{equation}
where $n$ is the normalization and $v_{\rm mag}$ is the velocity magnitude drawn from the Maxwellian distribution. The distribution of the kick directions is isotropic. We study three prescriptions for natal asymmetric kicks to calculate the kick normalization ($n$):
(\textit{i}) The natal kicks are normalized by the mass of the created BH using the following normalization,
    \begin{equation}
        n = \frac{1.4 \rm M_{\odot}}{M_{\rm BH}}
    ,\end{equation}
    where $1.4$~M$_{\odot}$ represents the mass of a typical NS and $M_{\rm BH}$ is the BH mass. With increasing BH masses, the imparted kicks are weaker. \boldred{The NS kicks are not rescaled.} (\textit{ii}) The natal kicks are normalized by the fall-back fraction using the following normalization,
    \begin{equation}
        n = 1 - f_{\rm fb},
    \end{equation}
    where $f_{\rm fb}$ is the fall-back fraction, which is the fraction of the stellar material that is not driven away by the SN explosion and falls back on the newly formed CO. Following the \citet{2020MNRAS.499.2803P} mechanism, $f_{\rm fb}=1.0$ for all BHs. Following the \citet{2012ApJ...749...91F} mechanism, $f_{\rm fb}=1.0$ for stellar core masses (of the pre-collapse star) greater than $11$~M$_{\odot}$ and for masses lower than $11.0$~M$_{\odot}$, $f_{\rm fb} < 1.0$. \boldred{Hence, whenever $f_{\rm fb} = 1.0$, there is no baryonic mass loss considered, while still accounting for neutrino losses \citep[which is capped at 0.5~M$_{\odot}$][]{2023ApJS..264...45F}.}
    \boldred{For NSs, when the SN mechanism follows the prescription from \citet{2012ApJ...749...91F}, the fall-back fraction for NSs is $f_{\rm fb} < 1.0$ (however, for NSs from electron-capture SN, $f_{\rm fb}=0$). When the SN mechanism follows the prescription from \citet{2020MNRAS.499.2803P}, it is assumed that there is no fall-back on the NS \citep[$f_{\rm fb}=0.0$;][]{2023ApJS..264...45F}.}    
     (\textit{iii}) The natal kicks are not normalized for BHs ($f_{\rm fb}=0$, therefore, $n=1$). Hence, the BHs receive \boldgreen{strong} kicks. Neutron star kicks are normalized using the fall-back fraction.

\subsection{Circularization of the orbit at the onset of Roche-lobe overflow}
\label{sec:circi_efefct_des}
We check for potential initiation of the RLO in a binary at the periastron of the orbit. At the onset of RLO, we assume that the orbit is circularized as tidal forces circularize the orbit at short timescales \citep[about $10^4$~years for a semidetached binary with a giant donor;][]{1989A&A...223..112Z, 1995A&A...296..709V}. Following the circularization, {\tt POSYDON} evolves the mass-transfer phase utilizing the precomputed model grids of circular binaries. The separation of the circularized binary will determine the stability and duration of the mass-transfer phase, and thus its observability as an XRB, thereby affecting the XLF. We investigated two options for the circularized binary separation.\ First, the final separation of the circular binary was taken to be the ``periastron'' of the eccentric orbit at the onset of RLO. The orbit will lose both energy and angular momentum as it circularizes and the new circularized binary separation ($a_{\rm circ}$) will be    
\begin{equation}\label{eq:peri}
        a_{\rm circ} = a(1-e),
\end{equation}
where $a$ is the semimajor axis and $e$ is the eccentricity of the eccentric orbit.

Second, the binary conserves its angular momentum and loses only energy as it circularizes, and the new circularized orbital separation is larger than the periastron distance of the originally eccentric orbit. For an eccentric binary with a semimajor axis $a$ and an eccentricity $e$, the new circular binary separation $a_{\rm circ}$) is calculated imposing the circular angular momentum is equal to \boldred{t}he eccentric angular momentum:
    \begin{equation}
        a_{\rm circ} = a(1-e^2).
        \label{eq:alt}
    \end{equation}

\boldred{In practice, both of these approaches are approximations of a potentially much more complicated process, encompassing the range of outcomes of the detailed treatment, and thus will give us some insight into the potential signatures that this process imprints on the XLF. In order to effectively check the validity of these assumptions, we would need a better treatment of eccentric orbits in mass-transferring binaries, which we do not currently have. A comprehensive theoretical framework of the secular evolution of eccentric, mass-transferring binaries has been recently developed \citep[][]{2007ApJ...667.1170S,2009ApJ...702.1387S, 2010ApJ...724..546S, 2016ApJ...825...70D, 2016ApJ...825...71D} but has not been applied yet in the context of detailed binary evolution models.}

\subsection{Common-envelope evolution}
\label{sec:ce_alpha}
During a mass-transfer phase, if a dynamical instability is encountered, a CE phase is initiated, where the companion of the expanding star is engulfed by its envelope. The star that is spiraling inside the envelope loses orbital energy as frictional heat is deposited into the envelope, which increases the internal temperature of the envelope and it expands. When enough energy is deposited into the envelope during the inspiral, the envelope is expelled \citep[refer to][for a review of the CE]{2013A&ARv..21...59I}.
We parameterize CE using the so-called $\alpha_{\rm CE}$--$\lambda_{\rm CE}$ prescription \citep{1984ApJ...277..355W, 1988ApJ...329..764L} to predict the outcome of the CE phase. This prescription calculates the change in the orbital energies before and after the CE ($E^{i,f}_{\rm orb}$) and compares it to the binding energy ($E_{\rm bind}$) of the envelope. It follows as
\begin{align}\label{eq:ce_eq}
    E_{\rm bind} &= \alpha_{\rm CE}(E^{i}_{\rm orb} - E^{f}_{\rm orb}) \\
    \Rightarrow G\frac{M_{1}M_{1, \rm env}}{\lambda_{\rm CE} R_{1}} &= \alpha_{\rm CE} \bigg( -\frac{GM_{1}M_{2}}{2a_{\rm i}} + \frac{GM_{1,\rm c}M_{2}}{2a_{\rm f}} \bigg),
\end{align}where $\alpha_{\rm CE}$ (known as the CE efficiency) is the fraction of orbital energy that is injected into the envelope, $\lambda_{\rm CE}$ is a factor that accounts for the structure of the envelope of the mass-losing star \citep{1990ApJ...358..189D} known as the binding energy parameter, $M_1$ and $M_2$ are the masses of the two stars (with the subscripts ``env'' and ``c'' denoting the envelope and core mass of the respective star). If the predicted final orbital separation after the envelope ejection $a_{\rm f}$ is such that either the stellar core of the donor or the accretor is overfilling their respective Roche lobes, we consider the envelope ejection to be unsuccessful assuming the two bodies would merge. If neither \boldred{is} filling their Roche lobe, we consider the envelope ejection to be successful resulting in a detached, circular, and close binary.

For a high CE efficiency parameter ($\alpha_{\rm CE} \sim 1.0$) since most of the available orbital energy can be used to unbind the CE, the envelope can be ejected with the binary shrinking less than for lower efficiency. Ultimately, an efficient CE leads to relatively wide post-CE orbits and allows lower mass companions to eject the envelope of the CO progenitor star.

We investigate the following two values for $\alpha_{\rm CE}$.\ First, to investigate the case when the energy conversion is not perfectly efficient, we study a population with $\alpha_{\rm CE}=0.3$ to see the effect of a low efficiency on the CE phase, as some studies show $\alpha_{\rm CE}$ in the range $0.1$ to $0.5$ \citep{2010A&A...520A..86Z, 2010MNRAS.403..179D, 2015MNRAS.450L..39N}.
Second, to investigate the effect of full availability of orbital energy to eject the envelope, we used $\alpha_{\rm CE} = 1.0$. 

A value of $\alpha_{\rm CE}>1.0$ might point to extra sources of energy to eject the envelope due to uncertainties in the CE energy budget. Equivalently, it might also imply that the binding energy of the envelope is lower than that estimated from single-star models, as the stellar structure will differ for stars undergoing CE compared to undisturbed single stars \citep{2019ApJ...883L..45F, 2021A&A...645A..54K, 2021A&A...650A.107M}, \boldred{\citep{2023ApJ...942L..32R}}. In our study, we adopt the second interpretation and explore different estimates of the binding energy further below. Additionally, for CE donors in the Hertzsprung gap, there is no well-formed, steep boundary between the core and envelope. In this case, the inspiral never stops and the CE results in a binary merger \citep{2007ApJ...662..504B}. This is the ``pessimistic'' approach when estimating the outcome of the CE \citep{2000ARA&A..38..113T, 2004ApJ...601.1058I}.



The binding energy ($E_{\rm bind}$) of a stellar envelope is often characterized by the binding energy parameter ($\lambda_{\rm CE}$), which in turn is dependent on how the He-rich core of the donor is defined. To begin with, $\lambda_{\rm CE}$ is defined as follows,
\begin{equation}
    \lambda_{\rm CE} = -\frac{GMM_{\rm env}}{E_{\rm bind}R},
\end{equation}
where $M$ and $M_{\rm env}$ are the stellar mass and envelope mass of the donor, and $R$ is the stellar radius. The envelope binding energy, and hence also the $\lambda_{\rm CE}$, are dependent on where we set the core-envelope boundary in the mass profile of the donor star and its value changes as the star evolves \citep{1990ApJ...358..189D}. In the context of the CE phase, the core-envelope boundary separates the part of the star that will be ejected, the envelope, from the remaining stripped core, which will contract and allow the binary to detach. The position of the boundary is often parameterized as the first point in mass where the fraction of H drops below an adopted threshold. 
The core-envelope boundary is not a well-constrained value, depending highly on the evolutionary stage of the donor and investigations show that the choice of the boundary will have different results on the CE outcome \citep{2000A&A...360.1043D,2001A&A...369..170T,2011ASPC..447...91I,2016A&A...596A..58K}. While traditionally this boundary is set where the hydrogen (H) abundance ($X_H$) drops at least below $10\%$, recent studies of COs inspiraling within the envelope of a supergiant star suggested that the core-envelope boundary might be located further out  \citep{2019ApJ...883L..45F, 2019A&A...628A..19Q, 2021A&A...645A..54K, 2021A&A...650A.107M}. The binding energy of the envelope will be higher (or lower) if we set the boundary at a deeper (or outer) mass coordinate. Therefore, by changing the boundary we implicitly change the $\lambda_{\rm CE}$, and the values for the boundary that we study are at H-abundance of $0.01$ and $0.3$. The use of stellar profiles from detailed grids in {\tt POSYDON} aids in a physically motivated estimation of $\lambda_{\rm CE}$.  

\subsection{Criteria for observable wind-fed accretion}\label{sec:wind_hmxb_crit}
For wind-fed HMXBs, X-ray emission is produced when the accretor captures and accretes a fraction of the donor's stellar wind. One can estimate the amount of accreted material using the Bondi-Hoyle approximation, which assumes that the accretor moves through the locally isotropic wind of the mass-losing companion \citep[][]{1944MNRAS.104..273B}. In this mechanism, the amount of accreted material is proportional to the rate at which the companion star is losing mass in stellar winds, inversely proportional to the orbital separation squared, and also inversely proportional to the fourth power of the wind velocity. This makes it more efficient for binaries with giant-star companions, which tend to have sufficiently intense and slow stellar winds.

In contrast to NS accretors, which have a hard surface, material accreting onto a BH has to first form an accretion disk before falling into the BH's horizon, in order to efficiently dissipate its gravitational energy in the form of electromagnetic radiation and become a bright X-ray source. \citet{2021PASA...38...56H} investigated the BH HMXB Cygnus~X-1, which does not follow the simplified picture presented by the Bondi-Hoyle model, as it has a focused accretion stream \citep{2005ApJ...620..398M, 2009ApJ...690..330H}. The accretion stream is produced due to the high Roche-lobe filling factor of the companion and is aided by the rotation of the donor, which reduces wind terminal velocity and promotes the capturing of the wind leaving the equator.
 
When this stream has high enough angular momentum, material forms an accretion disk before falling into the BH, producing X-rays in the process. \citet{2021PASA...38...56H} \boldgreen{estimated} that for an HMXB to have observable X-ray emission from wind-fed accretion, the donor should fill its Roche lobe by at least $80$ to $90\%$. We identify observable wind-fed XRBs as XRBs (having MS donors) where the donor fills its Roche lobe by at least $80\%$. In our population study, we explore two cases.\ First, we used the \citet{2021PASA...38...56H} criteria and identified wind-fed XRBs where the donor fills its Roche lobe by at least $80\%$. Second, we did not use this \boldred{criterion} for defining the creation of an accretion stream.


\begin{table}[]
\begin{tabular}{l|l}
\hline\hline
Parameters                     & \multicolumn{1}{c}{Parameter options} \\ \hline
Remnant mass prescription                  & \citet{2020MNRAS.499.2803P}                                   \\
Natal kick normalization    & BH mass normalized kicks                                \\
Orbit circularization at RLO & Periastron                                   \\
CE efficiency ($\alpha_{\rm CE}$)                  & 1.0                                    \\
CE core-envelope boundary   & At $X_{\rm H} = 0.3$                                   \\
Observable wind-fed disk          & No criterion
\\\hline
\end{tabular}
\caption{Parameters of the default population.}
\label{table:params_def}
\end{table}

\section{Results}\label{sec:results}

Each of the parameters explored (Table~\ref{table:params}) probes different aspects of binary evolution and might have an effect on the resulting population of simulated HMXBs, and thus the synthetic XLF. We present all the 96 constructed synthetic XLFs (cumulative and differential) in Appendix~\ref{appendix:graph}. To investigate these effects, we compared the XLFs from the simulated populations to a ``default'' XLF that follows the parameters shown in Table~\ref{table:params_def} and corresponds to model~15 in Appendix~\ref{appendix:graph}. In constructing the synthetic XLFs we need to take \boldred{into} account any directional X-ray emission. For instance, Eq. \ref{eq:beaming} describes the geometrical beaming of the X-ray emission in \boldred{the} case of super-Eddington mass transfer, which implies that some XRBs would be less visible due to non-preferential line-of-sight angles with respect to the rotational axis of the accretion disk. Therefore, in constructing the XLF we down-weight the beamed sources using the beaming factor described in Eq.~~\ref{eq:beaming}. We down-weighted the Be-XRB sources in accordance with a duty cycle of $10\%$. Additionally, we approximated all calculated bolometric X-ray luminosities to X-ray luminosities in the {\it Chandra} band in the range 0.5 to 8~keV.   

\begin{figure}[!ht]
\centering
\includegraphics[width=\linewidth]{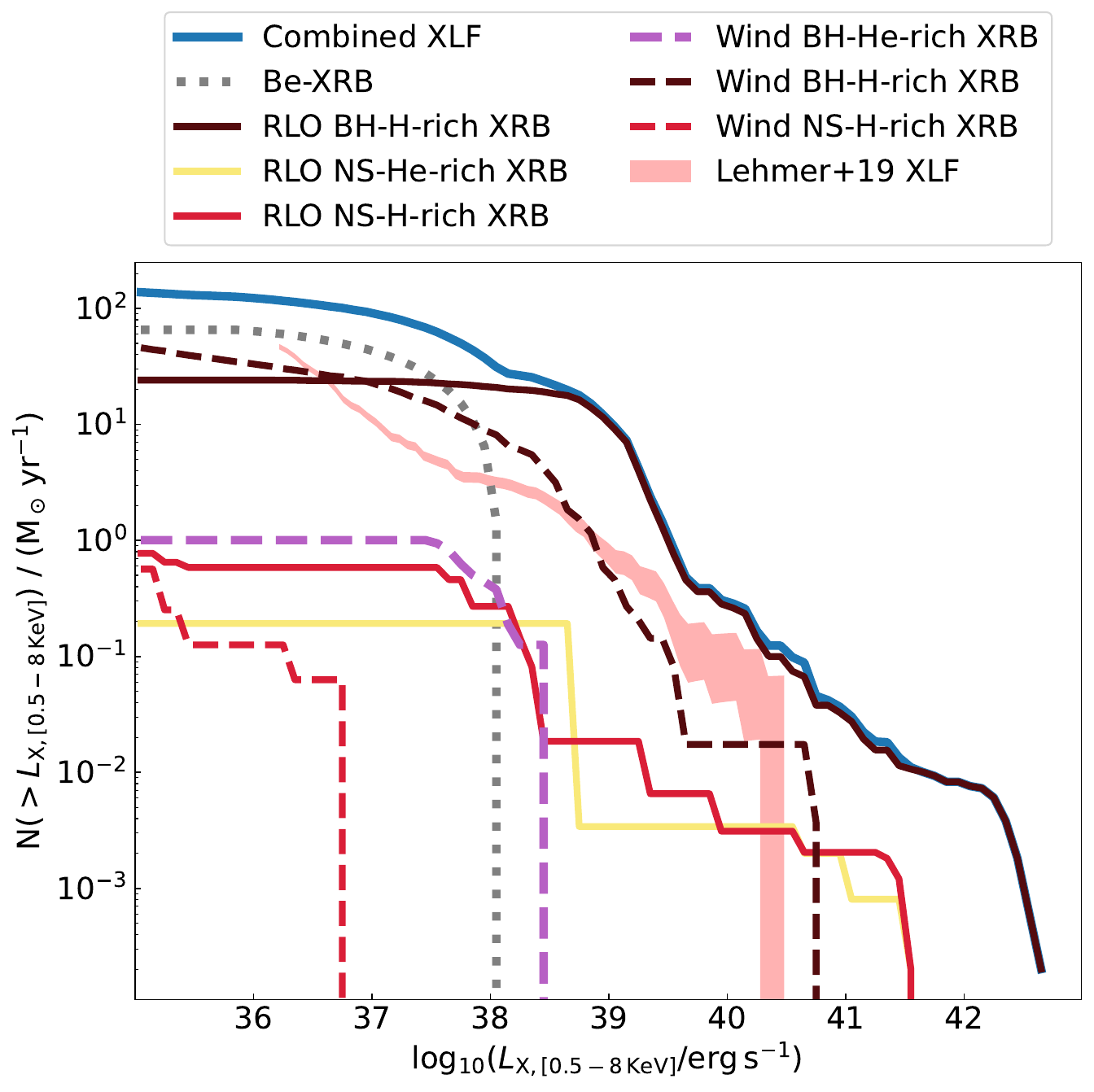}
\caption{XLF of the default population (solid blue line) compared to the observed XLF from \citet[pink shaded region]{2019ApJS..243....3L}. The population is further split by the type of XRB (RLO or wind), type of accretor (BH or NS), and type of donor (H-rich or He-rich). Be XRBs are shown with a dotted gray line.}
\label{fig:default_donor_type}
\end{figure}

The resulting synthetic XLF of the default population (shown as N(>$L_{X,[0.5 - 8 \rm Kev]}$)/SFR) split to show different types of mass transfer, COs, and donor types, is shown in Fig.~\ref{fig:default_donor_type}. The figure also compares the synthetic XLF to the observed HMXB XLF from \citet{2019ApJS..243....3L}. \boldred{The assumptions we made in the simulations regarding the SFH and metallicity are similar to the measured properties of the host galaxies of the XRBs that were used to construct the observed XLF (see the discussion in Sect.~\ref{sec:ini_properties}).} Around the luminosity of $10^{38}$~erg~s$^{-1}$, we see an overabundance of XRBs in the synthetic XLF compared to observations, by a factor of up to 10. However, the synthetic XLF shows a break at an approximate luminosity of $10^{38}$~erg~s$^{-1}$ where the shape of the XLF changes, a feature also present in the observed XLF. For X-ray luminosities below the break, Be XRBs dominate the population (dotted gray line) and for luminosities above the break, it is dominated by RLO BH XRBs with H-rich donors (shown as the solid brown line). Among the Be XRBs, $\sim 96\%$ of them have  NS accretors, while only $\sim 4\%$ appear to be BH Be XRBs (that is, a ratio of NS Be XRBs to BH Be XRBs of $24$). This model prediction is consistent with other theoretical models that found the ratio of NS Be XRBs to BH Be XRBs in the range $10$ to $50$ \citep{2009ApJ...707..870B, 2014ApJ...796...37S}, depending on the specifications of the physical model assumed. 

There is evidence that Be XRBs are formed from binary interactions that occurred previously in the evolution, with the B-type stars spinning up due to stable transfer of matter from the NS-progenitors \citep{2014ApJ...796...37S,2020MNRAS.498.4705V}. The number of NS Be XRBs is most likely affected by the slope of the initial-mass function; a high number of lower-mass stars produced would lead to a higher number of lower-mass COs \citep{2009ApJ...707..870B}. Another effect that could dictate the Be-XRBs demographic is the CE phase, as binaries forming BHs have more massive primaries that initiate CE earlier (often when they are in the Hertzsprung gap) than binaries with lower-mass primaries, which merge due to the application of the pessimistic approach during the CE phase \citep[see ][]{2009ApJ...707..870B, 2012ApJ...759...52D}. However, in our models, when we compare against a population run with the optimistic approach (where we allow all donor types to successfully eject the CE), we see no significant differences in XRB demographics. The total default XLF has a steeper slope above the break, which is qualitatively similar to observations, primarily dominated by the population of RLO BH XRBs with H-rich donors. As an additional check, we run a population with an initial binary fraction of 0.5 (that would affect the normalization) and see no significant difference.

\begin{figure}[!ht]
\centering
\includegraphics[width=\linewidth]{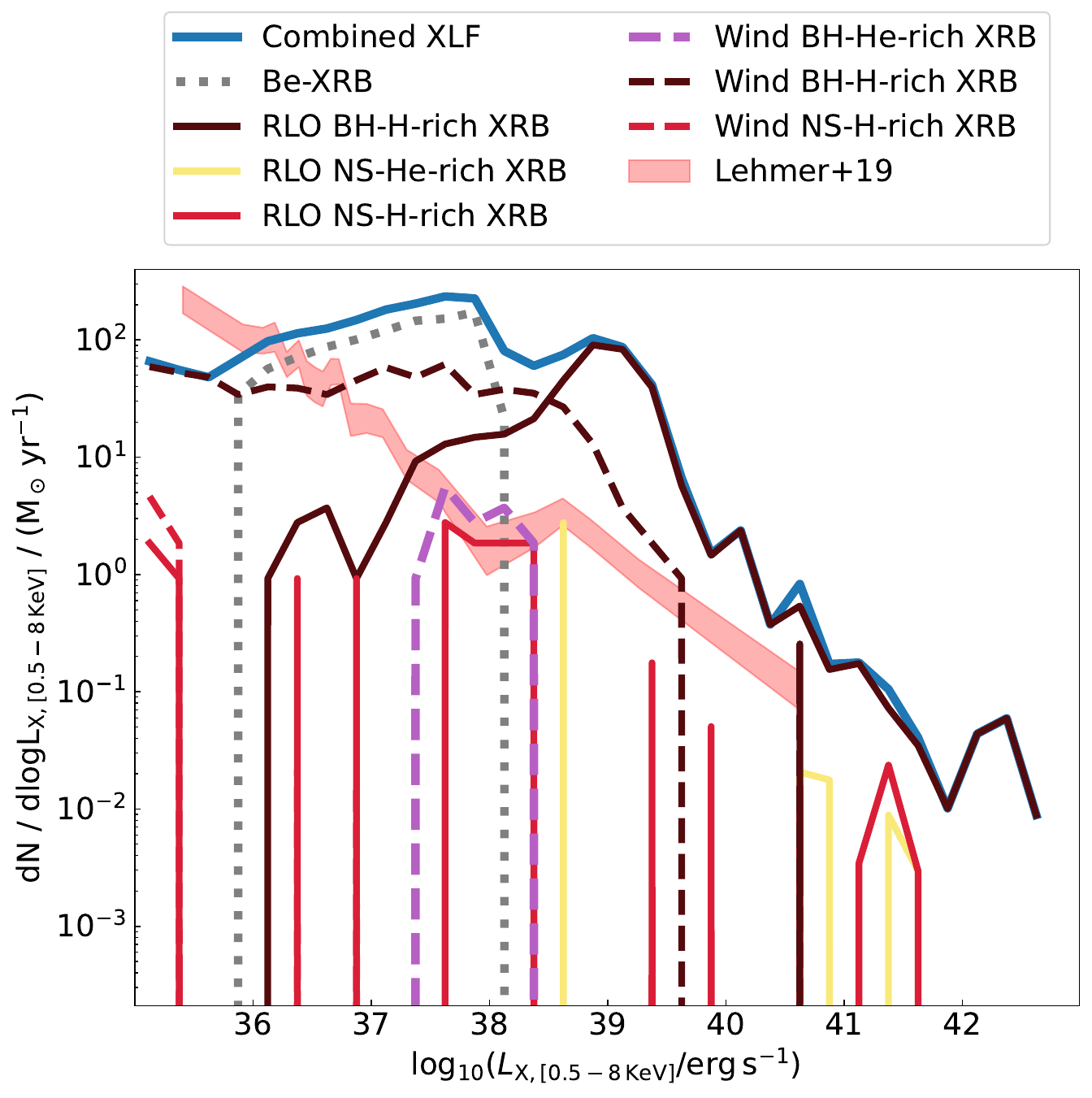}
\caption{Differential form of the synthetic XLFs, showing the default population (solid blue line) compared to the observed XLF from \citet[pink shaded region]{2019ApJS..243....3L}. The population is further split by the type of XRB (RLO or wind), type of accretor (BH or NS), and type of donor (H-rich or He-rich). Be XRBs are shown with a dotted gray line.}
\label{fig:15_diff}
\end{figure}

Looking at the combined XLF in Fig.~\ref{fig:default_donor_type} for X-ray luminosities about $10^{38}$ to $10^{39}$~erg~s$^{-1}$, the number of XRBs exceeds the observed sample by factors up to $\sim 10$. We refer to this excess as the ``XLF bump'' in the rest of this study. Looking at the subpopulations, the bump seems to be primarily coming from the RLO BH XRBs with H-rich donors. At lower luminosities (below $10^{38}$~erg~s$^{-1}$) XRBs are dominated mainly by Be XRBs, which are highly dependent on our assumptions for Be XRBs, with some contribution from wind BH XRBs with H-rich donors. There is also a slight difference at the higher luminosity end of the XLF with respect to the observations, a result of the scarce observational data at these luminosities. The observed XRBs reach up to $3\times 10^{40}$~erg~s$^{-1}$ while out simulations produce XRBs with luminosities up to $ 3\times 10^{42}$~erg~s$^{-1}$. Since at a given luminosity, the cumulative XLF carries the uncertainties of the following luminosities, we refer to Fig.~\ref{fig:15_diff}, which shows the differential form of the XLF. We see that the populations of Be XRBs, RLO BH XRBs, and wind-fed BH XRBs (both with H-rich donors) dominate the XLF bump. Since we do not model part of the binary populations (initial zero-age MS binary consists of mainly massive stars), luminosities below $\lesssim 10^{36}$~erg~s$^{-1}$ are under-filled. 

\begin{figure}[!ht]
\centering
\includegraphics[width=\linewidth]{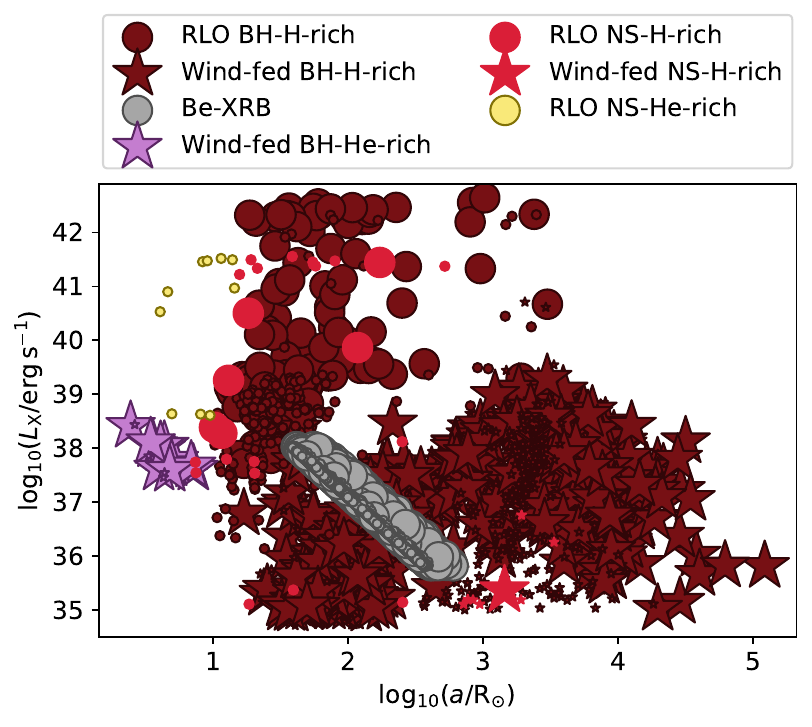}
\caption{\boldred{Distribution of the calculated X-ray luminosities of the simulated XRBs in the default population with respect to the orbital separation. The population is further split by the type of XRB (RLO or wind), type of accretor (BH or NS), and type of donor (H-rich or He-rich). Be XRBs are shown with gray circles. Large and small symbols denote massive (with $\gtrsim 8.0$~M$_{\odot}$) and less massive (with $< 8.0$~M$_{\odot}$) XRBs, respectively. }}
\label{fig:sep_vs_Lx}
\end{figure}
\boldred{Additionally, we can look at the distribution of the XRB luminosities with respect to the binary separation in Fig.~\ref{fig:sep_vs_Lx}, distinguishing between massive donors ($\gtrsim 8.0$~M$_{\odot}$) and less massive donors ($< 8.0$~M$_{\odot}$). We see that the different parts of the parameter space are occupied by different types of XRBs, which reflects the dependence of their luminosities on their respective formation channels. Generally, the XRBs undergoing RLO with H-rich donors have close orbits centered around 10~days and bright luminosities ($\gtrsim 5\times 10^{37}$~erg~s$^{-1}$), while wind-fed XRBs with H-rich donors have wider orbits going as high as 10$^5$~R$_{\odot}$ and lower luminosities ($\lesssim 5\times 10^{38}$~erg~s$^{-1}$). XRBs with He-rich donors are present at close orbits ($\lesssim 10$~R$_{\odot}$) because He stars are more compact compared to H-rich stars, which enables the binaries to evolve to close orbits.}

\boldred{In Fig.~\ref{fig:sep_vs_Lx}, the less massive donors in the population of RLO BH XRBs are at the lower end of the distribution, going as bright as $10^{39}$~erg~s$^{-1}$, as the massive donors can easily undergo high mass-transfer rates without exceeding the super-Eddington limit (or exceeding it by a relatively small amount); this makes their RLO phase more stable at higher luminosities. In the population with wind-fed BH XRBs, the less massive donors occupy the entire luminosity range. However, they are mostly present around large separations of around $10^3$~R$_{\odot}$, with the donors being mostly in the post-MS, giant phase, during which they can drive strong winds. At closer orbits ($\lesssim 10^3$~R$_{\odot}$), the systems undergoing wind-fed accretion are mostly massive MS stars that have strong stellar winds. The estimated X-ray luminosities for Be XRBs follow Eq.~\ref{eq:be_xrb}, which expresses the observed correlation between the peak luminosity and orbital period, hence the tight correlation in Fig.~\ref{fig:sep_vs_Lx}.}

In the following sections, we explore how different assumptions (mentioned in Sect.~\ref{sec:numericaltools} and summarized in Table~\ref{table:params}) about physical processes in the formation and evolution of XRBs may leave an imprint on the XLF of the whole population. We also explore the best-fitting combination of parameters that reduces the XLF bump.

\begin{figure}[!ht]
\centering
\includegraphics[width=\linewidth]{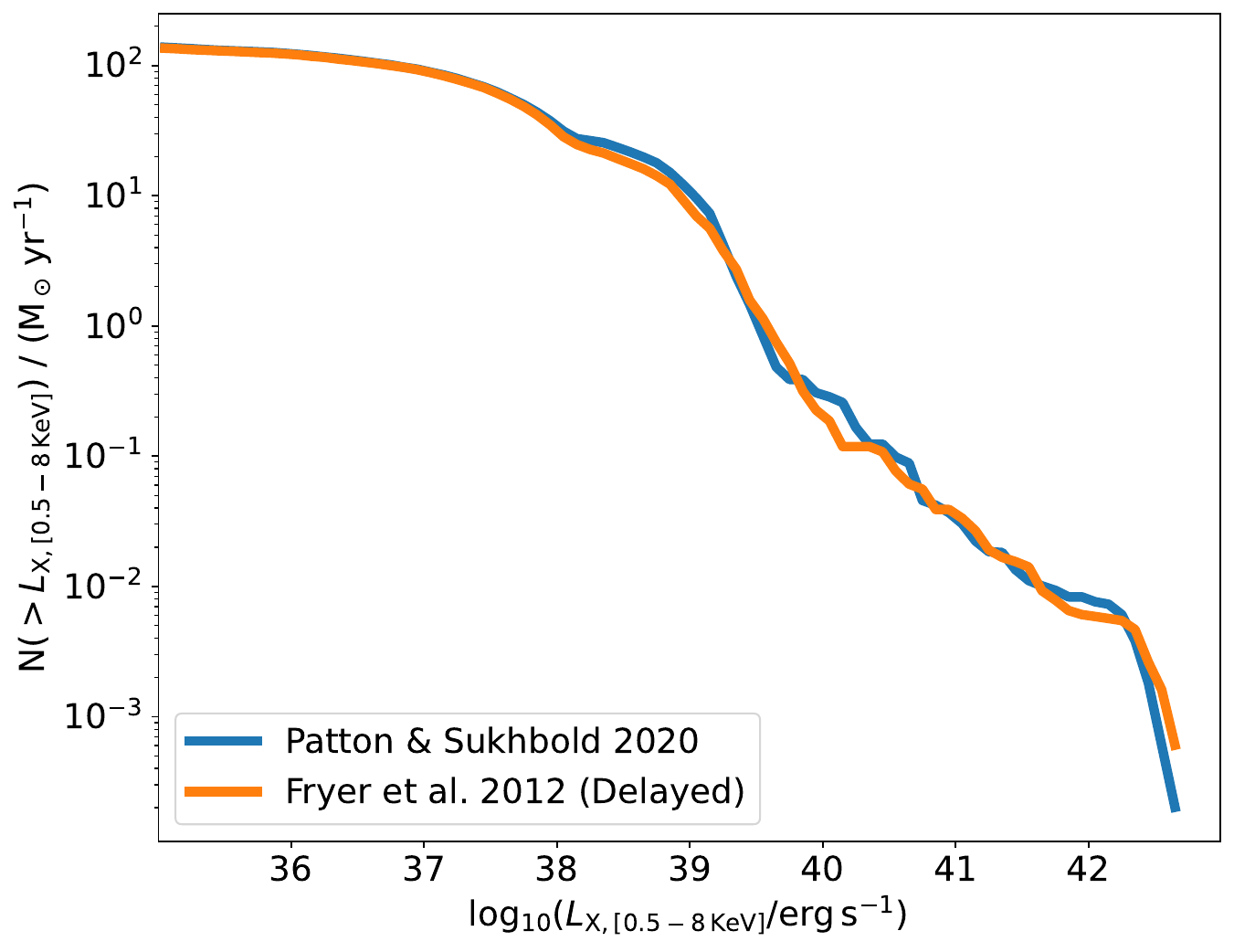}
\caption{XLFs of the default population with the parameters in Table~\ref{table:params_def} showing the \citet{2020MNRAS.499.2803P} SN mechanism (blue line) and the population with the \citet[delayed]{2012ApJ...749...91F} SN mechanism (orange line).} 
\label{fig:cc_pres_effect}
\end{figure}

\begin{figure}[!ht]
\centering
\includegraphics[width=\linewidth]{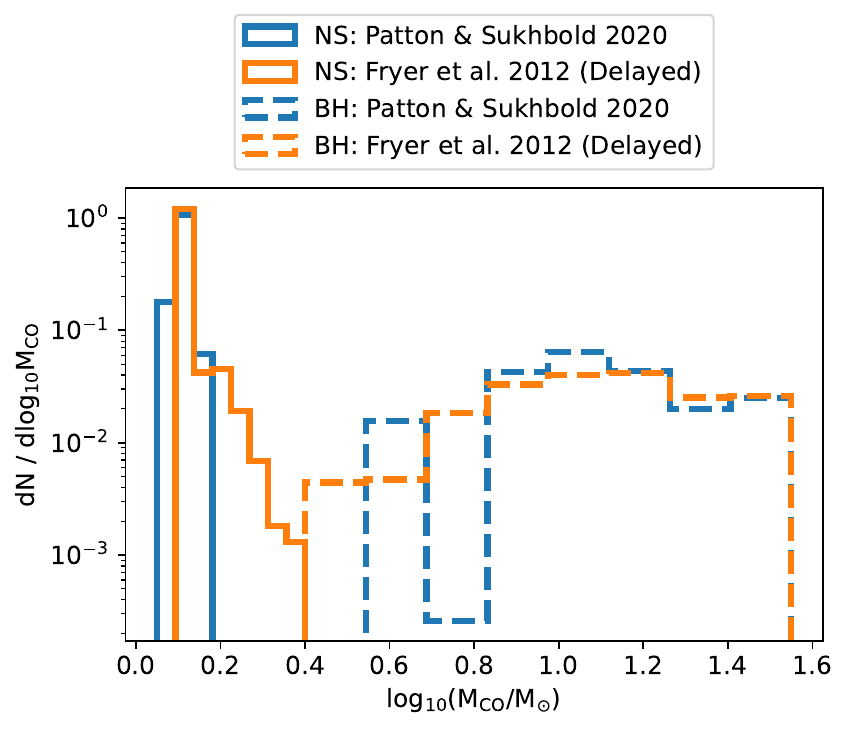}
\caption{Distribution of newly formed CO masses ($M_{\rm CO}$) for binaries that remain bound post-SN, for the default population with the parameters in Table~\ref{table:params_def}, including the \citet{2020MNRAS.499.2803P} SN mechanism (blue line) and the population with the \citet[delayed]{2012ApJ...749...91F} SN mechanism (orange line). The populations are further split by the type of CO, NS (solid line), and BH (dashed line). The distributions have been weighted with the same factor that normalizes the maximum counts per bin for the default population for NS formation (NS: \boldred{\citealp{2020MNRAS.499.2803P},} solid blue line) to 1.}%
\label{fig:cc_pres_ALL}
\end{figure}

\subsection{Effect of varying the remnant mass prescription}

We explored two prescriptions that describe the masses of SN remnants, from \citet{2020MNRAS.499.2803P} and \citet{2012ApJ...749...91F} (see our Sect.~\ref{sec:sn_mech}). We compare the synthetic XLFs of XRB populations simulated with two remnant formation prescriptions in Fig.~\ref{fig:cc_pres_effect}, one following \citet{2020MNRAS.499.2803P} (the default population) and the other using the delayed prescription from \citet{2012ApJ...749...91F} (which corresponds to model~7 in Appendix~\ref{appendix:graph}). We find no significant differences in the resulting XLFs between the two mechanisms. In order to see how these prescriptions affect the resulting populations we look into the mass distribution of the COs produced.

Figure~\ref{fig:cc_pres_ALL} shows the normalized distributions of CO masses (by type) at the end of the SN phase, for binaries that remain bound post-SN, for both prescriptions. The CO mass range covered by the two prescriptions is similar. However, \citet{2020MNRAS.499.2803P} produces a gap in the CO mass distribution between NSs and BHs and a narrow NS mass distribution, while the delayed prescription produces a continuous CO mass distribution. In both cases, the NS mass distribution peaks around $1.4$~M$_{\odot}$ \citep[similar to][]{2012ApJ...757...55O,2016ARA&A..54..401O}. Our results are also similar to the more detailed comparative study by \citet{2022MNRAS.511..903P}, and they indicate that the HMXB XLF does not provide sufficient information to constrain the mass spectrum of COs.

\begin{figure}[!ht]
\centering
\includegraphics[width=\linewidth]{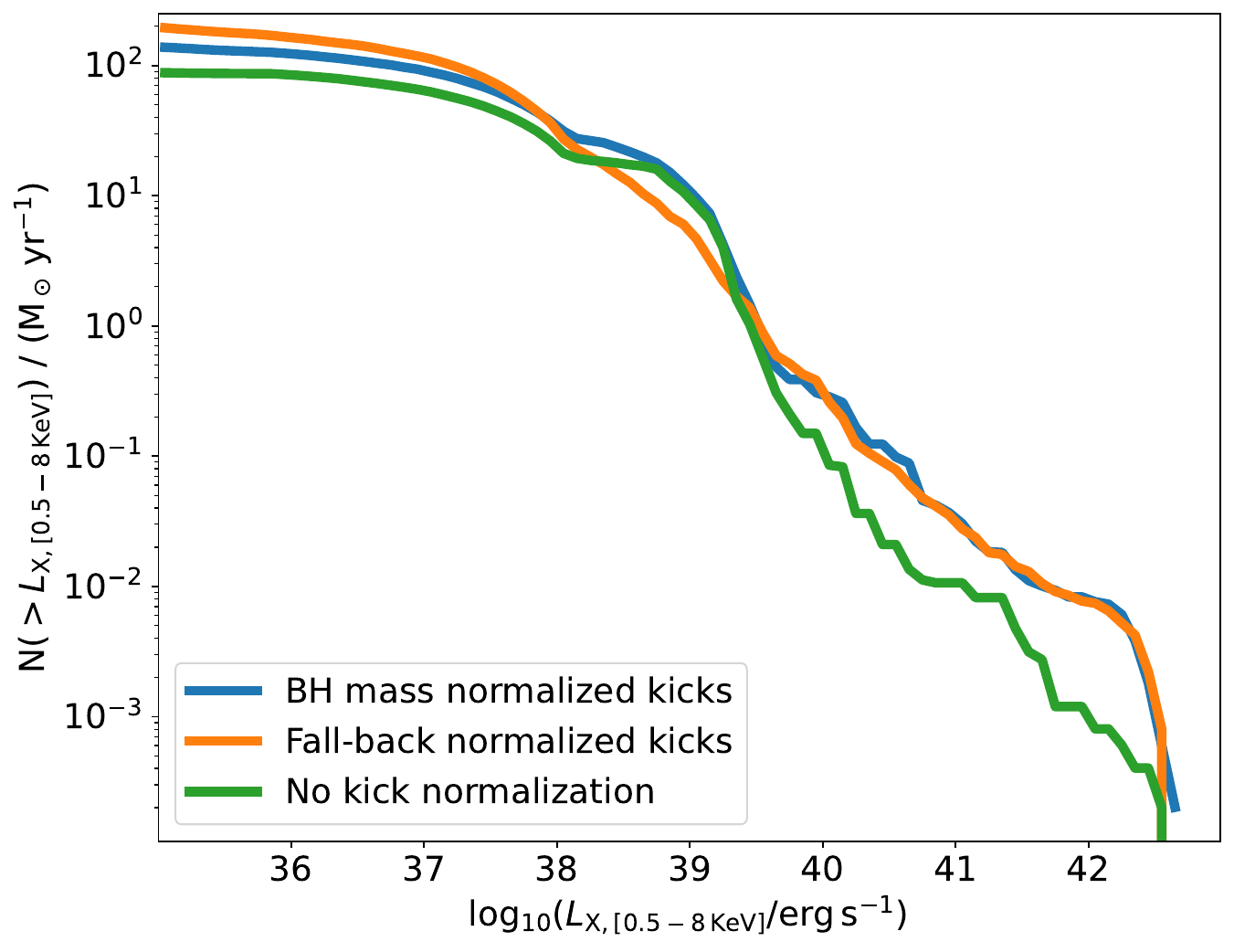}
\caption{XLFs of the default population with the parameters in Table~\ref{table:params_def} showing the population with BH mass normalized SN kicks (blue line), the population with fall-back normalized SN kicks (orange line), and the population with no normalization on the SN kicks (green line).}
\label{fig:kick_effect}
\end{figure}

\begin{figure}[!ht]
\centering
\includegraphics[width=\linewidth]{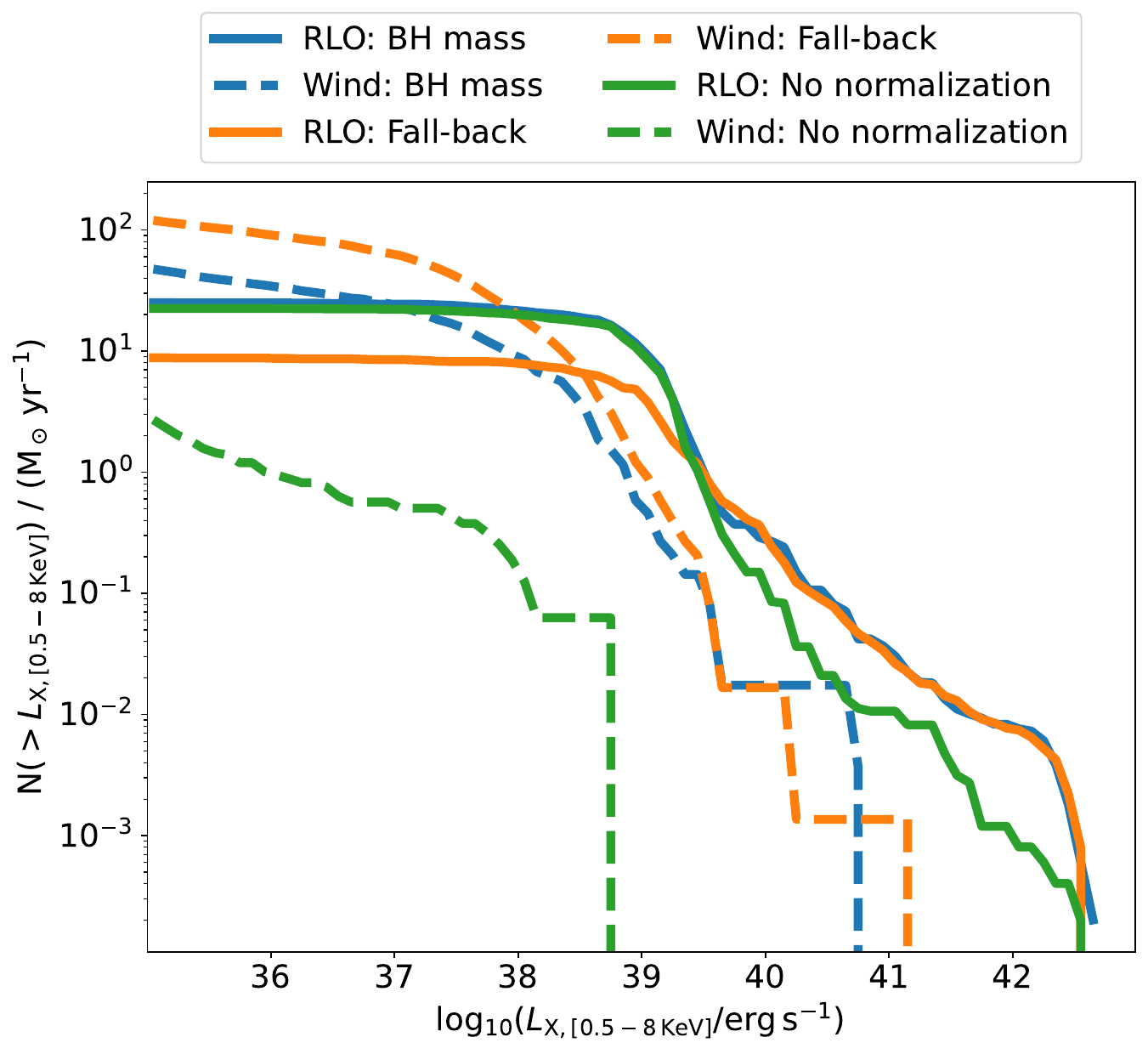}
\caption{XLFs of the default population with the parameters in Table~\ref{table:params_def} showing the population with BH mass normalized SN kicks (blue lines), the population with fall-back normalized SN kicks (orange lines), and the population with no normalization on the SN kicks (green lines). The populations follow the remnant mass prescription from \citet{2020MNRAS.499.2803P}. The populations are further split by the type of mass transfer that \boldred{occurred}, RLO (solid line) or wind-fed accretion (dashed line). For the sake of clarity, the Be-XRB population is not shown as it is not affected by the kick normalization and obscures the effects of the BH kick normalization.}
\label{fig:kick_effect_all}
\end{figure}

\begin{figure}[!ht]
\centering
\includegraphics[width=\linewidth]{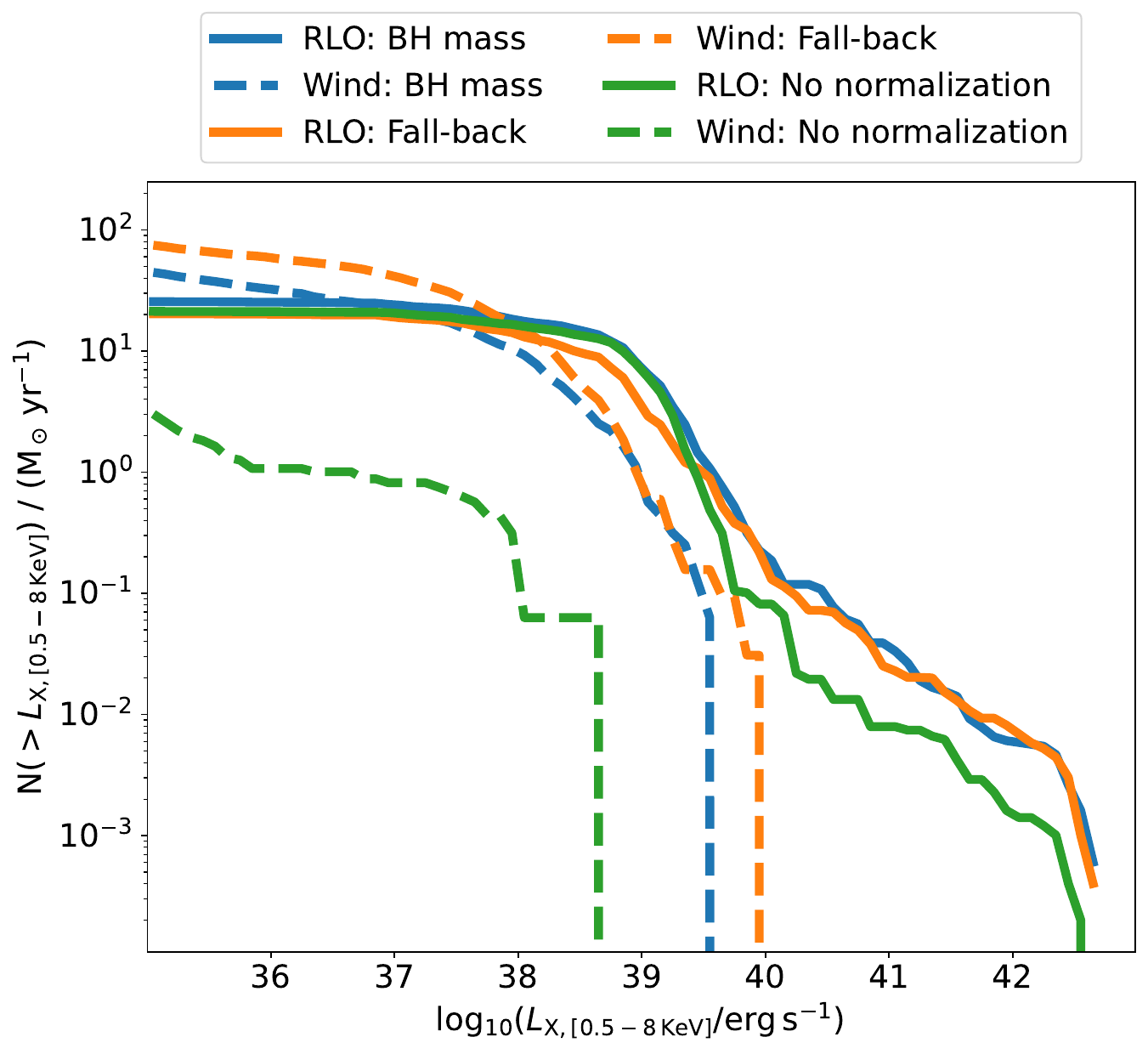}
\caption{XLFs of the default population with the parameters in Table~\ref{table:params_def} showing the population with BH mass normalized SN kicks (blue lines), the population with fall-back normalized SN kicks (orange lines), and the population with no normalization on the SN kicks (green lines). The populations follow the remnant mass prescription from \citet[delayed]{2012ApJ...749...91F}. The populations are further split by the type of mass transfer that \boldred{occurred}, RLO (solid line) or wind-fed accretion (dashed line). For the sake of clarity, we did not include Be XRBs in this figure as they are not affected by the kick normalization and obscure the effects of the BH kick normalization.}
\label{fig:kick_effect_all_FD}
\end{figure}

\begin{figure}[!ht]
\centering
\includegraphics[width=\linewidth]{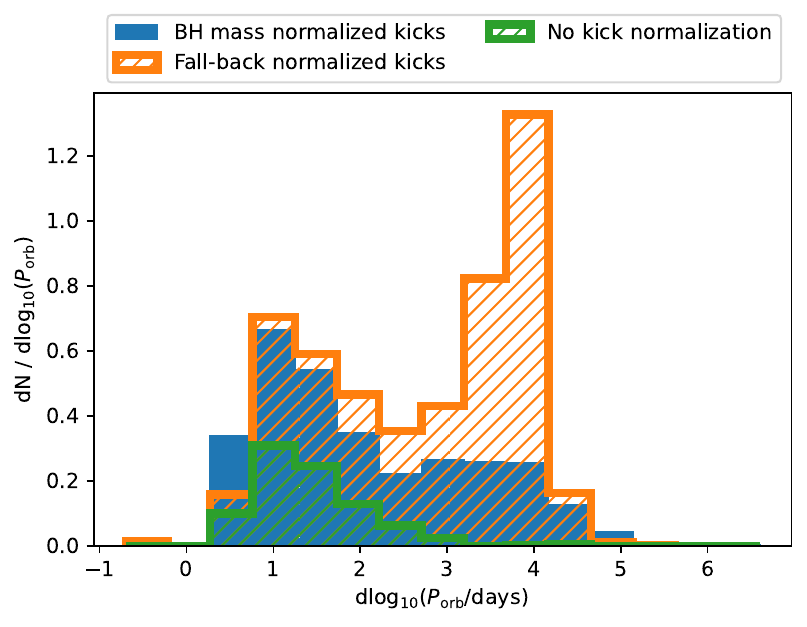}
\caption{Distributions of orbital periods ($P_{\rm orb}$) at the end of the SN for the progenitors of BH XRBs, for the default population with the parameters in Table~\ref{table:params_def}, including the population with BH mass normalized SN kicks (solid blue), the population with fall-back normalized SN kicks (hatched orange), and the population with no normalization on the SN kicks (hatched green). The distributions have been weighted with the same factor that normalizes the maximum counts per bin for the default population (BH mass normalized kicks shown in solid blue) to 1.}
\label{fig:kick_effect_PORB}
\end{figure}

\begin{figure}[!ht]
\centering
\includegraphics[width=\linewidth]{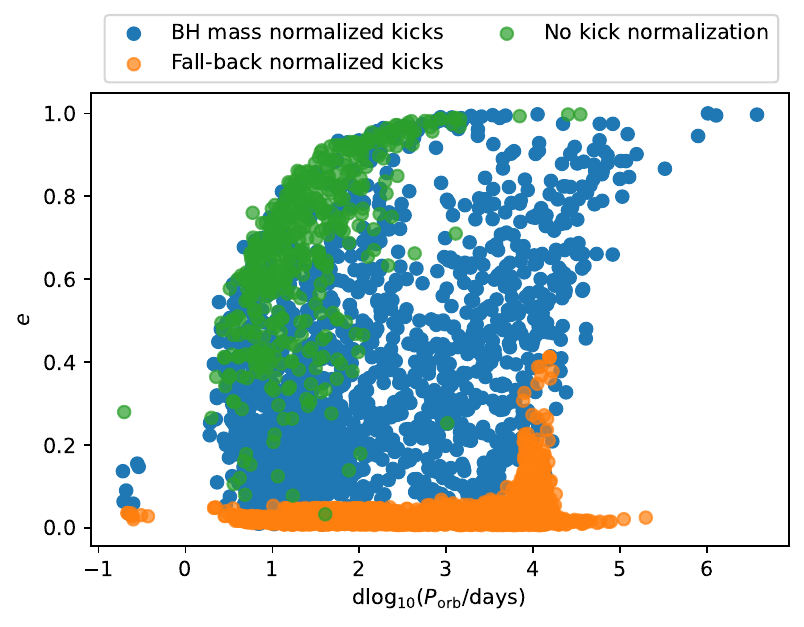}
\caption{Distribution of the eccentricity ($e$) versus $\log_{10}$ of the orbital period ($P_{\rm orb}$) for the same populations as in Fig.~\ref{fig:kick_effect_PORB}.}
\label{fig:kick_effect_ECC}
\end{figure}

\subsection{Effect of varying the natal kick normalization}

As detailed in Sect.~\ref{sec:sn_kick}, we investigated three different prescriptions for the \boldred{SN} kick normalizations (see Table~\ref{table:params}). The kick normalization will influence the future evolution of the first-born CO and change the fraction of binaries that disrupt or have significantly altered orbits. These effects will reflect in the synthetic XLFs. In the default \boldred{BH-XRB} population with mass-weighted kicks, the \boldgreen{strength of the} kicks decreases with increasing BH mass, implying that low-mass BHs get considerable kicks. For BH kicks normalized by the fall-back fraction (corresponding to model~31 in Appendix~\ref{appendix:graph}), the BHs either receive no kicks or \boldgreen{weak} kicks (depending on the SN mechanism and the progenitor carbon-oxygen mass). Out of the three prescriptions described in Sect.~\ref{sec:sn_kick}, the strongest BH kicks are for the case with no additional normalization (corresponding to model~47 in Appendix~\ref{appendix:graph}), where BHs receive kicks drawn directly from a Maxwellian distribution with \boldred{$\sigma=265\, \rm km~s^{-1}$}, the same as observationally constrained for NSs \citep{2005MNRAS.360..974H}. 

Figure~\ref{fig:kick_effect} shows the comparison of these three normalizations with respect to our default model. With increasing kick velocities, the number of XRBs below an X-ray luminosity of $3\times 10^{38}$~erg~s$^{-1}$ decreases. At the same time, an excess of XRBs emerges at X-ray luminosities in the range $3\times 10^{38}$ to $10^{39}$~erg~s$^{-1}$ for populations with BH mass normalized kicks (that is, the default model) and with kicks that are not normalized. Above $10^{39}$~erg~s$^{-1}$, the XLF slope for the population with no normalization on the kicks drops below the other two normalization prescriptions. Figure~\ref{fig:kick_effect_all} shows again the XLFs of the three models, now further split by the type of mass transfer the XRBs currently undergo (wind or RLO). The first striking difference is the lower number of the wind-fed XRBs for the population with no kick normalization shown by the green-dashed line (reduced by a factor of $20$ compared to the default population). Another difference is the lower number of RLO XRBs for the population with fall-back normalized kicks. The reasons for these differences can be understood on the basis of the XRB orbital period distribution right after the SN that forms the BH. 

One complication here is that the default remnant mass prescription, which follows \citet{2020MNRAS.499.2803P}, assigns $f_{\rm fb}=1.0$ for all BHs, resulting in no BH kicks. For the prescription from \citet{2012ApJ...749...91F}, low-mass BHs receive \boldgreen{weak} kicks. Figure~\ref{fig:kick_effect_all_FD} shows the XLFs of the three BH kick models, further split by the type of mass transfer. Comparing to Fig.~\ref{fig:kick_effect_all} we see mostly similarities, except the higher number of RLO XRBs for the population with no kick normalization (by a factor of $\sim 2$). However, the luminosity regime \boldred{in which} this effect takes place (below $10^{39}$~erg~s$^{-1}$) is dominated by wind XRBs, and hence this increase is not reflected in the simulated XLF. Generally, we focus more on the default remnant prescription when discussing BH kick normalizations.

Since the kick prescriptions we use affect only the BH XRBs, we only focus on BH binaries in the following discussion. Figure~\ref{fig:kick_effect_PORB} shows the distribution of the orbital periods for the three kick normalization prescriptions, right after the SN of the primary star for the progenitor binaries of BH XRBs. For BH mass-weighted kicks, low-mass BHs will receive a \boldgreen{weak} kick, inversely proportional to their mass, which will disrupt some of the binaries at large orbits with orbital velocities comparable to the kicks ($\gtrsim 100$~days in Fig.~\ref{fig:kick_effect_PORB}). Binaries at periods $\gtrsim 100$~days are surviving with fall-back weighted kicks (refer to Sect.~\ref{sec:sn_kick}) as they have \boldred{full} fall-back and receive no kicks. The effect of \boldgreen{strong} kicks disrupting binaries at larger orbits is even more pronounced in BH XRBs with no kick normalization where most binaries above periods of 100~days are disrupted (in addition to an overall lower number of surviving binaries). The disruption of wider orbits leads to a lower number of wind\boldred{-fed} BH XRBs. This argument also explains the higher number of wind-fed XRBs for fall-back weighted kicks compared to those for mass-weighted kicks, and the lower number of XRBs with no kick normalization compared to the other two populations at high-tail ends of the XLF. 

Looking at Fig.~\ref{fig:kick_effect_PORB} for orbital periods of less than 100~days, mass-weighted normalization and fall-back normalization (with no BH kicks, effectively) allow more BH binaries to survive the SN kick than un-normalized kicks. This occurs despite the mass-weighted and un-normalized kicks imparting stronger kicks on many BH binaries than the fall-back kicks, which are much weaker (for \boldred{\citealp{2012ApJ...749...91F}}), or no kicks (for \boldred{\citealp{2020MNRAS.499.2803P}}). The reason for this is the eccentricity imparted on the binaries surviving the SN kicks. Figure~\ref{fig:kick_effect_ECC} shows the distribution of the eccentricities of the BH XRBs that survive SNe. There, we see that the surviving binaries in mass-weighted kicks and un-normalized kicks have acquired increased eccentricities. When the secondary stars evolve in these binaries and fill their Roche lobe in a highly eccentric orbit, our default assumption of instantaneous circularization at the periastron distance leads to close circular RLO XRBs (refer to Eq. \ref{eq:peri}). This causes the overall higher number of close-orbit RLO XRBs in the populations with mass-weighted and un-normalized kicks compared to the fall-back XLF (Figs.~\ref{fig:kick_effect_all} and~\ref{fig:kick_effect_PORB}). This higher number of RLO XRBs, in combination with the suppression of wind-fed binaries, produces the bump in the XLF in the $3\times 10^{38}$ to $10^{39}$~\boldred{erg~s$^{-1}$} X-ray luminosity range (see Fig.~\ref{fig:kick_effect}) for the XRBs with BH mass weighted kicks and kicks with no normalization. \boldred{Additionally, there is a small population of SN-surviving BH-binaries at periods $\lesssim 0.3$~days (for all three normalization approaches) that arise from a double CE phase in the evolution prior to the BH formation, for binaries that started with an initial mass ratio $\sim 1$.}

\boldred{One thing to be noted is that in the population with fall-back normalized BH kicks and the \citet{2020MNRAS.499.2803P} SN mechanism, since there is no \boldred{baryonic} mass loss during the formation of the BH, the change in eccentricity is provided by the neutrino loss (which is limited up to $0.5$~M$_{\odot}$). Therefore, in the absence of kicks by mass loss, all eccentric systems would be assumed to result from neutrino losses. However, an additional effect might come into play as the resultant BH mass is taken as the mass of the pre-SN He-core, and whatever H-envelope is left on the star at this stage is assumed to be lost. If the star loses considerable mass in the form of its envelope, an eccentricity is introduced into the orbit. The difference between the pre- and post-SN masses of BH-resulting stars goes up to $8.0$~M$_{\odot}$ for \boldred{$13\%$} of the BH-forming binaries (with a peak around $2.0$~M$_{\odot}$), clearly more than that accounted for by neutrino mass losses. Hence, there donors retain their H-envelope \boldred{during their pre-SN evolution, which was then lost during} the SN event.} 

\boldred{Therefore,} in Fig.~\ref{fig:kick_effect_ECC}, we see \boldred{some} binaries with increased eccentricities (\boldred{going} up to 0.4) around the orbital period of $10^4$~days, \boldred{in the population with fall-back normalized kicks}. These systems end up as wide wind-fed XRBs in Fig.~\ref{fig:kick_effect_all} and their eccentricity comes from \boldred{the combined effect of H-envelope ejection and} neutrino loss. The BH masses in these systems span a range the 4.00 to 14.00~M$_{\odot}$, and their pre-SN progenitors were late giants with radii $\sim 1000$~R$_{\odot}$. \boldred{The reason these systems are centered around $10^4$~days is that they form the tail end of the population with eccentric post-SN orbits. These systems did not undergo an RLO phase previously \boldred{as they had wide orbital periods at zero-age MS $\lesssim 5\times 10^3$~days, which further widened with stellar winds and were on the order of $10^4$~days when the first SN occurred}. Orbits wider than $10^4$~days at SN were disrupted due to the kick from the H-envelope loss, and orbits narrower than these had interacting binaries that resulted in loss of the H-envelope (the pre-SN stars in these binaries are in their giant phase with radii in the order of $1000$~R$_{\odot}$). In general,} Our findings highlight the importance of our assumptions for the circularization of eccentric orbit when RLO occurs. We discuss this further in Sect.~\ref{sec:circ_effect}.

\begin{figure}[!ht]
\centering
\includegraphics[width=\linewidth]{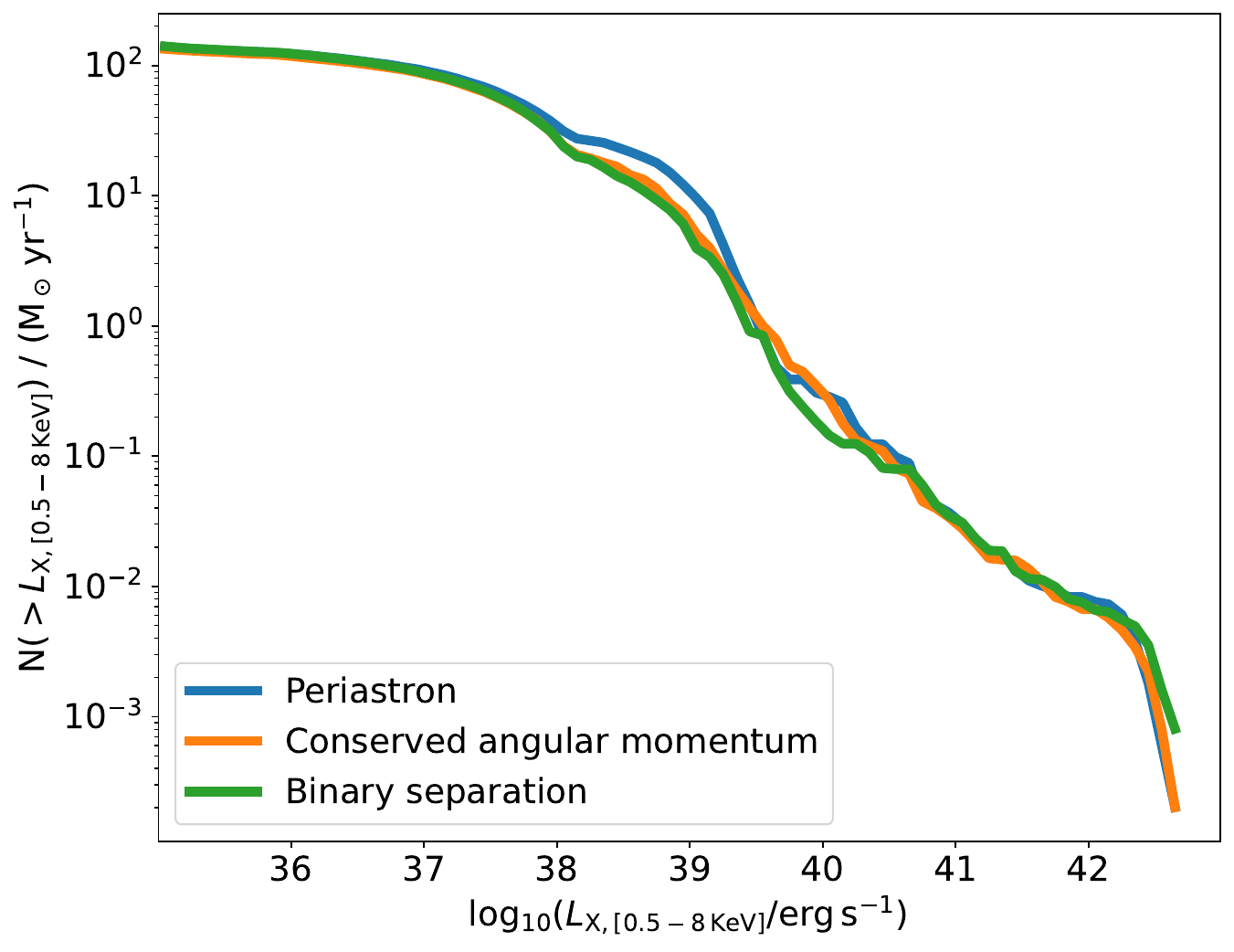}
\caption{XLFs of the default population with the parameters in Table~\ref{table:params_def} showing circularization at RLO onset at the periastron of the eccentric orbit (blue line), the population with the circularization at an orbit conserved angular momentum (orange line), and the population with circularization at the binary separation of the eccentric orbit (green line).}
\label{fig:circ_effect}
\end{figure}

\begin{figure}[!ht]
\centering
\includegraphics[width=\linewidth]{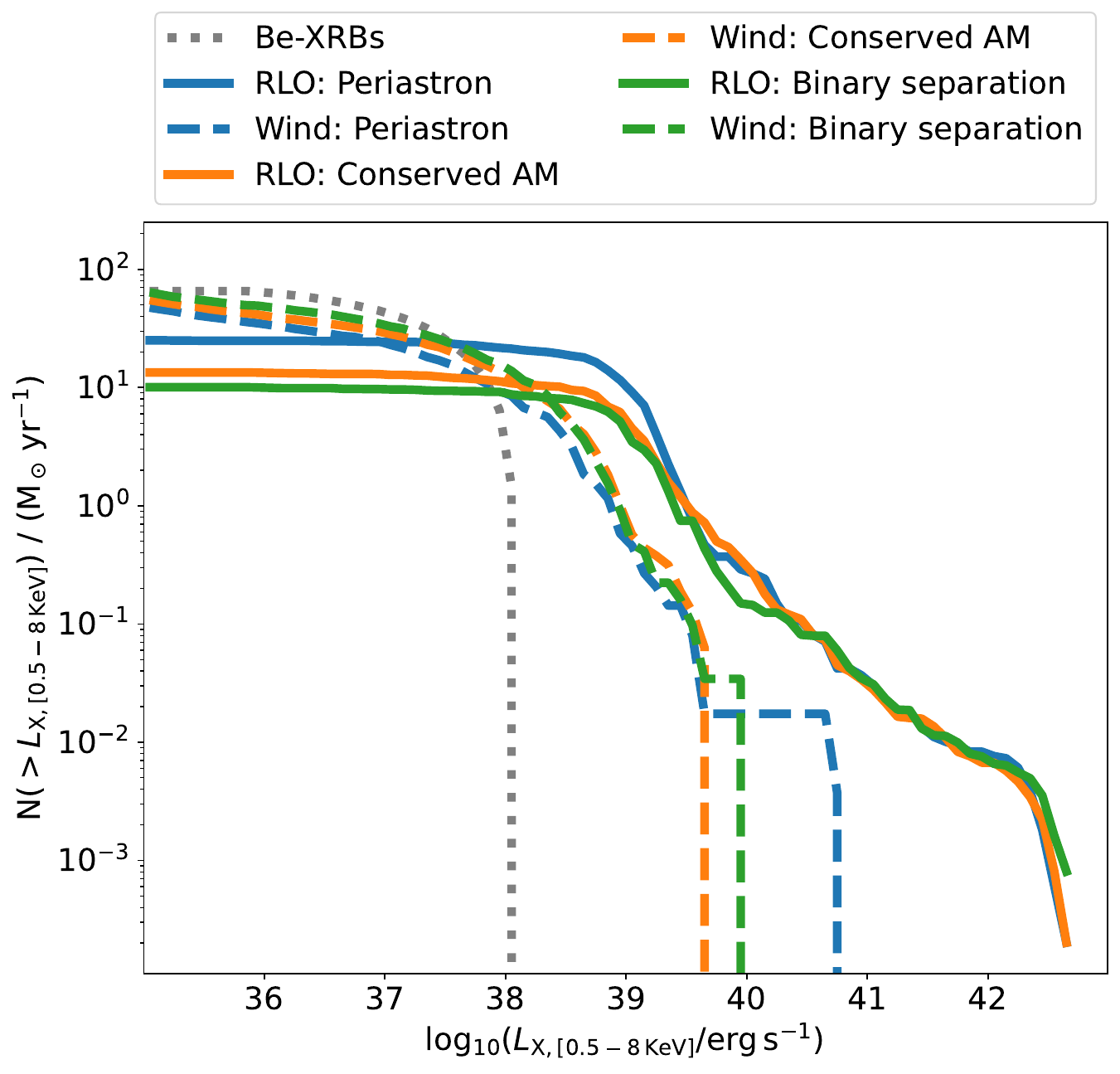}
\caption{XLFs of the default population with the parameters in Table~\ref{table:params_def} showing the population with pre-RLO circularization at the periastron (blue lines), at an orbit with conserved angular momentum (orange lines), and at the binary separation of the pre-RLO eccentric orbit (green lines). The populations are further split by the type of mass transfer that \boldred{occurred}, RLO (solid line), and wind-fed accretion (dashed line). Be XRBs are shown with a dotted gray line, \boldred{showing} no effect as they are not circularized. AM stands for angular momentum.}
\label{fig:circ_effect_all}
\end{figure}

\begin{figure}[!ht]
\centering
\includegraphics[width=\linewidth]{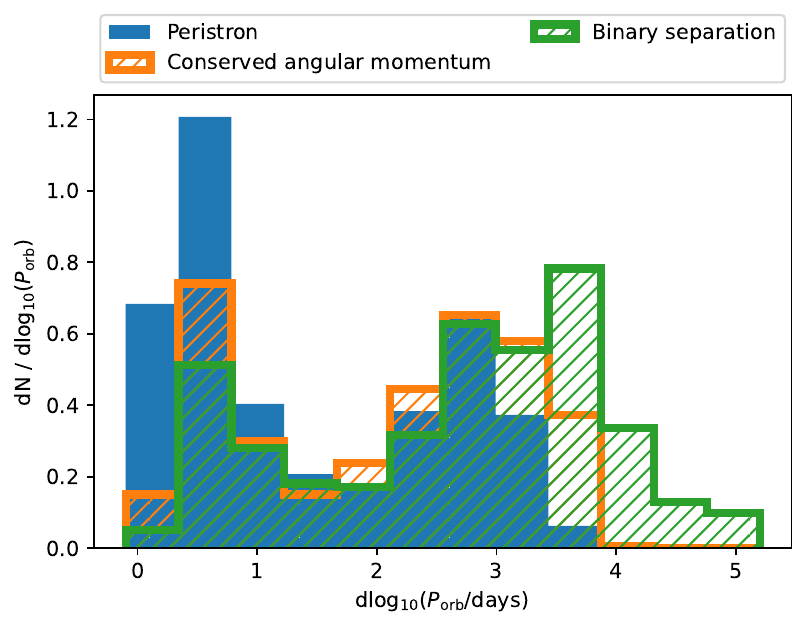}
\caption{Distributions of orbital periods ($P_{\rm orb}$) of circularized orbits at the onset of RLO for the resulting XRBs. \boldred{The presented systems were identified as initiating RLO at the periastron, and then the separation of the circularized orbit was adjusted following the respective model used. The figure shows} the default population with the parameters in Table~\ref{table:params_def} including circularization at RLO onset at the periastron of the eccentric orbit (solid blue), the population with the circularization at an orbit conserved angular momentum (hatched orange), and the population with circularization at the binary separation of the eccentric orbit (hatched green). The distributions have been weighted with the same factor that normalizes the maximum counts per bin for the default population (with orbits circularized at the periastron shown in solid blue) to 1.}
\label{fig:circ_effect_PORB}
\end{figure}

\subsection{Effect of varying the circularization of the orbit at the onset of Roche-lobe overflow}\label{sec:circ_effect}

In Sect.~\ref{sec:circi_efefct_des}, we mentioned that we can vary the assumptions for the circularization of the orbits at the onset of RLO. In this section we look at the effects of changes in orbit circularization on the XLF. Figure~\ref{fig:circ_effect} shows the synthetic XLFs for models following the different assumptions, and an additional model \boldred{that} is explained further below. A circular orbit with the same angular momentum as the eccentric orbit at the start of RLO would have a larger separation than an orbit with the separation equal to the periastron distance (as is in the default population), as is evident on comparing Eqs.~\ref{eq:peri} and \ref{eq:alt}). The population where we conserve the orbital angular momentum during the circularization at the onset of RLO corresponds to model~16 in Appendix~\ref{appendix:graph}. Intuitively, wider orbits should reduce the number of binaries undergoing RLO early on in the evolutionary life of the donor (in the long-lasting MS phase) and increase the binaries undergoing RLO with giant donors. Since the latter phase is a shorter-lived phase, the number of observed RLO XRBs would be reduced.

To investigate an even more extreme assumption of conserving the orbital energy during the circularization process, we run a single population where binaries that overfill their Roche lobe in eccentric orbits, result in circularized orbits with the same separations as the pre-RLO eccentric orbits (shown as the green line in Fig.~\ref{fig:circ_effect}). The number of XRBs around luminosities $10^{38}$ to $10^{39}$~erg~s$^{-1}$ reduces by a factor of up to 2 with the wide orbits, a difference that, when considering the total XLF, is not pronounced. This approach is similar to the assumption made in some population synthesis codes, for example {\tt BSE} \citep{2000MNRAS.315..543H}\footnote{The {\tt BSE} code does not actually assume that binaries that overfill their Roche-lob in eccentric orbits circularize instantaneously conserving their orbital energy but instead ignores altogether the RLO at periastron, and continues to evolve the binary as detached until the donor star fills its Roche lobe at a distance equal to the orbital separation.}. The rest of the parameters for this model are the same as the default population. We do not use this circularization option in the rest of the iterations of {\tt POSYDON} models.

Figure~\ref{fig:circ_effect} shows a slight reduction in the bump of the XLF around $\sim 10^{38}\,\rm erg\,s^{-1}$ for the models where we assume that binaries overfilling their RLO in eccentric orbits instantaneously circularize, conserving either angular momentum or orbital energy. To further investigate how the treatment of the circularization affects the populations, in Fig.~\ref{fig:circ_effect_all} we divide the XLFs of the three models by the type of mass transfer undergone by the XRBs (namely, wind-fed XRBs, RLO XRBs, or Be XRBs). There is a noticeable effect of increasing the circularized orbital separation in the RLO systems. Systems with wider orbits lead to fewer RLO XRBs. 

We can see the distributions of the circularized orbital periods of the resulting XRBs at RLO onset, for all three approaches in Fig.~\ref{fig:circ_effect_PORB}. \boldred{The presented systems were identified as initiating RLO at the periastron, and then the separation of the circularized orbit was adjusted following the respective model used}. The wider circularized orbits suppress the number of XRBs below periods of  $\sim 30$~days. This results in the decrease of RLO XRBs seen in the synthetic XLFs (Fig.~\ref{fig:circ_effect_all}). \boldred{On the other hand, looking at periods $\gtrsim 100$~days, circularization at the binary separation leads to a larger number of XRBs compared to} the other two populations, \boldred{as expected. About 90\% of the wide orbits (with periods $\gtrsim 100$~days) lead to wind-fed XRBs, for all three circularization options}. 

\begin{figure}[!ht]
\centering
\includegraphics[width=\linewidth]{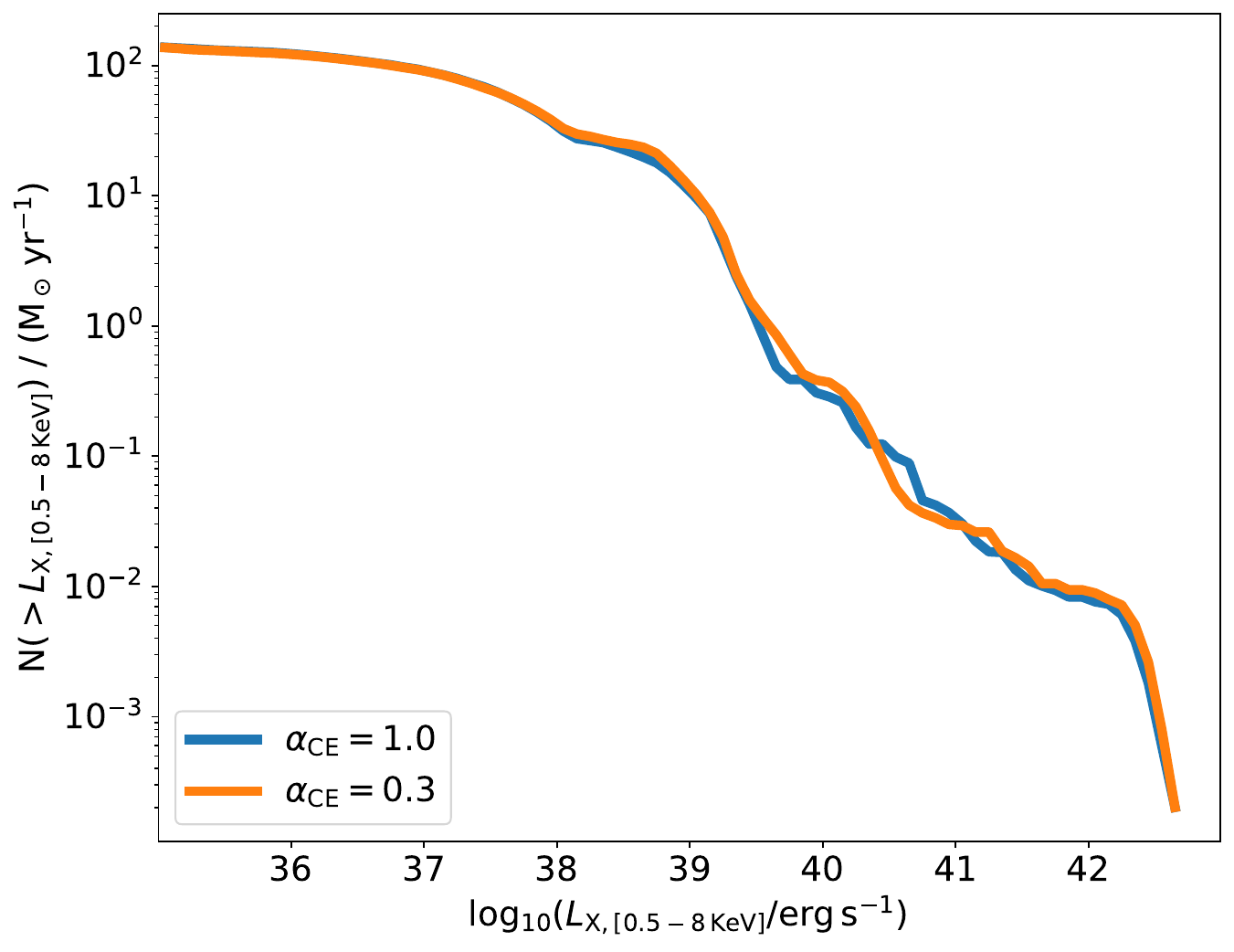}
\caption{XLF of the default population with the parameters in Table~\ref{table:params_def} showing the population with CE efficiency ($\alpha_{\rm CE}$) of 1.0 (blue line) and the population with $\alpha_{\rm CE}$ of 0.3 (orange line).}
\label{fig:alpha_effect}
\end{figure}

\begin{figure}[!ht]
\centering
\includegraphics[width=\linewidth]{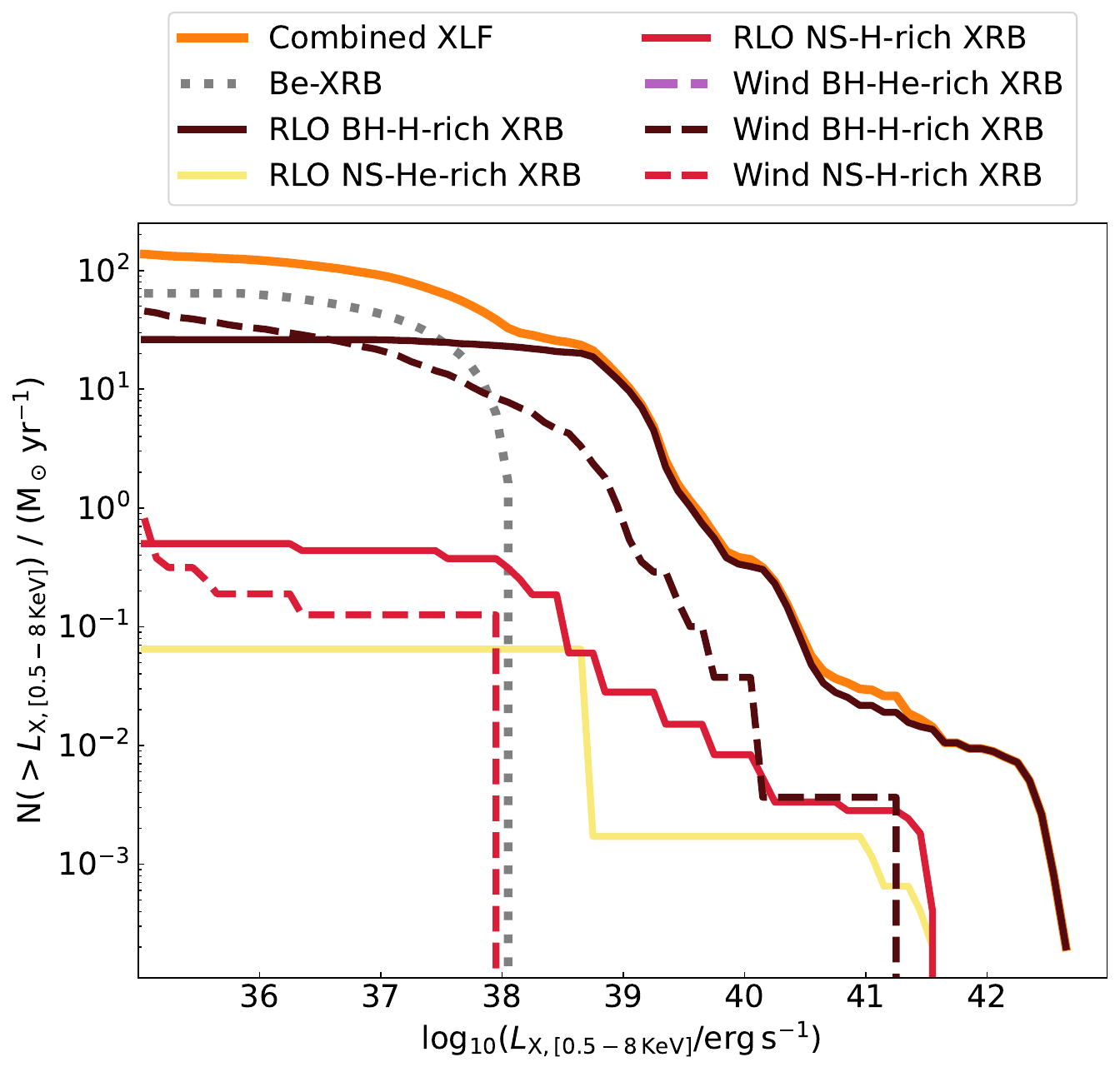}
\caption{XLF of the population with $\alpha_{\rm CE}=0.3$ (solid orange line). The population is further split by the type of XRB (RLO or wind), type of accretor (BH or NS), and type of donor (H-rich or He-rich). Be XRBs are shown with a dotted gray line.}
\label{fig:11_donor_type}
\end{figure}

\subsection{Common-envelope evolution}

\subsubsection{Effect of varying the common-envelope efficiency}\label{sec:alpha_effect}

We investigate the effect of different assumptions for the $\alpha_{\rm CE}$--$\lambda_{\rm CE}$ prescription on the synthetic XLF. Firstly, we look into two different values of $\alpha_{\rm CE}$. Figure~\ref{fig:alpha_effect} shows the synthetic XLFs of populations with $\alpha_{\rm CE}=1.0$ (default population) and $\alpha_{\rm CE}=0.3$ (corresponding to model~11 in Appendix~\ref{appendix:graph}). 
Overall, there is no significant difference between the two populations when looking at the total XLF.\ This is similar to the results by \citet{2011ASPC..447..121L} and \citet{2014ApJ...797...45Z}, who found that $\alpha_{\rm CE}$ does not significantly alter the HMXB numbers. However, in their study, a high value of $\alpha_{\rm CE}$ is needed to fit the XLF. Figure~\ref{fig:11_donor_type} shows the XLF of the population with $\alpha_{\rm CE}=0.3$, split into the different types of XRBs. Compared to Fig.~\ref{fig:default_donor_type} with $\alpha_{\rm CE}=1.0$, there is a decrease in XRBs with helium-rich (He-rich) donors, as those binaries have undergone a CE in their prior evolution and the $\alpha_{\rm CE}$ governs this phase. For instance, the population of wind-fed BH XRBs with He-rich donors disappears with a lower $\alpha_{\rm CE}$ (shown as the dashed purple line in Fig.~\ref{fig:default_donor_type}) and the population of RLO NS XRBs with He-rich donors decreases by a factor of 3 (shown as the solid yellow line). The lower the $\alpha_{\rm CE}$, the more the accretor inspirals in the envelope in order to eject it. Alternatively, higher $\alpha_{\rm CE}$ corresponds to a larger post-CE orbital separation and a higher probability of surviving the CE phase. However, the variation of $\alpha_{\rm CE}$ has little effect on the total synthetic XLF since, based on our models, BH XRBs with hydrogen-rich (H-rich) donors and Be XRBs, which dominate the XLF, do not go through a CE phase during their formation. 

\begin{figure}[!ht]
\centering
\includegraphics[width=\linewidth]{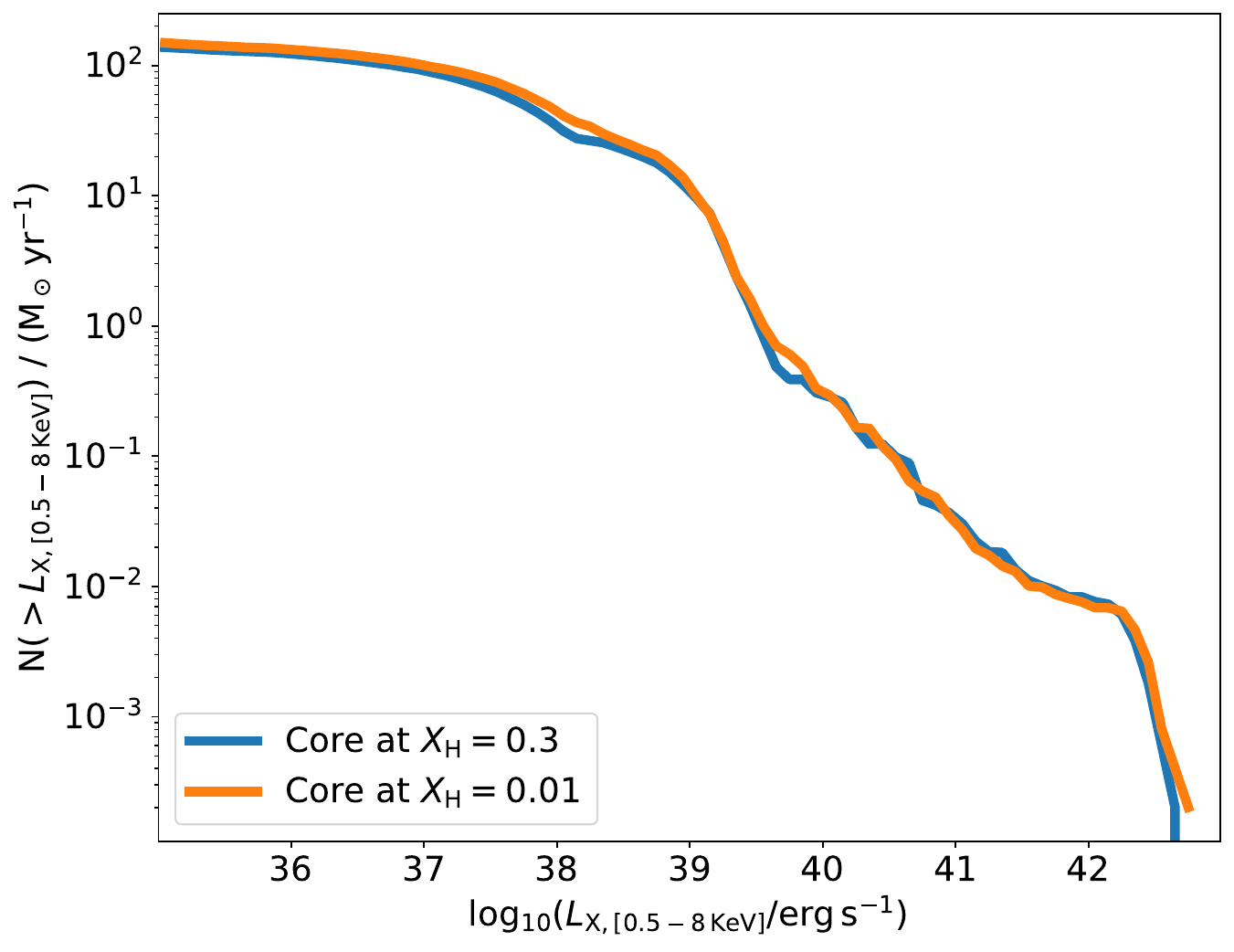}
\caption{XLFs of the default population with the parameters in Table~\ref{table:params_def} showing the population with core-envelope definition at $X_{\rm H} = 0.3$ (blue line) and the population with core-envelope definition at $X_{\rm H} = 0.01$ (orange line).}
\label{fig:lambda_effect}
\end{figure}

\begin{figure}[!ht]
\centering
\includegraphics[width=\linewidth]{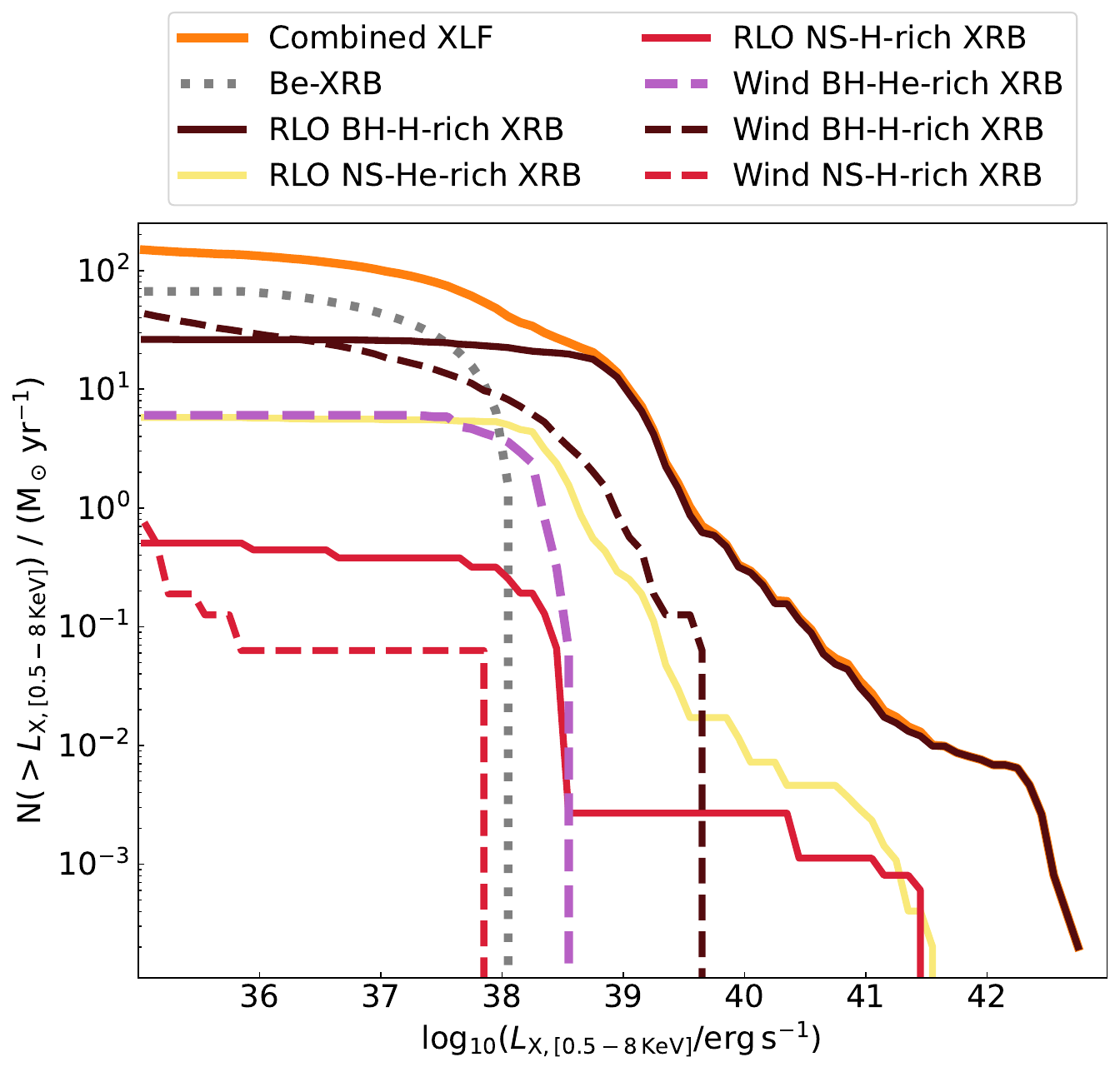}
\caption{XLF of the population with core-envelope definition at $X_{\rm H} = 0.01$  (solid orange line). The population is further split by the type of XRB (RLO or wind), type of accretor (BH or NS), and type of donor (H-rich or He-rich). Be XRBs are shown with a dotted gray line.}
\label{fig:13_donor_type}
\end{figure}

\subsubsection{Effect of varying the definition of the core-envelope boundary in CE evolution}\label{sec:lambda_effect}

The second CE parameter that we look into is the definition of the core-envelope boundary, via which we indirectly investigate the binding energy parameter $\lambda_{\rm CE}$.
Figure~\ref{fig:lambda_effect} compares the synthetic XLFs of the populations with the core-envelope boundary definition during the CE at $0.30$ (default population) and at $X_{\rm H} = 0.01$ (corresponding to model~13 in Appendix~\ref{appendix:graph}; refer to Sect.~\ref{sec:ce_alpha} for more details). Similarly to varying the CE efficiency, we do not see a distinct difference between the two XLFs. We can investigate the populations in detail by looking at Fig.~\ref{fig:13_donor_type} where we show the XLF of the model with core-envelope boundary definition at $X_{\rm H} = 0.01$, split by the different XRB subpopulations. XRBs with He-rich donors, which have undergone a CE in their prior evolution, increase in number when compared to the default model shown in Fig.~\ref{fig:default_donor_type}.  

\begin{figure}[!ht]
\centering
\includegraphics[width=\linewidth]{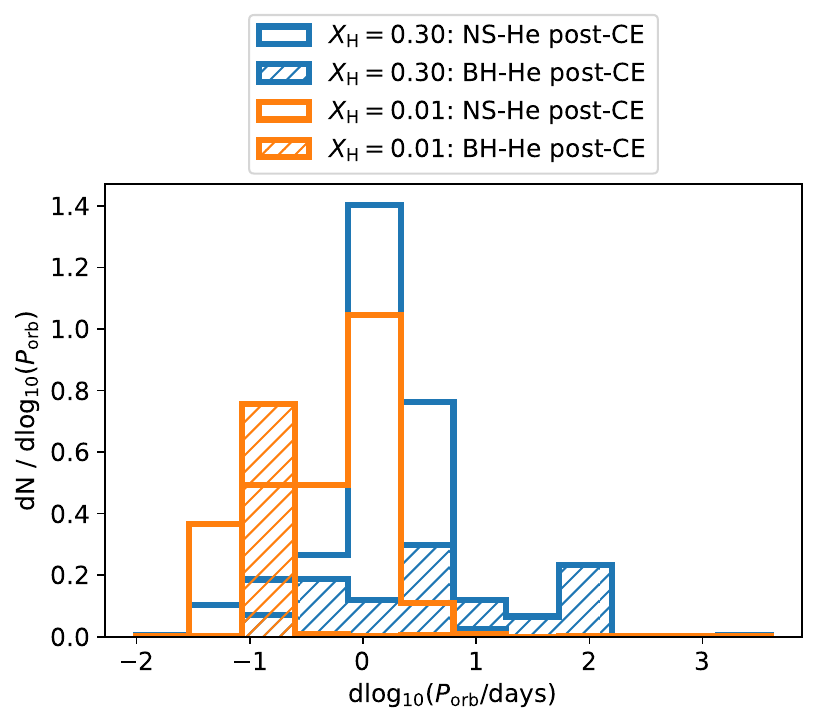}
\caption{Distribution of the post-CE orbital periods ($P_{\rm orb}$) for the default population with the parameters in Table~\ref{table:params_def} including the core-envelope definition at $X_{\rm H} = 0.30$ (blue histogram) and the population with the core-envelope definition at $X_{\rm H} = 0.01$ (orange histogram). The populations are further split by the type of CO in the surviving binary, showing NSs (step) and BHs (hatched). The distributions have been weighted with the same factor that normalizes the maximum counts per bin for the NS-He post-CE binary population to 1.}%
\label{fig:lambda_effect_PORB}
\end{figure}

Comparing the post-CE populations in Figs.~\ref{fig:default_donor_type} and \ref{fig:13_donor_type}, we get similar fractions of surviving NS and BH binaries for both definitions of the core-envelope boundary. The fraction of binaries with NSs and BHs that enter a CE and successfully eject it is $\sim 5$\% and $\sim 18\%$, respectively. However, some XRB subpopulations with He-rich donors show significant variations. Specifically, there is an increase in the number of RLO NS XRBs with He-rich donors (increase by a factor of $30$) and in the number of wind-fed BH XRBs with He-rich donors (increase by a factor of $6$), going from the core-envelope boundary definition during the CE at $X_{\rm H} = 0.3$ (Fig.~\ref{fig:default_donor_type}) to $X_{\rm H} = 0.01$ (Fig.~\ref{fig:13_donor_type}). Even then, this increase is not reflected in the overall synthetic XLFs as it is dominated by systems not undergoing CE. Additionally, the fractional increase in the two subpopulations is not the same, with RLO NS XRBs with He-donors increasing by a larger factor than wind-fed BH XRBs with He-rich donors. 

The orbital energy released into the CE by the inspiraling CO depends on the mass of the CO ($\propto M_{\rm CO}$), with BHs providing more energy than NSs. Hence, more BHs survive the CE phase than NSs (cf. $\sim 18\%$ versus $\sim 5\%$). The XRB population that results from the further evolution of binaries that survive the CE, is determined by the distribution of the orbital periods of the surviving binaries. Figure~\ref{fig:lambda_effect_PORB} shows the normalized distributions of the orbital periods post-CE for the models with the core-envelope boundary definitions at $X_{\rm H} = 0.30$ and at $0.01$, for different types of CO (NSs or BHs). We see that the distribution shifts toward lower periods when going from $X_{\rm H} = 0.3$ to $X_{\rm H} = 0.01$, with a greater difference for BH-He-star binaries than NS-He-star binaries (which maintains the peak of its distribution around $\sim 1$~day). The resulting closer orbits affect different types of binaries differently. 

For NS binaries, closer orbits (for the model with core-envelope boundary definition during the CE at $X_{\rm H} = 0.01$) result in a higher number of RLO systems, as seen in Fig.~\ref{fig:13_donor_type}. The reason is that the subsequent donors for these systems typically have masses $\lesssim 4$~M$_{\odot}$ and will expand overfilling their Roche lobes earlier in their evolutionary lifetime. This leads to more donors in their He-MS phase than in their He-giant phase. Mass-transfer phases with He-MS donors have \boldred{a} longer duration than with He-giant donors and therefore, there would be a significantly higher chance of observing them. Within our Milky Way, so far, there are no observations of RLO NS XRBs with He-rich donors. If they are present they would be expected to lie along the Galactic plane where younger populations dominate. However, the extinction along the Galactic plane would reduce any chances of observations. Since our models already predict a small number of these sources within our galaxy, even in the most optimistic case ($\lesssim 10$), the lack of observed systems cannot put a robust constraint on our models.

For BH binaries, closer orbits (for the model with core-envelope boundary definition at $X_{\rm H} = 0.01$) result in more efficient Bondi-Hoyle wind-fed accretion and hence a higher number of bright wind-fed BH XRBs with He-star donors. However, there are no RLO BH XRBs with He-rich donors as seen in Figs.~\ref{fig:default_donor_type} and \ref{fig:13_donor_type}. The reason is that the stripped He-rich stars in these post-CE binaries have masses $\gtrsim 4$~M$_{\odot}$, and do not expand much (to fill their Roche lobe) during their subsequent evolution \citep[refer to Fig.~8 in][; effect also seen by \citealt{1986A&A...167...61H}]{2023ApJS..264...45F}. Hence, no RLO BH XRBs with He-rich donors are produced in our synthetic populations for either core-envelope boundary definitions.

\begin{figure}[!ht]
\centering
\includegraphics[width=\linewidth]{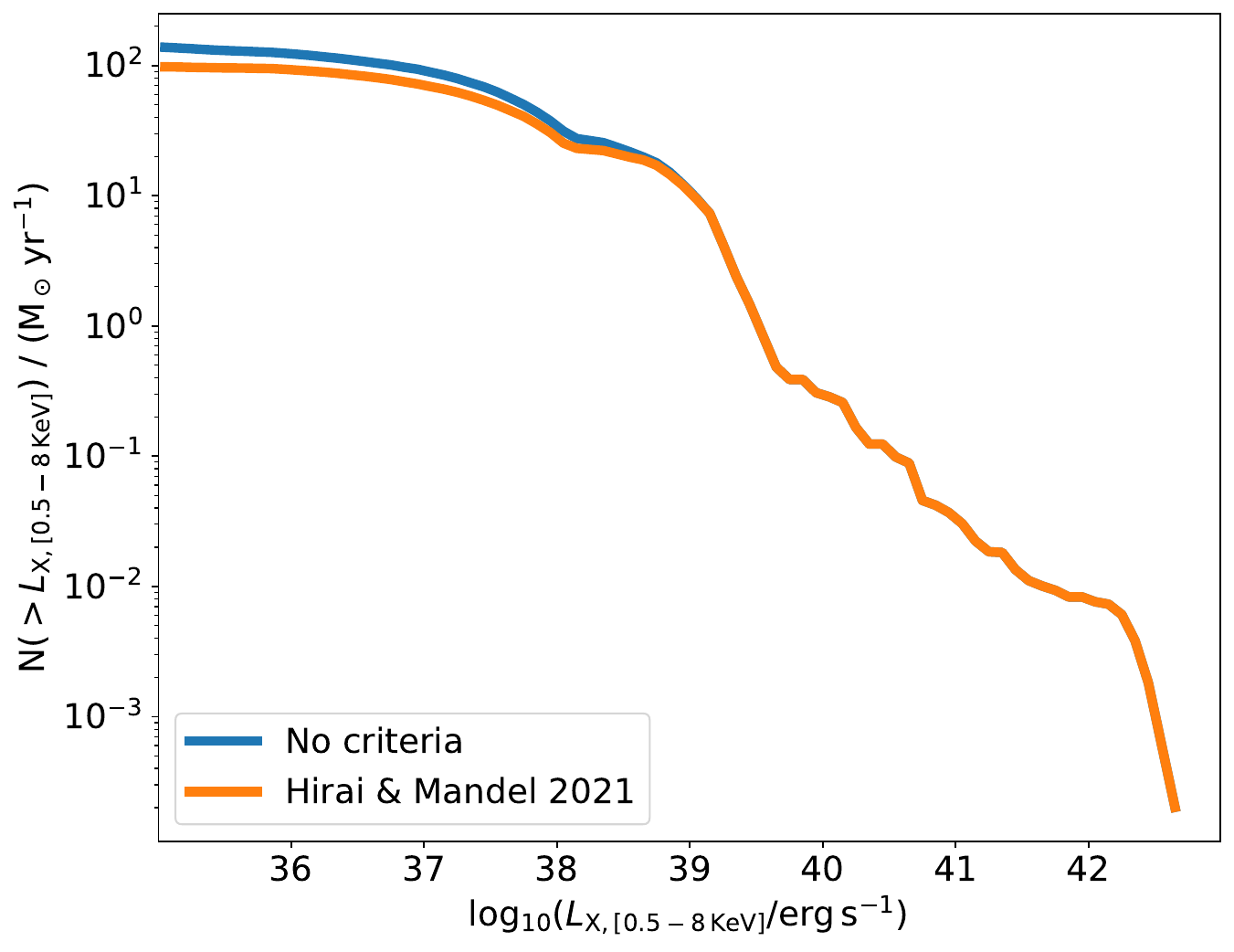}
\caption{XLFs of the default population with the parameters in Table~\ref{table:params_def} showing no criterion for wind-fed accretion (blue line) and the population with limited observability of wind-fed XRBs following \citet{2021PASA...38...56H} (orange line).}
\label{fig:wind_limited_effect}
\end{figure}

\begin{figure}[!ht]
\centering
\includegraphics[width=\linewidth]{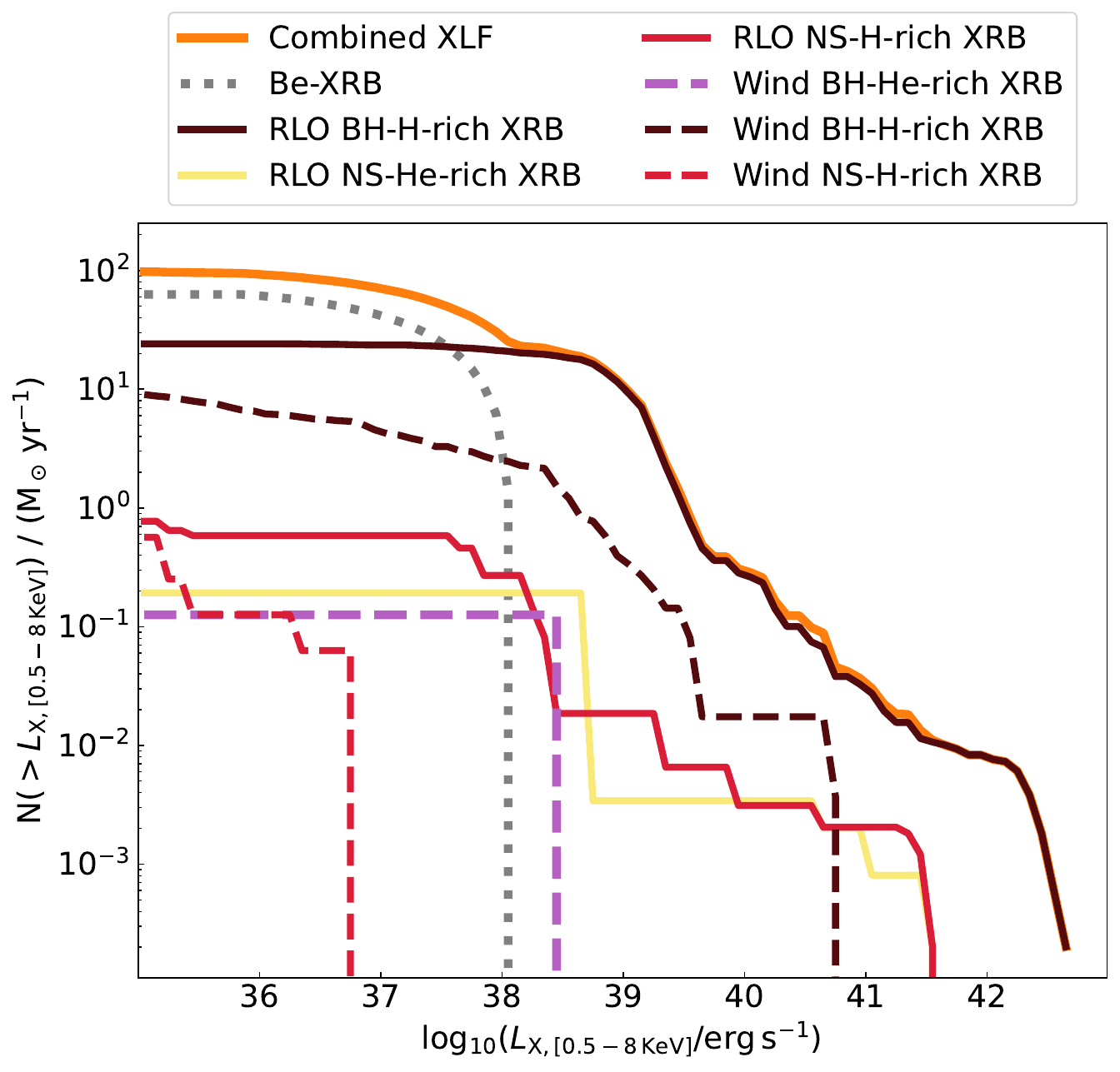}
\caption{XLF of population with limited observability of wind-fed XRBs following \citet{2021PASA...38...56H}.The population is further split by the type of XRB (RLO or wind), type of accretor (BH or NS) and type of donor (H-rich or He-rich). Be XRBs are shown with a dotted gray line}
\label{fig:default_donor_type_rlo_limited}
\end{figure}

\subsection{Effect of the criteria for observable wind-fed accretion}

Figure~\ref{fig:wind_limited_effect} presents the effect of limiting the observable X-ray in wind-fed HMXBs, following the study by \citet[][see Sect.~\ref{sec:wind_hmxb_crit} for more information]{2021PASA...38...56H}. We limit the subpopulation of wind-fed BH HMXBs with the criterion that filling of the donor Roche lobe by at least $80\%$ is needed to have observable X-ray luminosity. This criterion is applied only for systems with BH accretors undergoing wind-fed accretion, as the hard surface of a NS lifts the requirement of the accretion disk formation in order to produce X-ray emission. Since wind-fed BH XRBs dominate the XLF below luminosities $\sim 10^{38}$~erg~s$^{-1}$, the \citet{2021PASA...38...56H} criteria only affect the number of BH XRBs at the lower luminosity end of the XLF. Looking at the combined XLFs in Fig.~\ref{fig:wind_limited_effect}, XRBs with X-ray luminosities below $\sim 10^{38}$~erg~s$^{-1}$ are indeed reduced in number. 

The XLF of the model with the criterion by \citet{2021PASA...38...56H} is shown in Fig.~\ref{fig:default_donor_type_rlo_limited} (and corresponding to model~63 in Appendix~\ref{appendix:graph}), divided into the types of XRBs (showing the types of mass transfer, donor, and accretor). Compared to the default model (Fig.~\ref{fig:default_donor_type}) we can see a clear decrease in the number of wind-fed BH XRBs. Wind-fed BH XRBs with H-rich donors decreased by a factor of $5$, while wind-fed BH XRBs with He-rich donors decreased by a factor of $10$. The majority of the population of Be XRBs is not affected by the limited observability as the condition is applicable to only BH XRBs.


\section{Discussion}\label{sec:discussion}

\boldred{In addition to the physical assumptions investigated above, there are certain assumptions that have been implicitly made when carrying out the population synthesis study, the effect of which was not explored in detail. When the binary consists of two non-degenerate stars at the beginning of the evolution, any mass transfer that occurs is considered to be mostly nonconservative, as the spin-up of the accretor close to critical rotation prevents further accretion \citep{1981A&A...102...17P,2005A&A...435.1013P, 2012A&A...537A..29R, 2021ApJ...923..277R, 2022A&A...659A..98S}. The excess material is lost from the vicinity of the accretor, carrying away its angular momentum in the form of a fast isotropic wind. A more conservative mass-transfer phase would lead to more massive accretors with increased mass-transfer stability and consequently, more massive companions in the resulting XRBs. Additionally, the envelopes of these massive accretors would be easier to eject during a CE phase that might follow, increasing their chances of survival \citep{2023ApJ...942L..32R}. After either of the two stars has undergone a SN event and formed a CO, there might be a second RLO phase for which the mass-accretion rate is limited by the CO Eddington limit. The excess material, again, leaves the system taking away the angular momentum of the accretor. With a change in the critical mass-accretion rate, the stability of the mass transfer and the resulting population would be affected.}

\boldred{We assume stellar wind loss even during RLO, which would affect the orbit, particularly for massive donor stars with lower-mass accretors where the widening effect from strong winds would counter the contraction from RLO and stabilize the mass transfer \citep[refer to][for a detailed description of the wind-loss prescriptions used]{2023ApJS..264...45F}. Additionally, there could be increased RLO stability due to irradiation effects from the accretion luminosity of the CO driving extra mass loss from the donor surface \citep{1989ApJ...343..292R,1993ApJ...410..281T}, which is not considered in the present study. Finally, the assumption of the fast isotropic wind generated when the mass-transfer rate exceeds the Eddington limit, which takes away the accretor angular momentum without interacting with the rest of the binary is only an approximation that would break down in cases of close binaries where the wind velocity is comparable to the orbital velocity and the outgoing wind will interact with the orbit, thereby taking away some orbital angular momentum \citep{2011ASPC..445..355M, 2021arXiv210709675S}.}

\boldred{As seen in Sect.~\ref{sec:results},} many of the various physical parameters that were explored in this study do not leave a distinct imprint on the combined synthetic XLF. However, these parameters may significantly affect some XRB subpopulations that are subdominant overall. With the advent of multiwavelength observational data sets, which will not only put constraints on the X-ray luminosity distributions of the overall XRB populations but also on the properties of their donors \citep{2016MNRAS.459..528A,2018ApJ...862...28L, 2019ApJ...887...20A, 2021ApJ...906..120L} and accretors \citep{2014HEAD...1411006Z, 2014ApJ...797...79W, 2016MNRAS.458.3633M, 2016ApJ...824..107Y, 2018ApJ...862...28L, 2018ApJ...864..150V, 2021ApJ...920..120Y}, we will be able to look into the XLFs of subpopulations and set more stringent constraints in population synthesis models and the associated stellar and binary evolution physics.

Future studies comparing the HMXB synthetic populations (their donor and accretor characteristics) to observations would help us better understand the physics involved. Detailed characterization studies have been done for HMXBs in various metallicity and SFR environments, showing differences in various cases. For instance, in the Large Magellanic Cloud (with an average metallicity of $Z_{\rm solar}/2$), HMXBs are dominant in regions with recent star formation bursts \citep{2016MNRAS.459..528A}, while in the SMC (with an average metallicity of $Z_{\rm solar}/4$) HMXBs are found more in regions where star formation bursts occurred 25 to 60~Myr ago \citep{2010ApJ...716L.140A}. We have mentioned various other uncertainties involved in binary evolution, such as mass-transfer efficiency or angular momentum losses, that have not been included in this study as their accurate behavior is uncertain, and varying them to investigate their effect is computationally expensive when detailed binary sequences are involved, as it would require recomputing entire the binary track library.

In our default population (refer to Fig.~\ref{fig:default_donor_type}) there is an overabundance of XRBs compared to observations. In this section we discuss some of the effects of combinations of the different physical parameters that we have included in our study and possible causes for this discrepancy (for instance, transient behavior and assumptions of orbital eccentricity). We also identify four models that best describe the observed XLF. 

\subsection{Combinations of physical parameters}\label{sec:combi}

There is a certain level of degeneracy involved in the effects of the different parameters, on the XRB populations. So far we explored how the variation in a single parameter affects the XLF, but sometimes the combined effect of two or more parameters can have a more profound effect on the resulting synthetic XLF. Therefore, we discuss a few instances where the combined effects of two parameters have a significant change in the XLF and explore how some of the combinations reduce the so-called XLF bump. The combinations we discuss below do not correspond to the set of parameters that best reproduce the observed XLF, instead they showcase some examples of combined effects of the parameters on the synthetic XLF. In all the XLFs, the overabundance of XRBs below the luminosity of $10^{38}$~erg~s$^{-1}$ is caused by the population of Be XRBs and is highly dependent on the empirical relation used to calculate the luminosities, along with the assumptions regarding \boldred{the} identification of transient Be XRBs in the simulated populations.

\begin{figure}[!ht]
\centering
\includegraphics[width=\linewidth]{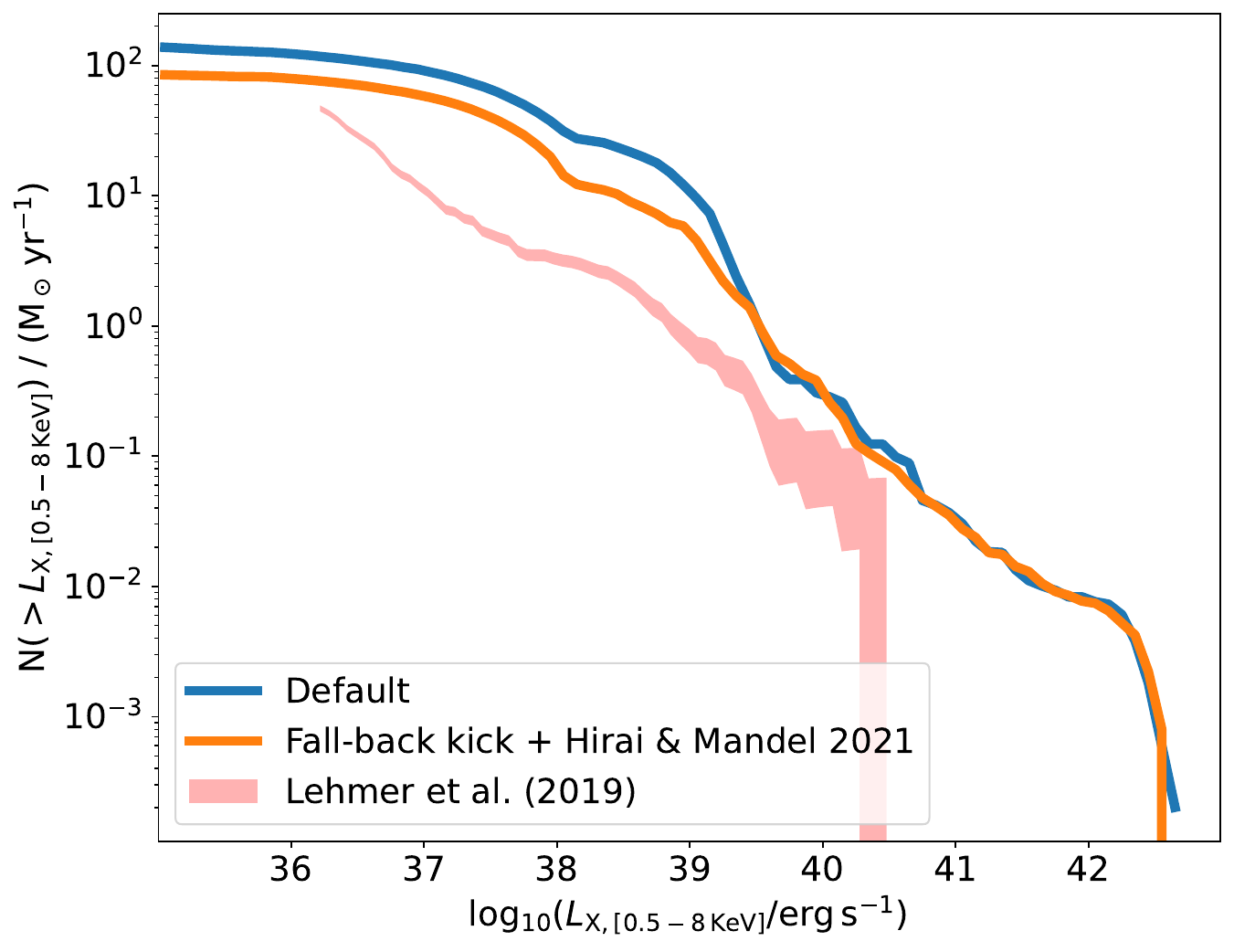} 
\caption{XLFs of populations showing the default population with the parameters in Table~\ref{table:params_def} (blue line) and the population with a combination of the fall-back SN kicks and the limited observability of wind BH XRBs following \citet{2021PASA...38...56H} (orange line). The two curves are compared to the observed XLF from \citet{2019ApJS..243....3L}.}
\label{fig:disc_fbk_wind}
\end{figure}

For instance, from the three prescriptions of BH kick normalizations, fall-back kicks have the lowest number of RLO XRBs \citep[for the default remnant mass prescription of][]{2020MNRAS.499.2803P} at the point where the XLF moves from being dominated by wind-fed XRBs to RLO XRBs, shown by the change of slope at $10^{39}$~erg~s$^{-1}$ for the solid orange line in Fig.~\ref{fig:kick_effect_all}. On the other hand, fall-back kicks result in a larger number of wind-fed XRBs for luminosities of less than $\sim 10^{38}$~erg~s$^{-1}$ than the other two kick prescriptions. Following \citet{2021PASA...38...56H}, the observability of some of these wind-fed XRBs may actually be limited (see Sect.~\ref{sec:wind_hmxb_crit}). Figure~\ref{fig:disc_fbk_wind} shows the result of these two effects combined, also comparing the resultant XLF to the default model (that has the parameter values shown in Table~\ref{table:params_def}), and to the observations from \citet[][pink shaded region]{2019ApJS..243....3L}. Wind-fed XRBs for X-ray luminosities below $10^{38}$~erg~s$^{-1}$ are reduced in number when the observable wind-fed accretion criterion is included. The combined effect of these two assumptions leads to a smoother XLF and a less pronounced  bump around $\sim 10^{38}\,\rm erg\, s^{-1}$ (up to a factor of $\sim 6$ compared to observations).

\begin{figure}[!ht]
\centering
\includegraphics[width=\linewidth]{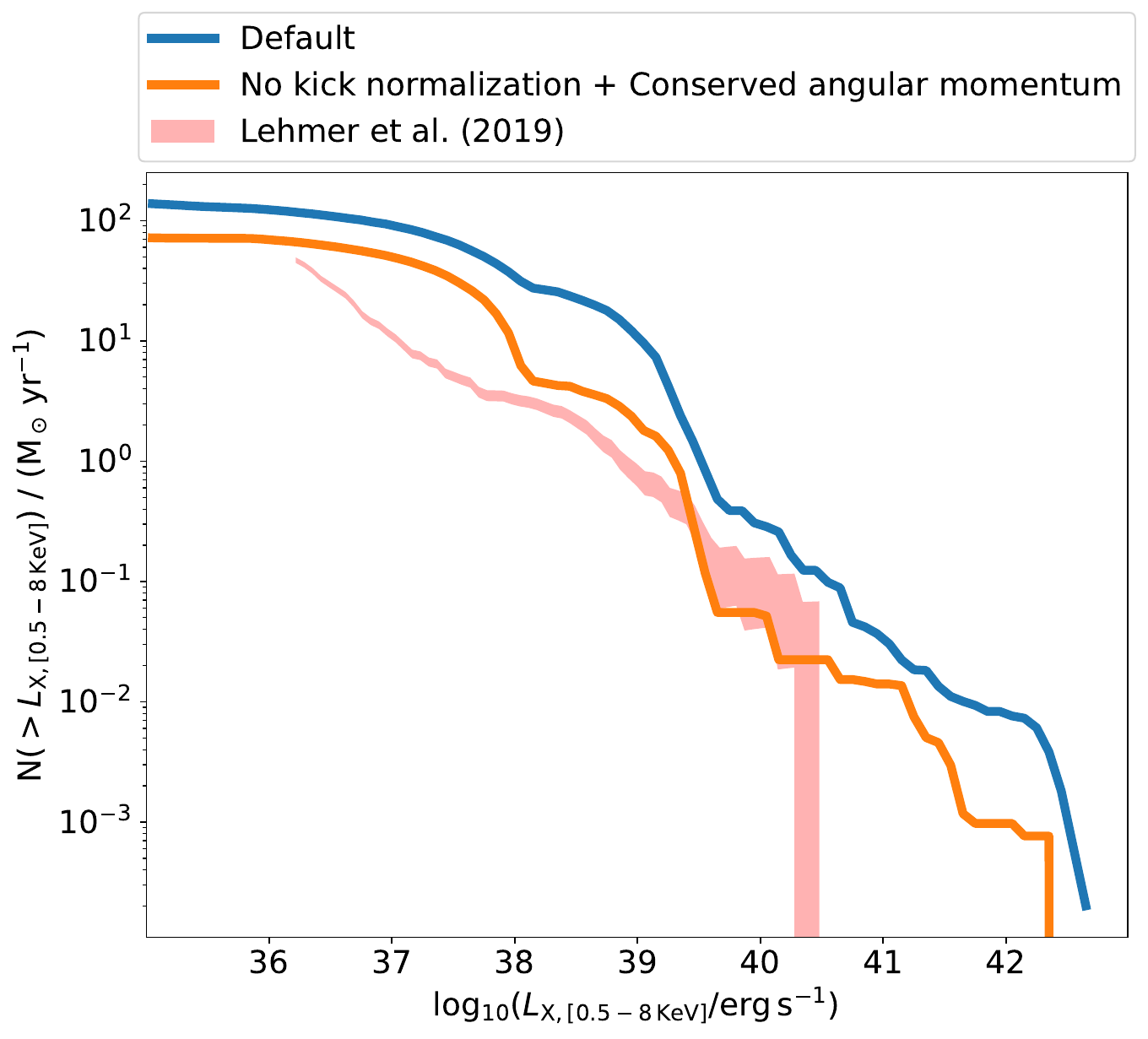} 
\caption{XLFs of populations showing the default population with the parameters in Table~\ref{table:params_def} (blue line) and the population when we do not normalize the BH kicks and conserve the angular momentum of the orbit at the onset of RLO (orange line). The two curves are compared to the observed XLF from \citet{2019ApJS..243....3L}.}
\label{fig:disc_nN_ALT}
\end{figure}

Similarly, the combined effect of BH kicks with no normalization (see Sect.~\ref{sec:sn_kick}) and eccentric orbits that circularize at the onset of RLO at final orbital separations wider than the periastron distance (e.g., by assuming that the angular momentum of the orbit at the onset of RLO is conserved; see Sect.~\ref{sec:circ_effect}) also results in a drastic change in the synthetic XLF. Figure~\ref{fig:disc_nN_ALT} shows the XLF for this case, comparing it to the default model and the observed XLF from \cite[][pink shaded region]{2019ApJS..243....3L}. The normalization of the synthetic XLF with the combined effects compares well to that of the observed XLF, particularly at $10^{38}$~erg~s$^{-1}$ where the slope changes for both the XLFs. However, the slope at luminosities greater than $10^{39}$~erg~s$^{-1}$ is steeper than the observed XLF. 

\subsection{High-mass X-ray binary transients}

Generally, known transient HMXBs are either Be XRBs or super-fast X-ray transients. Peak luminosities of Be XRBs go as high as $\sim 10^{38}$~erg~s$^{-1}$ \citep{2005A&AT...24..151R,2014ApJ...786..128C}, which is the low-luminosity end of the observed XLF from \citet{2019ApJS..243....3L}. In all the XLFs we have studied thus far, we see a consistent population of Be XRBs, as shown in Figs.~\ref{fig:default_donor_type}, \ref{fig:13_donor_type}, \ref{fig:11_donor_type}, and \ref{fig:default_donor_type_rlo_limited}. Be XRBs are not affected by most of the prescriptions that we are investigating. For instance, BH kicks and wind-fed BH-XRB observability are merely affecting the Be-XRB population,  $96\%$ of which are wide NS XRBs in our simulations.

We should stress once again, however, that our treatment of Be-XRB X-ray luminosities is only a rough estimate based on empirical data of the peak luminosities of observed sources in the galaxy, which might overestimate their brightness \citep[luminosities are calculated using the fitting formula from][]{2006ApJ...653.1410D}\boldred{, while still approximately matching the total number of Be XRBs to observations (as seen in Fig.~\ref{fig:default_donor_type})}. Thus, their contribution to the synthetic XLFs presented throughout the paper should only be considered as an order of magnitude estimate. Super-fast X-ray transients are transients with wind-fed OB-type supergiant donors \citep[peak luminosities $\sim 10^{36}$~erg~s$^{-1}$;][]{2006ESASP.604..165N}. However, we do not consider super-fast X-ray transients in our calculations. The mechanisms that cause these systems to be transients are not fully understood and due to their low luminosities (which is at the sensitivity limit of most extragalactic surveys outside the local group), they provide limited insights by comparisons between our synthetic XLFs and the observed XLFs. 

\begin{figure}[!ht]
\centering
\includegraphics[width=\linewidth]{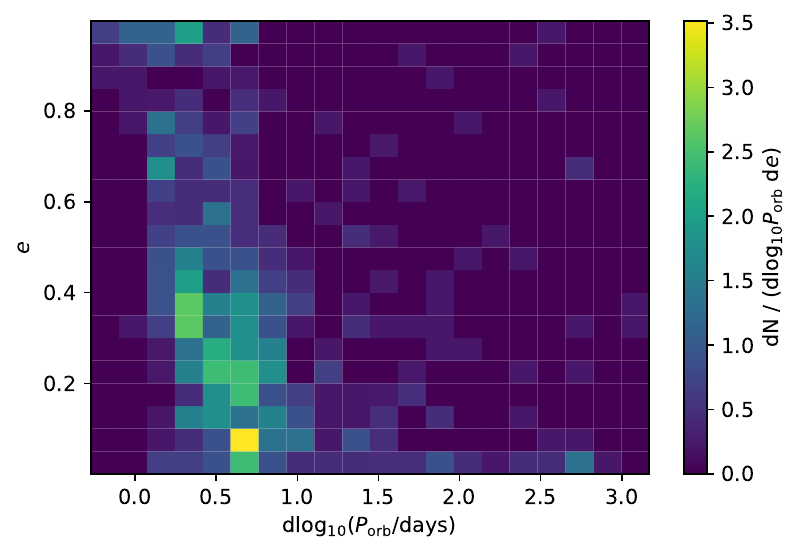}
\caption{Two-dimensional distributions of the eccentricity ($e$) and $\log_{10}$ of \boldred{the} orbital period ($P_{\rm orb}$) at the onset of RLO for the resulting XRBs, for the default model with parameters as described in Table~\ref{table:params_def}.}
\label{fig:ecc_at_rlo}
\end{figure}

\subsection{Instantaneous circularization at the onset of Roche-lobe overflow}

During the formation of the CO, symmetric mass loss and asymmetric natal kicks often impart a significant eccentricity on the post-SN binary orbit. Tidal interactions between the binary components tend to circularize the orbit \citep{1981A&A....99..126H, 1989A&A...223..112Z, 1995A&A...296..709V}, especially for the shorter period binaries. However, tides are often not efficient enough in doing so in the evolutionary timescales of the companion stars, which results in the donor stars overfilling their Roche lobes at the periastron of still eccentric orbits. Figure~\ref{fig:ecc_at_rlo} shows the two-dimensional distributions of orbital period and eccentricity of binaries composed of a CO and non-degenerate companion star at the onset of RLO, for the RLO XRBs present in our default model. \boldred{Also, 70\% of the BH XRBs with H-rich donors undergoing RLO (that dominate the XLF bump) have eccentricities greater than 0.2 at the onset of RLO, with the distribution peaking around a relatively low eccentricity of 0.2 to 0.3. Even so, the XLF bump will be affected by how eccentric binaries are treated and since transient XRBs would be less observed compared to persistent XRBs, the appearance of the XLF bump should reduce. Therefore,} for binaries with significant eccentricities at the onset of RLO, for example above 0.2, which constitute $68\%$ of the \boldred{total} binaries reaching an RLO XRB phase in our default model, the subsequent treatment of the eccentricity evolution during the \boldgreen{mass-transfer} phase become\boldred{s} important.

A commonly made assumption in binary population synthesis codes is that the orbit circularizes instantaneously at the onset of RLO \citep[e.g.,][]{2008ApJS..174..223B, 2023ApJS..264...45F}. Then different assumptions can be made to estimate the post-circularization orbital separation. We discussed some of these assumptions in Sects.~\ref{sec:circi_efefct_des} and \ref{sec:circ_effect}. We showed that they can have a significant impact on the properties of the resulting XRB population. The general trend we observed is that models assuming close circularized orbits at RLO onset (at the periastron) have a significant boost in the subpopulation of RLO XRBs with orbital periods of $\lesssim 10$~days, which also tend to be long-lived. In the synthetic XLFs this contributes to the excess or bump at X-ray luminosities in the range $10^{38}$ to $10^{39}\rm\,erg\,s^{-1}$.

In practice, all of the prescriptions used in our study of the circularization process upon the onset of mass transfer are crude simplifying assumptions, and the variations we observe in the resulting XRB population highlight the need for better treatment of this process. A theoretical framework for the secular evolution of mass-transferring eccentric binaries has been developed in recent years. Additionally, \citet{2007ApJ...667.1170S,2009ApJ...702.1387S, 2010ApJ...724..546S} showed that the circularization depends on the relative effect of the RLO phase and the tidal torques, and that mass transfer does not always circularize the orbit on a short enough timescale, in some cases even making the orbit more eccentric. \citet{2016ApJ...825...70D, 2016ApJ...825...71D}, \citet{2020MNRAS.496.3767V} and \citet{2021MNRAS.503.5569V} have developed more complete theories for the secular evolution of such eccentric mass-transferring binaries. However, although in principle feasible, the computational complexity has not yet allowed the implementation of these theories within the context of detailed binary evolution models. Our results, in addition to some observed mass-transferring binaries in eccentric orbits \citep{2014A&A...564A...1B, 2015A&ARv..23....2W}, highlight that the assumption of instantaneous circularization at the onset of RLO is a crude one, and further development of our modeling tools in that direction is imperative. A more physically realistic treatment of mass transfer in eccentric orbits will be included in future versions of {\tt POSYDON} \citep[][]{2023ApJS..264...45F}.

\subsection{Comparison to observations}
 
The primary scientific goal of this work is to study signatures that different physical processes during the formation of XRBs may imprint in the XLF of a population. Throughout the paper, we have used the observational derived XLF of HMXBs by \citet{2019ApJS..243....3L} as a qualitative benchmark in order to infer if the aforementioned signatures would be identifiable. Here, we make a first attempt to make a quantitative comparison between the synthetic XLFs calculated from the 96 models run with the different combinations of parameters (see Appendix~\ref{appendix:graph}) and the observationally derived one. 

Using the Kolmogorov-Smirnov test, we measured the probability that samples drawn from our {\tt POSYDON} populations (with $\sim 1.2 \times 10^4$ XRBs in each population) are drawn from the same underlying population as the samples drawn from the observed XLF ($10^4$ samples drawn). Even though this test is not sensitive at the tail ends of the distributions, we are more concerned with comparing the general shape and fit of the distribution. We use the one-sample Kolmogorov-Smirnov test to assess whether a synthetic XLF agrees with the observed XLF from \citet{2019ApJS..243....3L}. We find disagreement at a level of 0.01 for all XLFs from the simulated populations ($P_{\rm KS} \ll 0.01$ in all cases), which suggests that our synthetic XLFs are not drawn from the same underlying population as the observed XLF.

To identify the ``best-fitting'' model, we used two approaches. Firstly, we match the slope of the XLFs and disregard the differences in the XLF normalization. We identify models that would define the observed population (requiring a correction for the normalization). We calculate the ratios of the number of XRBs in different bins of the XLF and compare \boldred{them} to ratios in the observed XLF. Secondly, we match the normalization of the XLFs with the observed \boldred{ones} and find the best-fitting models. However, following this approach, \boldgreen{either} the shape or slope of the selected XLF might not match the observed XLF. To compare the normalizations, we do a chi-squared test using the number of XRBs in each bin of the synthetic XLF and the number of observed XRBs in each bin. 

\begin{figure}[!ht]
\centering
\includegraphics[width=\linewidth]{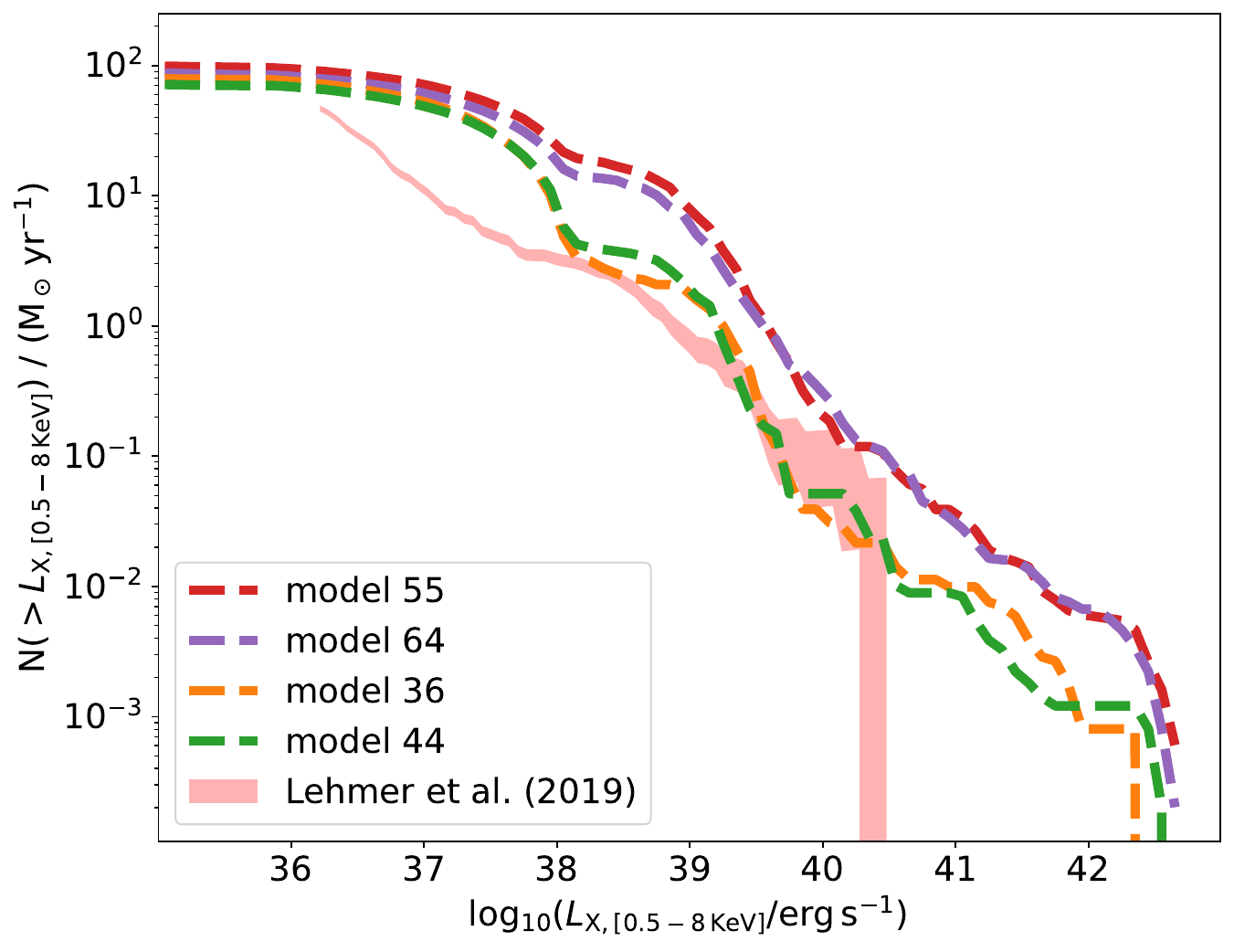} 
\caption{XLFs of populations showing the four best-fit models, which were run with the parameters described in Table~\ref{table:best_fit}. The synthetic curves are compared to the observed XLF from \citet{2019ApJS..243....3L}.}
\label{fig:best_fit}
\end{figure}

We present the two best-fitting models from each of one of the two approaches in Fig.~\ref{fig:best_fit}. The specific parameters for the models shown in the figure are described in Table~\ref{table:best_fit}. Models~55 and 64 were chosen based on the first approach (i.e., matching the overall shape of the observed XLF), but they overestimate the number of XRBs by up to one order of magnitude. Models~36 and 44 were chosen based on the second approach (i.e., matching the normalization of the observed XLF), and they reproduce the change of slope for the observations at luminosities of $10^{38}$~erg~s$^{-1}$. However, their overall shape is not consistent with observations. The two approaches selected two types of models with some key differences between the groups that cause the differences seen in Fig.~\ref{fig:best_fit}. Models~55 and 64 have BH-mass normalized kicks that result in weaker kicks than in models~36 and 44, \boldred{which} have no kick normalization. The stronger kicks in models~36 and 44 disrupt a part of the close orbit binaries when the SN occurs, resulting in \boldred{a} lower number of XRBs at $10^{38}$~erg~s$^{-1}$, where the XLF starts to be dominated by RLO systems.

\begin{table*}[]
\resizebox{\textwidth}{!}{
\begin{tabular}{l|llll}
\hline\hline
Parameters                          & model 55                                               & model 64                                     & model 36                                               & model 44                                     \\ \hline
Remnant mass prescription                  & \citet{2012ApJ...749...91F} (Delayed) & \citet{2020MNRAS.499.2803P} & \citet{2012ApJ...749...91F} (Delayed) & \citet{2020MNRAS.499.2803P} \\
Natal kick normalization    & BH mass normalized kicks                              & BH mass normalized kicks                    & No kick normalization                                 & No kick normalization                       \\
Orbit circularization at RLO & Periastron                                            & Conserved angular momentum                  & Conserved angular momentum                            & Conserved angular momentum                  \\
CE efficiency ($\alpha_{\rm CE}$)  & 1.0                                                   & 1.0                                         & 0.3                                                   & 0.3                                         \\
CE core-envelope boundary   & At $X_{\rm H} = 0.30$                              & At $X_{\rm H} = 0.30$                    & At $X_{\rm H} = 0.30$                              & At $X_{\rm H} = 0.30$                    \\
Observable wind-fed disk          & \citet{2021PASA...38...56H}           & \citet{2021PASA...38...56H}  & No criterion                                          & No criterion  
\\\hline
\end{tabular}}
\caption{Physics parameters corresponding to the best-fit models from this study. The best-fit models were selected based on two criteria, discussed in Sect.~\ref{sec:discussion}. }
\label{table:best_fit}
\end{table*}


\section{Conclusions}\label{sec:conclusion}

In this work we have investigated the effects of different physical assumptions in the context of XRB formation in young stellar populations. We ran populations through 96 models with different combinations of parameters (described in Table~\ref{table:params}). For this, we used {\tt POSYDON}, which is a new binary population synthesis code that utilizes detailed stellar structure and binary evolution models throughout the entire evolution of a binary. The populations were run assuming a constant SFR over a time duration of $100$~Myr. We created synthetic XLFs to study the simulated XRBs and compared them to observed HMXB XLFs from \citet{2019ApJS..243....3L} to obtain insights into the physics involved. The individual effects of the different physics parameters were discussed, and four best-suited models that most closely match the observations were chosen. We summarize our main findings in the following list:
\begin{itemize}
    \item The HMXB XLFs we generate\boldred{d} have a more complex shape than a single power law (for instance, see Fig.~\ref{fig:default_donor_type}). This qualitative effect is also seen in the empirical XLF from observations. In our simulations, the shape changes at a luminosity of $10^{38}$~erg~s$^{-1}$ (similar to observations) due to the population of RLO BH XRBs with H-rich donors. 
    \item There is an overabundance of XRBs compared to observations (up to a factor of $10$) for certain model parameter combinations at intermediate X-ray luminosities (referred to as the XLF bump, seen around luminosities of $10^{38}$ to $10^{39}$~erg~s$^{-1}$ in Fig.~\ref{fig:default_donor_type}), primarily due to RLO BH XRBs with H-rich donors. 
    \item Increasing the strength of SN kicks in BH XRBs does not necessarily disrupt close binaries. Instead\boldred{,} it results in a larger number of highly eccentric binaries that end up as bright RLO XRBs, compared to weaker kicks. However, strong kicks disrupt wide binaries, reducing the number of bright wind-fed BH XRBs (by a factor of $\sim 20$).
    \item Changing the core-envelope boundary does not have a significant effect on the surviving populations of NS and BH binaries. However, a more inefficient CE phase, due to either lower assumed $\alpha_{CE}$ values or higher envelope binding energies, would result in closer orbits leading to more XRBs.
    \item Including a limiting criterion for the observability of wind BH HMXBs, for example by requiring the formation of an accretion disk around the BH as in \citet{2021PASA...38...56H}, does not affect XRB luminosities greater than approximately $10^{38}$~erg~s$^{-1}$, and decreases the slope of the XLF at lower luminosities. However, this criterion does not affect most Be XRBs ($\sim 96\%$ of which are NS XRBs) as it applies to only wide BH XRBs.
    \item Commonly used assumptions such as instantaneous circularization at the periastron with the onset of RLO might not be accurate for binaries that have high eccentricities when the donor fills its Roche lobe (\boldred{$\sim 68\%$} have eccentricities above \boldred{$0.2$}). These eccentric sources might appear as transient HMXBs, which should be taken into account when simulating XLFs. 
    \item No combination of the explored physical parameters (combinations of parameters described in Table~\ref{table:params} and the XLFs shown in Fig.~\ref{fig:combined_24} and described in Table~\ref{table:all_params_1}) matches well both the shape and the normalization of the observed XLF simultaneously. However, in this work we did not explore any variations of the distributions of initial binary properties, such as variations in the initial mass function, orbital separation distributions, binary mass ratios, and the binary fraction,  which are expected to primarily change the overall normalization of the modeled XLFs, rather than their shape. 
\end{itemize}
Even though our synthetic XLFs do not always agree with observations, this work reveals the importance of large-scale parameter studies. Future comparative work that includes multiwavelength characterization studies of XRB populations across various metallicities and SFHs would help constrain the physics involved in binary evolution.

\begin{acknowledgements}
The authors thank the anonymous referee for their constructive comments that helped improve the manuscript. This work was supported by the Swiss National Science Foundation Professorship Grant (PP00P2\_176868; PI Fragos). This work was also supported by the European Union's Horizon 2020 research and innovation program under the Marie Sklodowska-Curie RISE action, grant agreement No 873089 (ASTROSTAT-II). \boldred{DM acknowledges support from the European Research Council (ERC) under the European Union’s Horizon 2020 research and innovation programme (grant agreement No. 101002352)}. KK acknowledges support from the Federal Commission for Scholarships for Foreign Students for the Swiss Government Excellence Scholarship (ESKAS No. 2021.0277). EZ acknowledges funding support from the European Research Council (ERC) under the European Union’s Horizon 2020 research and innovation programme (Grant agreement No. 772086).

\end{acknowledgements}

\bibliographystyle{aa}
\bibliography{main}

\begin{appendix}

\section{XLFs of the simulated population}
\label{appendix:graph}
We show all the cumulative XLFs from the 96 simulations carried out for this study in Figs.~\ref{fig:combined_24}, \ref{fig:combined_48}, \ref{fig:combined_72}, and \ref{fig:combined_96}, and all the corresponding differential forms of the XLFs in Figs.~\ref{fig:diff_combined_24}, \ref{fig:diff_combined_48}, \ref{fig:diff_combined_72}, and \ref{fig:diff_combined_96}. Further information about the combination of parameters that were used to run these populations are shared in Table~\ref{table:all_params_1}.

\begin{figure*}[htb]
\hspace*{1cm}\includegraphics[width=0.9\textwidth]{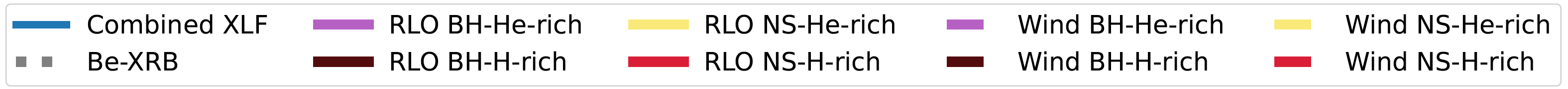}\\
\centering
\begin{tikzpicture}
\node (img1)  {\includegraphics[width=\textwidth]{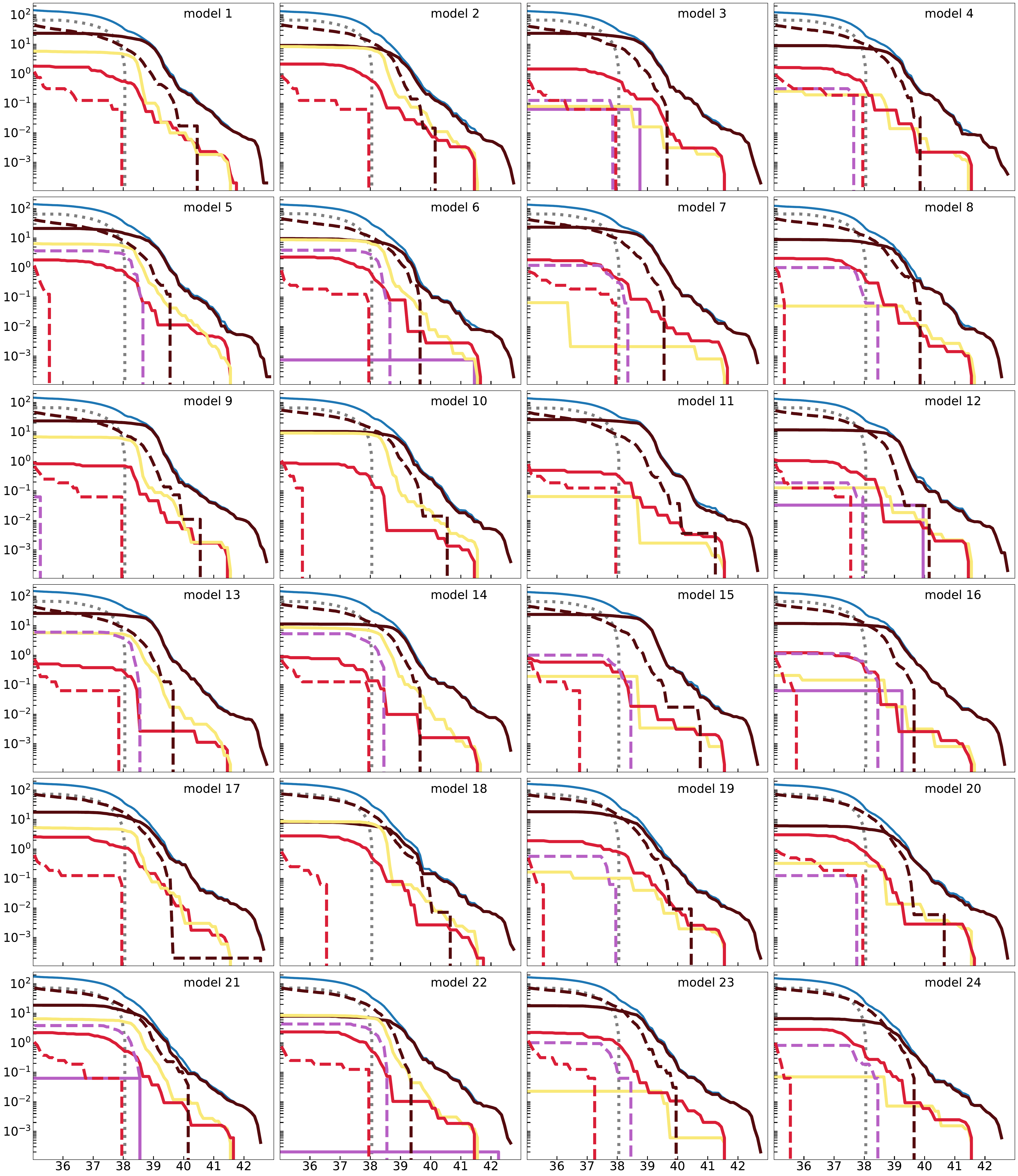}};
\node[node distance=0cm, yshift=-10.8cm] {log$_{10} (L_{\rm X,[0.5-8\,\rm KeV]}/\mathrm{erg\,s^{-1}}$)};
\node[node distance=0cm, rotate=90, anchor=center,yshift=9.4cm] {N$(>L_{\rm X,[0.5-8\,\rm KeV]})$ / ($\mathrm{M_\odot\,yr^{-1}}$)};
\end{tikzpicture}
\caption{Synthetic XLFs of populations 1 to 24 from Table~\ref{table:all_params_1}, showing the types of mass transfer that \boldred{occurred}, the donors, and the accretors.}
\label{fig:combined_24}
\end{figure*}

\begin{figure*}[htb]
\ContinuedFloat
\hspace*{1cm}\includegraphics[width=0.9\textwidth]{xlf_label.png}\\
\centering
\begin{tikzpicture}
\node (img1)  {\includegraphics[width=\textwidth]{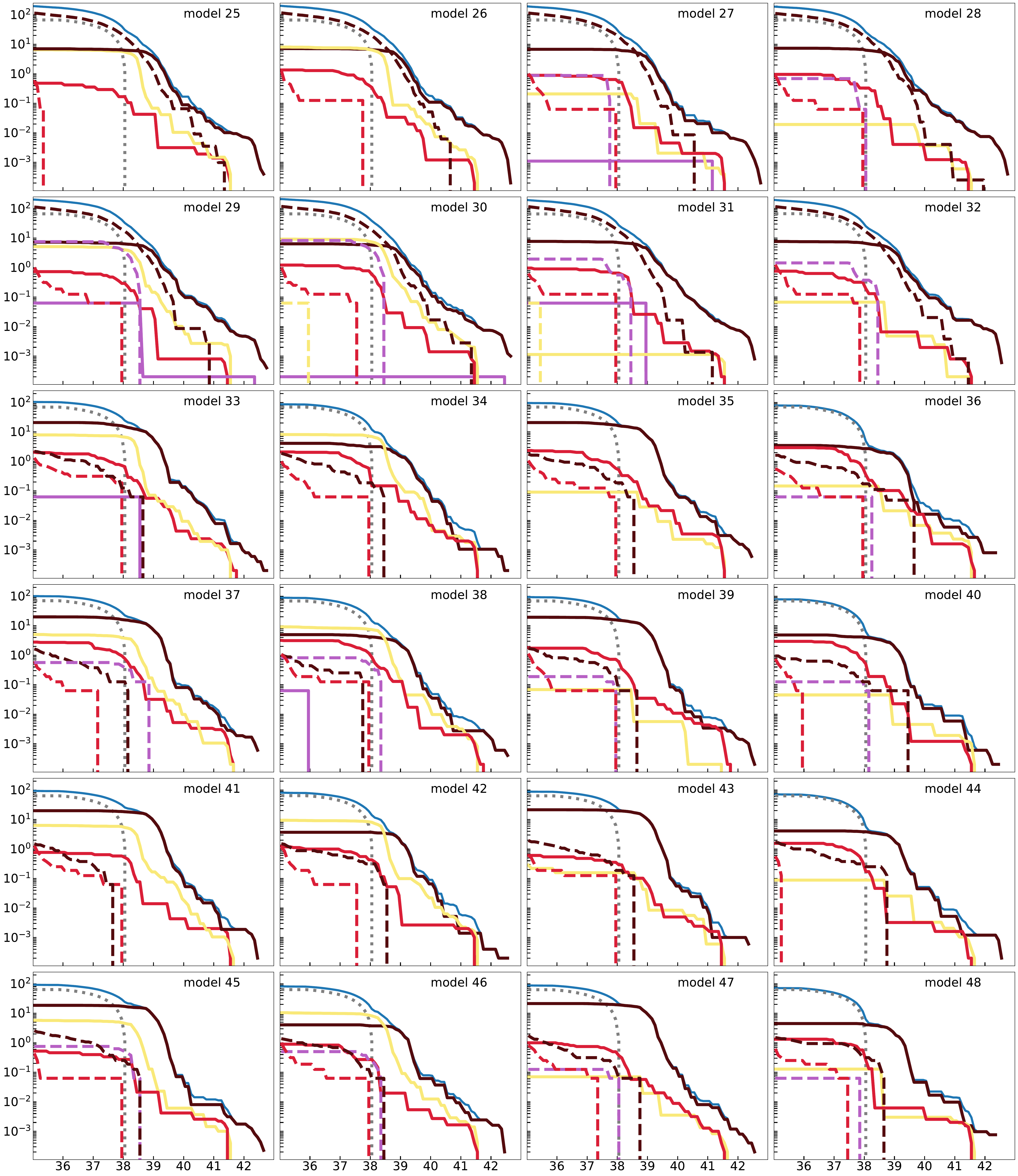}};
\node[node distance=0cm, yshift=-10.8cm] {log$_{10} (L_{\rm X,[0.5-8\,\rm KeV]}/\mathrm{erg\,s^{-1}}$)};
\node[node distance=0cm, rotate=90, anchor=center,yshift=9.4cm] {N$(>L_{\rm X,[0.5-8\,\rm KeV]})$ / ($\mathrm{M_\odot\,yr^{-1}}$)};
\end{tikzpicture}
\caption{Synthetic XLFs of populations 25 to 48 from Table~\ref{table:all_params_1}, showing the types of mass transfer that \boldred{occurred}, the donors, and the accretors.}
\label{fig:combined_48}
\end{figure*}

\begin{figure*}[htb]
\ContinuedFloat
\hspace*{1cm}\includegraphics[width=0.9\textwidth]{xlf_label.png}\\
\centering
\begin{tikzpicture}
\node (img1)  {\includegraphics[width=\textwidth]{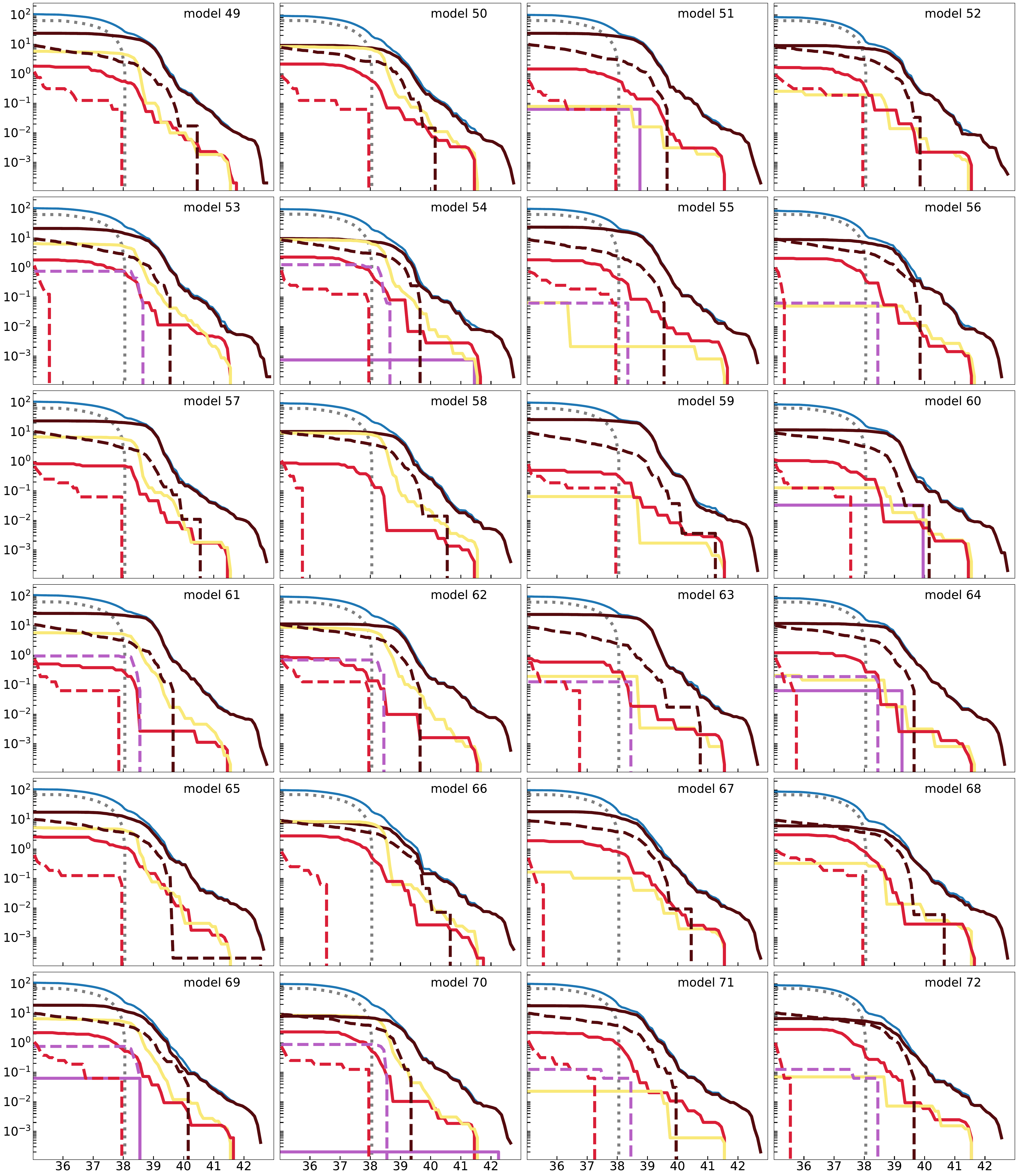}};
\node[node distance=0cm, yshift=-10.8cm] {log$_{10} (L_{\rm X,[0.5-8\,\rm KeV]}/\mathrm{erg\,s^{-1}}$)};
\node[node distance=0cm, rotate=90, anchor=center,yshift=9.4cm] {N$(>L_{\rm X,[0.5-8\,\rm KeV]})$ / ($\mathrm{M_\odot\,yr^{-1}}$)};
\end{tikzpicture}
\caption{Synthetic XLFs of populations 49 to 72 from Tables~\ref{table:all_params_1}, showing the types of mass transfer that \boldred{occurred}, the donors, and the accretors.}
\label{fig:combined_72}
\end{figure*}

\begin{figure*}[htb]
\ContinuedFloat
\hspace*{1cm}\includegraphics[width=0.9\textwidth]{xlf_label.png}\\
\centering
\begin{tikzpicture}
\node (img1)  {\includegraphics[width=\textwidth]{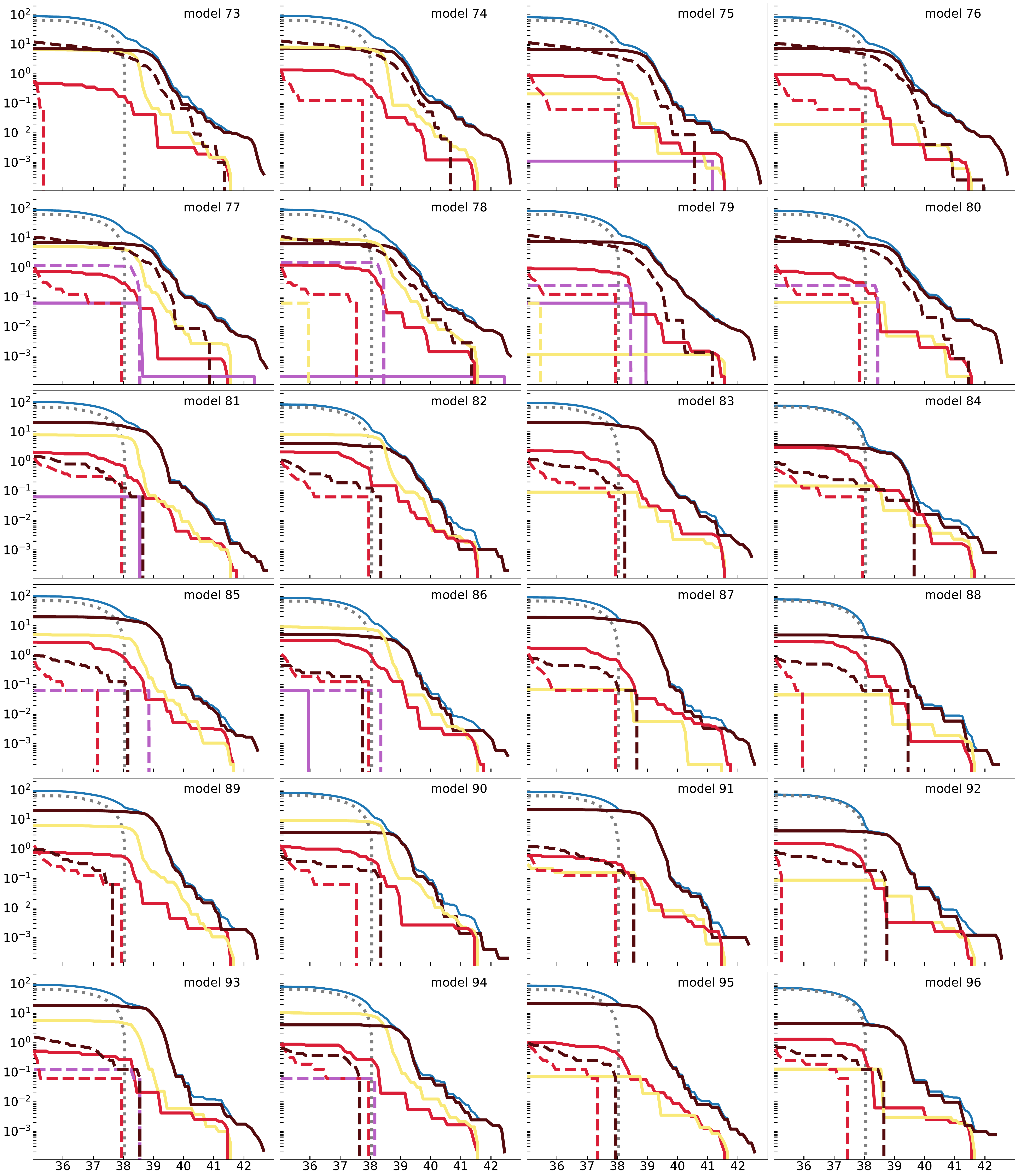}};
\node[node distance=0cm, yshift=-10.8cm] {log$_{10} (L_{\rm X,[0.5-8\,\rm KeV]}/\mathrm{erg\,s^{-1}}$)};
\node[node distance=0cm, rotate=90, anchor=center,yshift=9.4cm] {N$(>L_{\rm X,[0.5-8\,\rm KeV]})$ / ($\mathrm{M_\odot\,yr^{-1}}$)};
\end{tikzpicture}
\caption{Synthetic XLFs of populations 73 to 96 from Table~\ref{table:all_params_1}, showing the types of mass transfer that \boldred{occurred}, the donors, and the accretors.}
\label{fig:combined_96}
\end{figure*}


\begin{figure*}[htb]
\hspace*{1cm}\includegraphics[width=0.9\textwidth]{xlf_label.png}\\
\centering
\begin{tikzpicture}
\node (img1)  {\includegraphics[width=\textwidth]{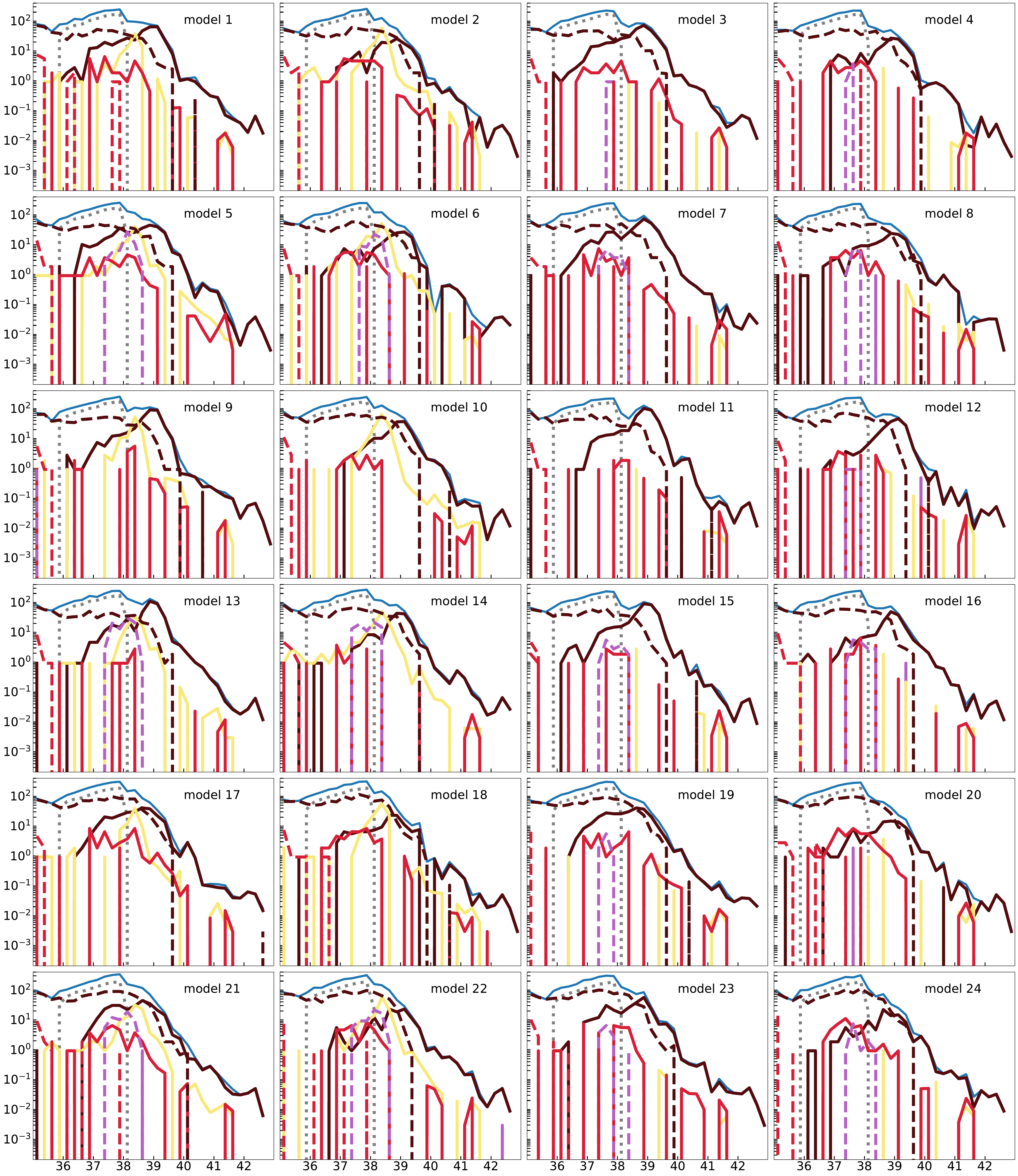}};
\node[node distance=0cm, yshift=-10.8cm] {log$_{10} (L_{\rm X,[0.5-8\,\rm KeV]}/\mathrm{erg\,s^{-1}}$)};
\node[node distance=0cm, rotate=90, anchor=center,yshift=9.4cm] {dN / dlog$_{10}L_{\rm X,[0.5-8\,\rm KeV]}$ / ($\mathrm{M_\odot\,yr^{-1}}$)};
\end{tikzpicture}
\caption{Synthetic differential forms of the XLFs for populations 1 to 24 from Table~\ref{table:all_params_1}, showing the types of mass transfer that \boldred{occurred}, the donors, and the accretors.}
\label{fig:diff_combined_24}
\end{figure*}

\begin{figure*}[htb]
\ContinuedFloat
\hspace*{1cm}\includegraphics[width=0.9\textwidth]{xlf_label.png}\\
\centering
\begin{tikzpicture}
\node (img1)  {\includegraphics[width=\textwidth]{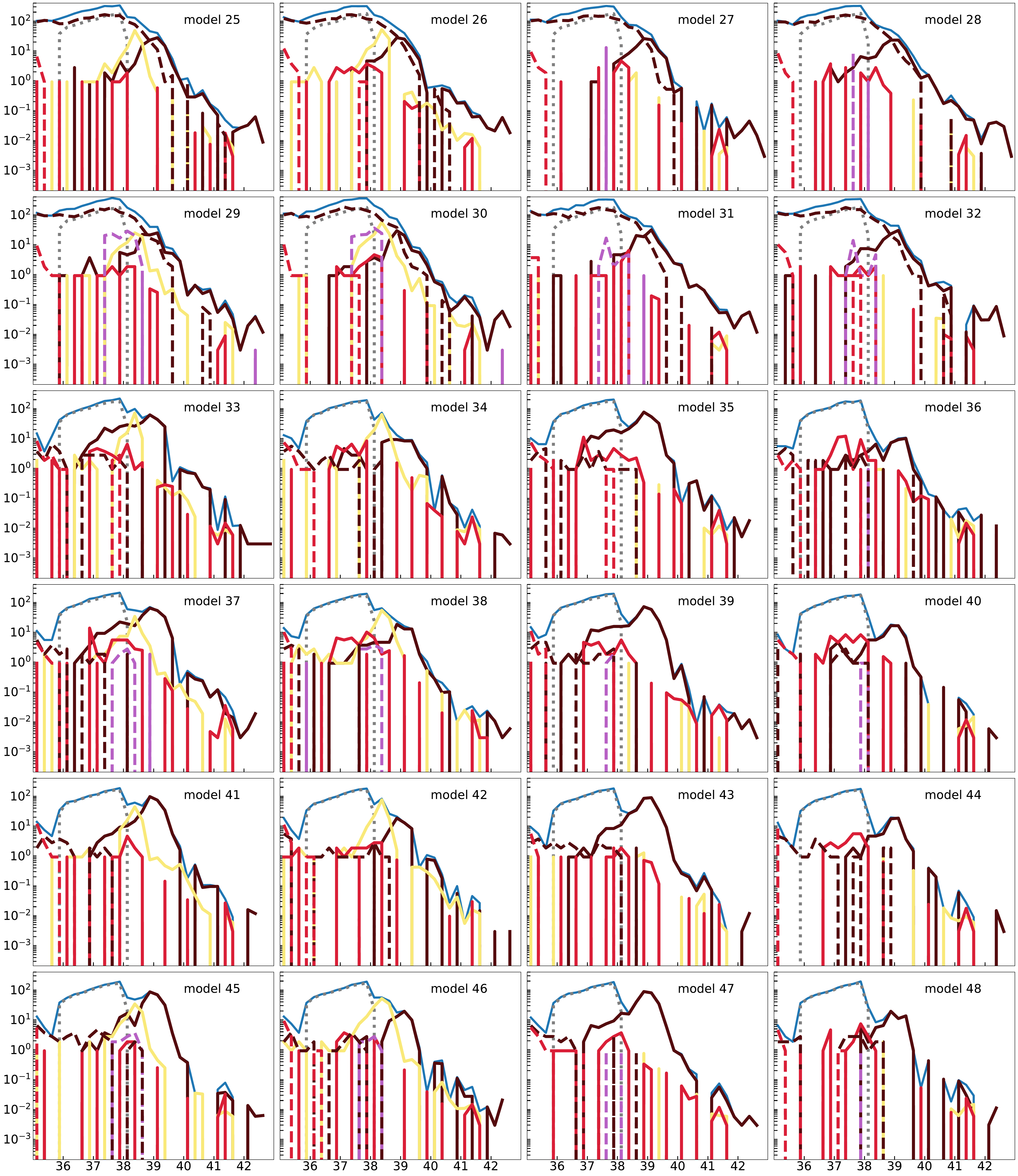}};
\node[node distance=0cm, yshift=-10.8cm] {log$_{10} (L_{\rm X,[0.5-8\,\rm KeV]}/\mathrm{erg\,s^{-1}}$)};
\node[node distance=0cm, rotate=90, anchor=center,yshift=9.4cm] {dN / dlogL$_{\rm X,[0.5-8\,\rm KeV]}$ / ($\mathrm{M_\odot\,yr^{-1}}$)};
\end{tikzpicture}
\caption{Synthetic differential forms of the XLFs for populations 25 to 48 from Table~\ref{table:all_params_1}, showing the types of mass transfer that \boldred{occurred}, the donors, and the accretors.}
\label{fig:diff_combined_48}
\end{figure*}

\begin{figure*}[htb]
\ContinuedFloat
\hspace*{1cm}\includegraphics[width=0.9\textwidth]{xlf_label.png}\\
\centering
\begin{tikzpicture}
\node (img1)  {\includegraphics[width=\textwidth]{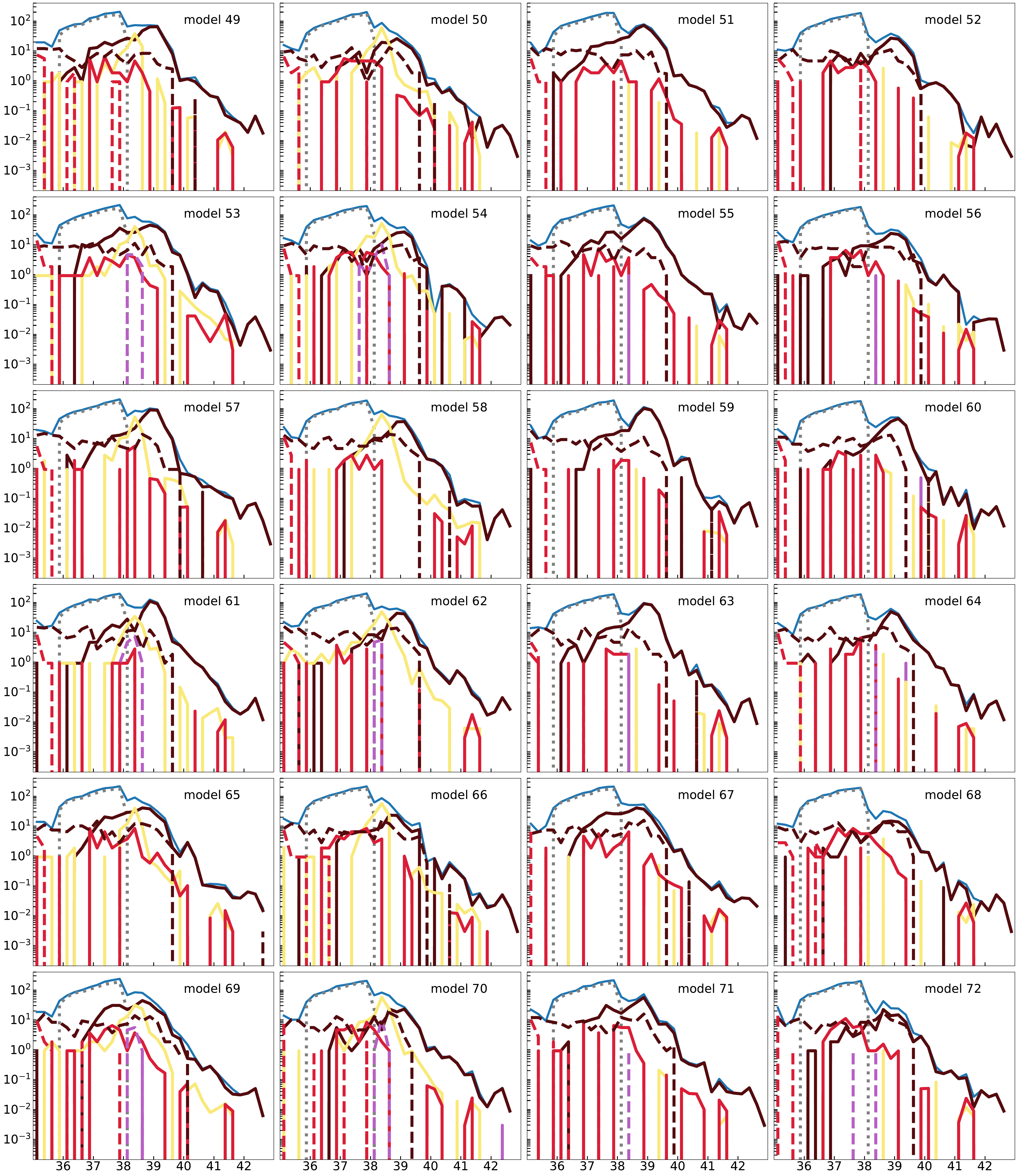}};
\node[node distance=0cm, yshift=-10.8cm] {log$_{10} (L_{\rm X,[0.5-8\,\rm KeV]}/\mathrm{erg\,s^{-1}}$)};
\node[node distance=0cm, rotate=90, anchor=center,yshift=9.4cm] {dN / dlog$_{10}L_{\rm X,[0.5-8\,\rm KeV]}$ / ($\mathrm{M_\odot\,yr^{-1}}$)};
\end{tikzpicture}
\caption{Synthetic differential forms of the XLFs for populations 49 to 72 from Tables~\ref{table:all_params_1}, showing the types of mass transfer that \boldred{occurred}, the donors, and the accretors.}
\label{fig:diff_combined_72}
\end{figure*}

\begin{figure*}[htb]
\ContinuedFloat
\hspace*{1cm}\includegraphics[width=0.9\textwidth]{xlf_label.png}\\
\centering
\begin{tikzpicture}
\node (img1)  {\includegraphics[width=\textwidth]{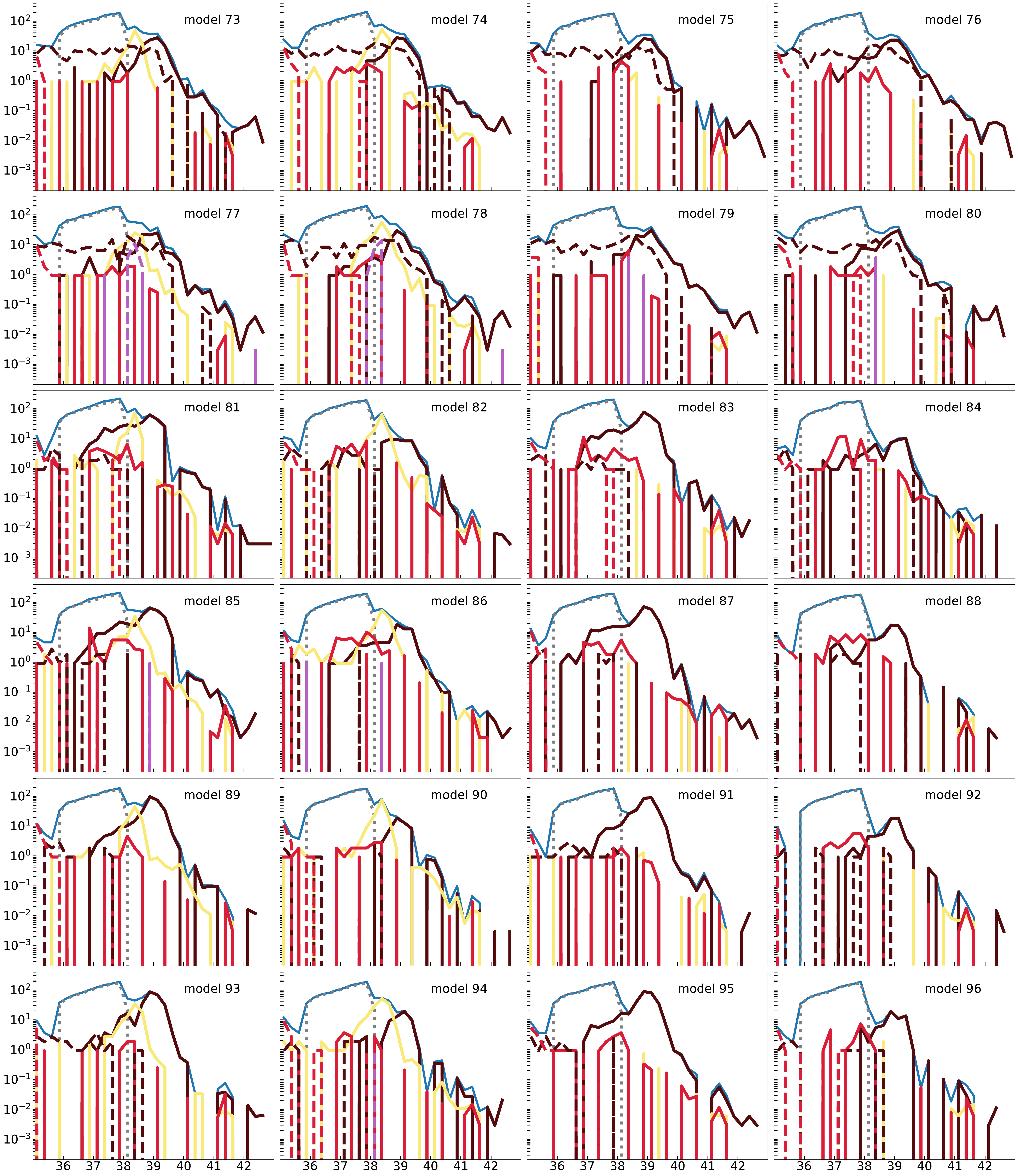}};
\node[node distance=0cm, yshift=-10.8cm] {log$_{10} (L_{\rm X,[0.5-8\,\rm KeV]}/\mathrm{erg\,s^{-1}}$)};
\node[node distance=0cm, rotate=90, anchor=center,yshift=9.4cm] {dN / dlog$_{10}L_{\rm X,[0.5-8\,\rm KeV]}$ / ($\mathrm{M_\odot\,yr^{-1}}$)};
\end{tikzpicture}
\caption{Synthetic differential forms of the XLFs for populations 73 to 96 from Table~\ref{table:all_params_1}, showing the types of mass transfer that \boldred{occurred}, the donors, and the accretors.}
\label{fig:diff_combined_96}
\end{figure*}

\begin{table*}[]
\begin{tabular}{c|cccccc}
\hline \hline
models & Remnant mass  & Natal kick norm. & Circularization at RLO &  $\alpha_{\rm CE}$ & CE core-envelope boundary &  Wind-fed disk \\ \hline

model 1 & FD2012 & BH mass & Periastron & 0.3 & At $X_{\rm H} = 0.01$ & No criterion \\
model 2 & FD2012 & BH mass & Conserved AM & 0.3 & At $X_{\rm H} = 0.01$ &  No criterion \\
model 3 & FD2012 & BH mass & Periastron & 0.3 & At $X_{\rm H} = 0.30$ & No criterion \\
model 4 & FD2012 & BH mass &  Conserved AM & 0.3 & At $X_{\rm H} = 0.30$ &  No criterion \\
model 5 & FD2012 & BH mass &  Periastron & 1.0 & At $X_{\rm H} = 0.01$ & No criterion \\
model 6 & FD2012 & BH mass &  Conserved AM & 1.0 & At $X_{\rm H} = 0.01$ & No criterion \\
model 7 & FD2012 & BH mass &  Periastron & 1.0 & At $X_{\rm H} = 0.30$ & No criterion \\
model 8 & FD2012 & BH mass & Conserved AM & 1.0 & At $X_{\rm H} = 0.30$ & No criterion \\
model 9 & PS2020 & BH mass & Periastron & 0.3 & At $X_{\rm H} = 0.01$ & No criterion \\
model 10 & PS2020 & BH mass & Conserved AM & 0.3 & At $X_{\rm H} = 0.01$ & No criterion \\
model 11 & PS2020 & BH mass & Periastron & 0.3 & At $X_{\rm H} = 0.30$ & No criterion \\
model 12 & PS2020 & BH mass & Conserved AM & 0.3 & At $X_{\rm H} = 0.30$ & No criterion \\
model 13 & PS2020 & BH mass & Periastron & 1.0 & At $X_{\rm H} = 0.01$ & No criterion \\
model 14 & PS2020 & BH mass & Conserved AM & 1.0 & At $X_{\rm H} = 0.01$ & No criterion \\
model 15 & PS2020 & BH mass & Periastron & 1.0 & At $X_{\rm H} = 0.30$ & No criterion \\
model 16 & PS2020 & BH mass & Conserved AM & 1.0 & At $X_{\rm H} = 0.30$ & No criterion \\
model 17 & FD2012 & Fall-back  & Periastron & 0.3 & At $X_{\rm H} = 0.01$ & No criterion \\
model 18 & FD2012 & Fall-back  & Conserved AM & 0.3 & At $X_{\rm H} = 0.01$ & No criterion \\
model 19 & FD2012 & Fall-back  & Periastron & 0.3 & At $X_{\rm H} = 0.30$ & No criterion \\
model 20 & FD2012 & Fall-back  & Conserved AM & 0.3 & At $X_{\rm H} = 0.30$ & No criterion \\
model 21 & FD2012 & Fall-back  & Periastron & 1.0 & At $X_{\rm H} = 0.01$ & No criterion \\
model 22 & FD2012 & Fall-back  & Conserved AM & 1.0 & At $X_{\rm H} = 0.01$ & No criterion \\
model 23 & FD2012 & Fall-back  & Periastron & 1.0 & At $X_{\rm H} = 0.30$ & No criterion \\
model 24 & FD2012 & Fall-back  & Conserved AM & 1.0 & At $X_{\rm H} = 0.30$ & No criterion \\
model 25 & PS2020 & Fall-back & Periastron & 0.3 & At $X_{\rm H} = 0.01$ & No criterion \\
model 26 & PS2020 & Fall-back & Conserved AM & 0.3 & At $X_{\rm H} = 0.01$ & No criterion \\
model 27 & PS2020 & Fall-back & Periastron & 0.3 & At $X_{\rm H} = 0.30$ & No criterion \\
model 28 & PS2020 & Fall-back & Conserved AM & 0.3 & At $X_{\rm H} = 0.30$ & No criterion \\
model 29 & PS2020 & Fall-back & Periastron & 1.0 & At $X_{\rm H} = 0.01$ & No criterion \\
model 30 & PS2020 & Fall-back & Conserved AM & 1.0 & At $X_{\rm H} = 0.01$ & No criterion \\
model 31 & PS2020 & Fall-back & Periastron & 1.0 & At $X_{\rm H} = 0.30$ & No criterion \\
model 32 & PS2020 & Fall-back & Conserved AM & 1.0 & At $X_{\rm H} = 0.30$ & No criterion \\

model 33 & FD2012 & No norm. & Periastron & 0.3 & At $X_{\rm H} = 0.01$ & No criterion \\
model 34 & FD2012 & No norm. & Conserved AM & 0.3 & At $X_{\rm H} = 0.01$ & No criterion \\
model 35 & FD2012 & No norm. & Periastron & 0.3 & At $X_{\rm H} = 0.30$ & No criterion \\
model 36 & FD2012 & No norm. & Conserved AM & 0.3 & At $X_{\rm H} = 0.30$ & No criterion \\
model 37 & FD2012 & No norm. & Periastron & 1.0 & At $X_{\rm H} = 0.01$ & No criterion \\
model 38 & FD2012 & No norm. & Conserved AM & 1.0 & At $X_{\rm H} = 0.01$ & No criterion \\
model 39 & FD2012 & No norm. & Periastron & 1.0 & At $X_{\rm H} = 0.30$ & No criterion \\
model 40 & FD2012 & No norm. & Conserved AM & 1.0 & At $X_{\rm H} = 0.30$ & No criterion \\
model 41 & PS2020 & No norm.  & Periastron & 0.3 & At $X_{\rm H} = 0.01$ & No criterion \\
model 42 & PS2020 & No norm.  & Conserved AM & 0.3 & At $X_{\rm H} = 0.01$ & No criterion \\
model 43 & PS2020 & No norm.  & Periastron & 0.3 & At $X_{\rm H} = 0.30$ & No criterion \\
model 44 & PS2020 & No norm.  & Conserved AM & 0.3 & At $X_{\rm H} = 0.30$ & No criterion \\
model 45 & PS2020 & No norm.  & Periastron & 1.0 & At $X_{\rm H} = 0.01$ & No criterion \\
model 46 & PS2020 & No norm.  & Conserved AM & 1.0 & At $X_{\rm H} = 0.01$ & No criterion \\
model 47 & PS2020 & No norm.  & Periastron & 1.0 & At $X_{\rm H} = 0.30$ & No criterion \\
model 48 & PS2020 & No norm.  & Conserved AM & 1.0 & At $X_{\rm H} = 0.30$ & No criterion \\

model 49 & FD2012 & BH mass & Periastron & 0.3 & At $X_{\rm H} = 0.01$ & HM2021 \\
model 50 & FD2012 & BH mass &  Conserved AM & 0.3 & At $X_{\rm H} = 0.01$ & HM2021 \\
model 51 & FD2012 & BH mass &  Periastron & 0.3 & At $X_{\rm H} = 0.30$ & HM2021 \\
model 52 & FD2012 & BH mass &  Conserved AM & 0.3 & At $X_{\rm H} = 0.30$ & HM2021 \\
model 53 & FD2012 & BH mass &  Periastron & 1.0 & At $X_{\rm H} = 0.01$ & HM2021 \\
model 54 & FD2012 & BH mass &  Conserved AM & 1.0 & At $X_{\rm H} = 0.01$ & HM2021 \\
model 55 & FD2012 & BH mass & Periastron & 1.0 & At $X_{\rm H} = 0.30$ & HM2021 \\
model 56 & FD2012 & BH mass & Conserved AM & 1.0 & At $X_{\rm H} = 0.30$ & HM2021 \\
model 57 & PS2020 & BH mass & Periastron & 0.3 & At $X_{\rm H} = 0.01$ & HM2021 \\
model 58 & PS2020 & BH mass & Conserved AM & 0.3 & At $X_{\rm H} = 0.01$ & HM2021 \\
model 59 & PS2020 & BH mass & Periastron & 0.3 & At $X_{\rm H} = 0.30$ & HM2021 \\
model 60 & PS2020 & BH mass & Conserved AM & 0.3 & At $X_{\rm H} = 0.30$ & HM2021 \\
model 61 & PS2020 & BH mass & Periastron & 1.0 & At $X_{\rm H} = 0.01$ & HM2021 \\
model 62 & PS2020 & BH mass & Conserved AM & 1.0 & At $X_{\rm H} = 0.01$ & HM2021 \\
model 63 & PS2020 & BH mass & Periastron & 1.0 & At $X_{\rm H} = 0.30$ & HM2021 \\
model 64 & PS2020 & BH mass & Conserved AM & 1.0 & At $X_{\rm H} = 0.30$ & HM2021 \\

\\ \hline
\end{tabular}
\caption{Populations run with all the different combinations of parameters (refer to Table~\ref{table:params}) used for this study. FD2012 refers to the delayed SN prescription from \citet{2012ApJ...749...91F} and PS2020 refers to the SN prescription from \citet{2020MNRAS.499.2803P} (refer to Sect.~\ref{sec:sn_mech}). HM2021 refers to the observable wind-fed accretion criterion by \citet{2021PASA...38...56H} and AM stands for angular momentum.}
\label{table:all_params_1}
\end{table*}

\begin{table*}[]
\ContinuedFloat  
\begin{tabular}{c|cccccc}
\hline \hline
models & Remnant mass  & Natal kick norm. & Circularization at RLO &  $\alpha_{\rm CE}$ & CE core-envelope boundary &  Wind-fed disk \\ \hline

model 65 & FD2012 & Fall-back  & Periastron & 0.3 & At $X_{\rm H} = 0.01$ & HM2021 \\
model 66 & FD2012 & Fall-back  &  Conserved AM & 0.3 & At $X_{\rm H} = 0.01$ & HM2021 \\
model 67 & FD2012 & Fall-back  & Periastron & 0.3 & At $X_{\rm H} = 0.30$ & HM2021 \\
model 68 & FD2012 & Fall-back  &  Conserved AM & 0.3 & At $X_{\rm H} = 0.30$ & HM2021 \\
model 69 & FD2012 & Fall-back  & Periastron & 1.0 & At $X_{\rm H} = 0.01$ & HM2021 \\
model 70 & FD2012 & Fall-back  &  Conserved AM & 1.0 & At $X_{\rm H} = 0.01$ & HM2021 \\
model 71 & FD2012 & Fall-back  & Periastron & 1.0 & At $X_{\rm H} = 0.30$ & HM2021 \\
model 72 & FD2012 & Fall-back  &  Conserved AM & 1.0 & At $X_{\rm H} = 0.30$ & HM2021 \\
model 73 & PS2020 & Fall-back & Periastron & 0.3 & At $X_{\rm H} = 0.01$ & HM2021 \\
model 74 & PS2020 & Fall-back &  Conserved AM & 0.3 & At $X_{\rm H} = 0.01$ & HM2021 \\
model 75 & PS2020 & Fall-back & Periastron & 0.3 & At $X_{\rm H} = 0.30$ & HM2021 \\
model 76 & PS2020 & Fall-back &  Conserved AM & 0.3 & At $X_{\rm H} = 0.30$ & HM2021 \\
model 77 & PS2020 & Fall-back & Periastron & 1.0 & At $X_{\rm H} = 0.01$ & HM2021 \\
model 78 & PS2020 & Fall-back &  Conserved AM & 1.0 & At $X_{\rm H} = 0.01$ & HM2021 \\
model 79 & PS2020 & Fall-back & Periastron & 1.0 & At $X_{\rm H} = 0.30$ & HM2021 \\
model 80 & PS2020 & Fall-back &  Conserved AM & 1.0 & At $X_{\rm H} = 0.30$ & HM2021 \\
model 81 & FD2012 & No norm. & Periastron & 0.3 & At $X_{\rm H} = 0.01$ & HM2021 \\
model 82 & FD2012 & No norm. & Conserved AM & 0.3 & At $X_{\rm H} = 0.01$ & HM2021 \\
model 83 & FD2012 & No norm. & Periastron & 0.3 & At $X_{\rm H} = 0.30$ & HM2021 \\
model 84 & FD2012 & No norm. &  Conserved AM & 0.3 & At $X_{\rm H} = 0.30$ & HM2021 \\
model 85 & FD2012 & No norm. & Periastron & 1.0 & At $X_{\rm H} = 0.01$ & HM2021 \\
model 86 & FD2012 & No norm. & Conserved AM & 1.0 & At $X_{\rm H} = 0.01$ & HM2021 \\
model 87 & FD2012 & No norm. & Periastron & 1.0 & At $X_{\rm H} = 0.30$ & HM2021 \\
model 88 & FD2012 & No norm. & Conserved AM & 1.0 & At $X_{\rm H} = 0.30$ & HM2021 \\
model 89 & PS2020 & No norm.  & Periastron & 0.3 & At $X_{\rm H} = 0.01$ & HM2021 \\
model 90 & PS2020 & No norm.  & Conserved AM & 0.3 & At $X_{\rm H} = 0.01$ & HM2021 \\
model 91 & PS2020 & No norm.  & Periastron & 0.3 & At $X_{\rm H} = 0.30$ & HM2021 \\
model 92 & PS2020 & No norm.  &  Conserved AM & 0.3 & At $X_{\rm H} = 0.30$ & HM2021 \\
model 93 & PS2020 & No norm.  & Periastron & 1.0 & At $X_{\rm H} = 0.01$ & HM2021 \\
model 94 & PS2020 & No norm.  &  Conserved AM & 1.0 & At $X_{\rm H} = 0.01$ & HM2021 \\
model 95 & PS2020 & No norm.  & Periastron & 1.0 & At $X_{\rm H} = 0.30$ & HM2021 \\
model 96 & PS2020 & No norm.  &  Conserved AM & 1.0 & At $X_{\rm H} = 0.30$ & HM2021 \\
\\ \hline
\end{tabular}
\caption{Populations run with all the different combinations of parameters (refer to Table~\ref{table:params}) used for this study. FD2012 refers to the delayed SN prescription from \citet{2012ApJ...749...91F} and PS2020 refers to the SN prescription from \citet{2020MNRAS.499.2803P} (refer to Sect.~\ref{sec:sn_mech}). HM2021 refers to the observable wind-fed accretion criterion by \citet{2021PASA...38...56H} and AM stands for angular momentum.}
\end{table*}

\end{appendix}
\end{document}